\documentclass[12pt]{article}
\usepackage{amsmath,cite,epsfig,a4,times,euscript,listings,amssymb}
\usepackage{caption,subcaption}
\usepackage[text-hat-accent]{euler}
\usepackage{graphics}
\usepackage[usenames]{color}
\renewcommand{\baselinestretch}{1.1}
\newcommand{\myTitle}[1]{\begin{center}{\bf\Huge #1}\\[5ex]\end{center}}
\newcommand{\myAuthor}[1]{\begin{center}{\Large #1}\\[2ex]\end{center}}
\newcommand{\myAffiliation}[1]{\\[1ex]{\it\large #1}}

\newcommand{\myDate}{\begin{center}{\large\today}\\[5ex]\end{center}}
\newcommand{\myAbstract}[1]{\begin{center}\renewcommand{\baselinestretch}{1}{\bf Abstract}\\[2ex]\parbox{0.8\linewidth}{\small\hspace{15pt} #1}\end{center}\vspace{\baselineskip}}
\newcommand{\myReport}[1]{\hspace{\fill} #1}
\newcommand{\myPreprint}[1]{}
\newcommand{\myKeywords}[1]{}

\newcommand{\myScript}[1]{\EuScript{#1}}

\newcommand{\deflen}[2]{%      
    \expandafter\newlength\csname #1\endcsname
    \expandafter\setlength\csname #1\endcsname{#2}%
}

\newcommand{\Appendix}[1]{Appendix~\ref{#1}}   
\newcommand{\Section}[1]{Section~\ref{#1}}

\newcommand{\Figure}[1]{Fig.~\ref{#1}}
\newcommand{\Equation}[1]{Eq.~(\ref{#1})}
\newcommand{\ie}{{\it i.e.}}

\newcommand{\eg}{{\it e.g.}}
\newcommand{\Ord}{\myScript{O}}

\newcommand{\imag}{\mathrm{i}}

\newcommand{\vepv}{\epsilon}
\newcommand{\piep}{\pi_{\epsilon}}
\newcommand{\oeps}{o_\epsilon}
\newcommand{\aeps}{a_\epsilon}
\newcommand{\gQCD}{g_\mathrm{s}}
\newcommand{\lop}[2]{#1\!\cdot\!#2}
\newcommand{\Nc}{N_\mathrm{c}}
\newcommand{\CF}{C_F}

\newcommand{\xP}{x}
\newcommand{\xM}{\bar{x}}
\newcommand{\zP}{z}

\newcommand{\sTot}{\nu^2}
\newcommand{\sqrtS}{\nu}
\newcommand{\pP}{P}
\newcommand{\pM}{\bar{P}}
\newcommand{\EP}{E}
\newcommand{\EM}{\bar{E}}
\newcommand{\yrap}{y}

\newcommand{\alphaS}{\alpha_{\mathrm{s}}}

\newcommand{\theperp}{\scriptscriptstyle\perp}

\newcommand{\rperp}{r_{\scriptscriptstyle\perp}}

\newcommand{\qperp}{q_{\scriptscriptstyle\perp}}
\newcommand{\kperp}{k_{\scriptscriptstyle\perp}}
\newcommand{\xperp}{x_{\scriptscriptstyle\perp}}
\newcommand{\Kperp}{K_{\scriptscriptstyle\perp}}
\newcommand{\pperp}[1]{p_{#1\scriptscriptstyle\perp}}
\newcommand{\Born}{\mathrm{B}}
\newcommand{\Real}{\mathrm{R}}
\newcommand{\Virt}{\mathrm{V}}
\newcommand{\Coll}{\mathrm{C}}

\newcommand{\CFF}{\myScript{C}}
\newcommand{\inlbl}{i}
\newcommand{\inbar}{\overline{\imath}}
\newcommand{\JetB}{J_{\Born}}
\newcommand{\JetR}{J_{\Real}}
\newcommand{\mycite}[1]{\cite{#1}}
\newcommand{\kin}{k_{\inlbl}}
\newcommand{\kinBar}{k_{\inbar}}
\newcommand{\pSetN}{\{p\}_{n}}
\newcommand{\pSetNpls}{\{p\}_{n+1}}

\newcommand{\DIV}{\mathrm{div}}

\newcommand{\FIN}{\mathrm{fin}}
\newcommand{\univ}{\mathrm{univ}}
\newcommand{\unf}{\mathrm{unf}}
\renewcommand{\fam}{\mathrm{fam}}
\newcommand{\coll}{\mathrm{coll}}
\newcommand{\LOOP}{\mathrm{loop}}

\newcommand{\UVsubt}{\mathrm{UV}\textrm{-}\mathrm{sub}}
\newcommand{\Vcoef}{\myScript{V}}

\newcommand{\Matel}[1]{\big|\overline{M}_{#1}\big|^2}
\newcommand{\NLO}{\mathrm{NLO}}
\newcommand{\LO}{\mathrm{LO}}
\newcommand{\muY}{\mu_Y}
\newcommand{\muF}{\mu_F}
\newcommand{\muFbar}{\mu_{\bar{F}}}
\newcommand{\kstar}{k_{\star}}
\newcommand{\Fstar}{F}
\newcommand{\VplusR}{[\myScript{V}\!+\!\myScript{R}]}
\newcommand{\VplusRminusC}{[\myScript{V}\!+\!\myScript{R}\!-\!\CFF]}
\newcommand{\Divunf}{\VplusR_{\inlbl}^{\target}}
\newcommand{\ColorC}[1]{C_{#1}}
\newcommand{\target}{\mathrm{targ}}
\newcommand{\project}{\mathrm{proj}}
\newcommand{\resolved}{\mathrm{resolved}}
\newcommand{\unresolv}{\mathrm{div}}

\newcommand{\Green}{\mathrm{Green}}
\newcommand{\ColFac}{\mathrm{CF}}
\newcommand{\intdLxx}{\int_{\xP,\xM}\hspace{-1.5ex}d^2\!\myScript{L}\big(\{f_{\inlbl}\},\{f_{\inbar}\}\big)}
\newcommand{\intdLxkx}{\int_{\xP,\kperp,\xM}\hspace{-4.0ex}d^4\!\myScript{L}\big(F,\{f_{\inbar}\}\big)}
\newcommand{\intdLxk}{\int_{\xP,\kperp}\hspace{-3ex}d^3\!\myScript{L}(F)}
\newcommand{\lambdaZRO}{\lambda_0}
\newcommand{\lambdaONE}{\lambda_1}

\newcommand{\lambdaI}{\lambda_j}

\newcommand{\DELTAZRO}{\delta_0}
\newcommand{\DELTAONE}{\delta_1}

\newcommand{\DELTAI}{\delta_j}
\newcommand{\sigmahat}{\hat{\sigma}}
\newcommand{\kernel}{{K}}
\newcommand{\GreenF}{{G}}
\newcommand{\Kcusp}{\myScript{K}}
\newcommand{\kernelBFKL}{{K_{\mathrm{BFKL}}}}
\newcommand{\kernelBFKLfin}{{K_{\mathrm{BFKL}}^{\mathrm{fin}}}}

\newcommand{\DeltaI}{\Delta I}
\newcommand{\DeltaU}{\Delta U}
\newcommand{\DeltaN}{\Delta\myScript{N}}
\newcommand{\Impact}{I}

\newcommand{\myem}[1]{\underline{#1}}

\begin{document}

\myReport{IFJPAN-IV-2025-1}
%\myPreprint{}\\[2ex]

\myTitle{%
Hybrid high-energy factorization\\[-0.5ex]
and evolution at NLO\\[-0.5ex]
from the high-energy limit of\\[0.5ex]
collinear factorization
}

\myAuthor{%
Andreas~van~Hameren$^{a}\footnote{hameren@ifj.edu.pl}$
and Maxim~Nefedov$^{b}\footnote{maxim.nefedov@desy.de}$%
\myAffiliation{%
$^a$~Institute of Nuclear Physics Polish Academy of Sciences,\\
PL-31342 Krak\'ow, Poland%
\\[1ex]%
$^b$~Physics Department, Ben-Gurion University of the Negev,\\[0.5ex]
Beer Sheva 84105, Israel%
}
}

\myDate

\myAbstract{%
We derive the scheme of NLO computations of generic observables in high-energy hadron-hadron collisions within the framework of high-energy factorization (HEF) with one off-shell initial-state parton, by taking a high-energy limit of the NLO computation in collinear factorization (CF). The NLO terms belonging to the projectile and target are identified and the ambiguity of projectile-target separation is related with the Collins-Soper scale ($\muY$). The NLO unintegrated PDF(UPDF) is constructed in terms of the usual PDFs, and its $\muY$-evolution reproduces the Collins-Soper-Sterman equation in the TMD limit ($|\kperp|\ll \muY$). The resummation of high-energy logarithms is taken care of by the BFKL-Collins-Ellis evolution of the Green's function in the UPDF.
}

\myKeywords{QCD}

%\begin{document}
%

\newpage%
\setcounter{tocdepth}{4}
\tableofcontents

\newpage

\section{\label{intro}Introduction}

The High-Energy Factorization (HEF) formalism had been originally introduced~\mycite{Collins:1991ty,Catani:1990eg,Catani:1992rn,Catani:1994sq} to resum higher-order corrections to partonic coefficient functions and anomalous dimensions of Collinear Factorization (CF), which are enhanced by high-energy logarithms.
For the coefficient functions of CF, those are the logarithms of center-of-mass energy squared ($\hat{s}$) of the partons initiating the hard process with the hard scale $\mu$ in the regime $\Lambda_{\rm QCD}^2 \ll \mu^2 \ll \hat{s}$.
The HEF resummation naturally leads to the notion of ``unintegrated PDF'' (UPDF), which depends not only on the usual longitudinal momentum  fraction ($x$) but also on the transverse momentum ($\kperp$) of the parton initiating the hard process.
The UPDF is a process-independent quantity and should be supplemented by the corresponding ``off-shell'' (or ``space-like'') matrix element of the hard process in question, which also depends on $x$ and $\kperp$.
The behavior of the off-shell matrix element at $|\kperp|^2 \sim \mu^2$ actually determines the coefficients in front of $\alphaS^n \ln^{n-1} (\hat{s}/\mu^2)$ (Leading Logarithmic Approximation, LLA) corrections to the coefficient function of CF, which are resummed by the HEF formalism.

Off-shell Green's functions are not gauge-invariant in QCD, so the corresponding off-shell ($\kperp$-dependent) matrix element should be properly defined.
Such definition relies on gauge-invariant factorization of QCD matrix elements in the Regge limit, when the partonic energy $\hat{s}$ is much larger than any other scale.
The auxiliary parton method~\mycite{vanHameren:2012if} directly implements such factorization for tree-level matrix elements.
The auxiliary-parton definition of the off-shell matrix elements is the most convenient one for implementation into a matrix-element and Monte-Carlo event generator for the LO/LLA HEF computations~\mycite{vanHameren:2014iua,Bury:2015dla,vanHameren:2015bba,vanHameren:2016kkz}
%, such as \KaTie~\mycite{vanHameren:2016kkz}.
%

On more formal level, to define the off-shell matrix elements for HEF one looks for a general operator definition of a Reggeized gluon -- an effective gauge-invariant degree of freedom of QCD in the Regge limit.
Currently several such operator definitions had been studied in the literature, including the Lipatov's gauge-invariant effective field theory (EFT) for Multi-Regge processes in QCD~\mycite{Lipatov:1995pn}, definitions employing Wilson lines~\mycite{Caron-Huot:2013fea,Kotko:2014aba} and definitions emerging within Soft-Collinear Effective theory with Glauber operators~\mycite{Rothstein:2016bsq,Gao:2024qsg}.
The EFT definition had been used in several phenomenological studies within the ``Parton Reggeization Approach''~\mycite{Nefedov:2013ywa,Karpishkov:2017kph} and for the computation of one~\mycite{Chachamis:2012cc,Nefedov:2019mrg,Nefedov:2024swu} and two-loop~\mycite{Chachamis:2013hma} corrections, while the approach of~\mycite{Caron-Huot:2013fea} is more popular in the multiloop community~\mycite{Caola:2021izf,Buccioni:2024gzo}.
The auxiliary parton method is equivalent to these definitions at tree level.

Beyond tree level, a significant progress in understanding of the auxiliary parton approach has been achieved.
The structure of one-loop amplitudes with one off-shell parton was studied~\mycite{vanHameren:2017hxx,Blanco:2020akb,Blanco:2022iai} and the infra-red(IR) divergences in the Regge limit were separated into {\it familiar} and {\it unfamiliar} ones.
The {\it familiar} contribution to the one-loop matrix element with an off-shell parton has the same structure of IR divergences as in the $\kperp\to 0$ limit, while the unfamiliar part has no smooth $\kperp\to 0$ limit, contains additional {\it process-independent} IR divergences and high-energy logarithms.
In~\mycite{vanHameren:2022mtk}, the similar classification of IR divergences arising in the off-shell NLO cross section with an additional parton emission into {\it familiar} and {\it unfamiliar} ones was done.
Recently, the subtraction formalism at NLO has been developed to isolate the IR-divergences from an arbitrary $\kperp$-dependent partonic process at NLO~\mycite{Giachino:2023loc}.

In the present paper, we put together real-emission and virtual contributions in the auxiliary-parton method and separate the resulting expressions for the NLO cross section in CF into {\it projectile} and {\it target} contributions, which seem roughly equivalent to {\it familiar} and {\it unfamiliar} contributions mentioned above, but differ from them in important details.
The {\it projectile} contribution is process-dependent and provides a precise prescription for future NLO computations in HEF with one off-shell parton, also referred to as ``hybrid formalism'' \mycite{Dumitru:2005gt,Marquet:2007vb,Deak:2009xt}\footnote{We would like to emphasize, that in our conventions, the {\it projectile} hadron has {\it negative} rapidity, while the {\it target} flies forward, which is opposite to the convention used in most of the low-$x$ physics literature on the hybrid formalism.}.
The {\it target} contributions are process-independent and provide the expression for UPDF at NLO in $\alphaS$ in terms of usual PDFs.

The crucial tool, which allows us to realize the separation of {\it projectile} and {\it target} contributions is the embedding of the HEF computation into a particular class of infra-red and collinearly (IRC) safe observables in CF, which is discussed in the \Section{sec:CF-HEF} below.
This embedding allows us to realize the IRC-safe definitions of $x$ and $\kperp$ in terms of observable quantities (\Section{sec:x-and-kT}), which are in principle valid in all orders in $\alphaS$, and to parametrize the ambiguity of the separation between projectile and the target in terms of the rapidity separator $Y_\mu$, related with the Collins-Soper scale $\muY$\footnote{Such separation in context of low-$x$ physics for the first time was introduced in Ref.~\mycite{Nefedov:2020ecb}, later in~\mycite{Nefedov:2021vvy} it was shown that evolution in physical rapidity $Y_\mu$ together with the conservation of target light-cone momentum component leads to the LLA Sudakov form factor for the UPDF, which is in agreement with findings of the present paper }.
Requiring the HEF cross section to be independent on the scale $\muY$, we derive the generalized Collins-Soper-Sterman (CSS) equation (\Equation{eq:evol-eq-muY-fin}) for the $\muY$-evolution of UPDF.
This equation reduces to the usual (CSS) equation of Transverse-Momentum Dependent (TMD) factorization~\mycite{Collins:2011ca} in the limit $|\kperp|\ll \muY$.

Here it is a good place to emphasize the connection of our approach with current developments in low-$x$ physics which include the saturation phenomenon.
In a series of recent papers~\mycite{Mueller:2013wwa,Mueller:2012uf, Sun:2015doa,Sun:2014gfa,Mueller:2016xoc,Stasto:2018rci,vanHameren:2020rqt,Taels:2022tza,Caucal:2023nci,Caucal:2024bae,Altinoluk:2024vgg,Duan:2024qev} it has been realized on the basis of one-loop calculations, that the unintegrated PDF in the low-$x$ formalism acquires a Sudakov form factor in the regime $|\kperp|\ll \mu$.
In our paper we confirm this conclusion but also derive the generalized $\muY$-evolution equation which is valid outside of the TMD regime.
We find that the integral of UPDF over $\kperp$ evolves with the scale $\muY$ according to a DGLAP-like equation (\Equation{eq:evol-eqn-xT0-DGLAP}) below, which contains the $z\to 0$ and $z\to 1$ of the exact $P_{gg}$ DGLAP splitting function.
The latter feature relates our equation with the CCFM equation~\mycite{Ciafaloni:1987ur,Catani:1989sg,Catani:1989yc,Marchesini:1994wr}

Finally, in \Section{sec:UPDF}, we separate the NLO result for the target contribution into the NLO {\it impact-factor} and {\it Green's function} parts.
For the Green's function we derive an evolution equation (\Equation{eq:BFKL-CE-0}) which resums large $\ln(1/x) \sim \ln(\hat{s}/\mu^2)$-enchanced corrections to the coefficient function of CF and is equivalent to the BFKL-Collins-Ellis~\mycite{Collins:1991ty} equation.
Convolution of the PDF, NLO partonic impact-factor, and Green's function (\Equation{eq:UPDF-PDF-match}) provides the initial condition for the UPDF at the scale $\muY=|\kperp|$ which can be evolved up or down to the scale $\muY$ of the process under consideration.

The present paper has the following structure.
First, for the reader's convenience, we summarize our notation in \Section{Notation}.
In \Section{sec:ME-limits}, we summarize the limits of tree-level matrix elements with an auxiliary parton and up to one additional real emission, which form a basis of the auxiliary parton approach.
Then in \Section{sec:CF-HEF}, we describe in physical terms the embedding of the NLO HEF calculation into the NLO CF calculation and identify the $\kperp$-factorizable part of the NLO CF calculation.
In \Section{sec:Born}, we describe the relation between LO CF and LO HEF calculations.
In \Section{Sec:virtual}, we separate the virtual contribution into projectile, target and Green's function parts.
In \Section{sec:NLO-proj}, we put together real and virtual NLO CF contributions related with the projectile and corresponding subtraction term which is needed for the correct projectile-target separation.
From the finite NLO CF projectile contribution one obtains the NLO cross section in HEF, and requiring its $\muY$-independence we derive the $\muY$-evolution equation in \Section{sec:UPDF-evol}.
In a similar fashion we add-up together the NLO CF contributions belonging to the target in \Section{sec:target-NLO} and obtain the UPDF at NLO in $\alphaS$.
In \Section{sec:UPDF}, we derive the matching formula for the UPDF at the scale $\muY=|\kperp|$ and check that the NLO UPDF satisfy the $\muY$-evolution equation.
Finally, in \Section{sec:conclusions} we summarize our conclusions and outlook.

Appendices A, B and C, summarize respectively definitions of various constants, splitting functions and computing the convolution integrals used in the present paper. In the appendix D the details of the derivation of the Multi-Regge limit of tree level QCD matrix elements, described in Sec.~\ref{Sec:BFKLlimit}, are collected. In the appendix E the details of the derivation of the $\xperp$-space version of the evolution equation, discussed in Sec.~\ref{sec:xT-space-evol}. The appendix F studies the same evolution equation in $(\xperp,N)$-space, confirming that there is no overlap between resummation of high-energy logarithms and $\muY$-evolution.

\section{\label{Notation}Notation}
%%%%%%%%%%%%%%%%%%%%%%%%%%%%%%%%%%%%%%%%
%%%%%%%%%%%%%%%%%%%%%%%%%%%%%%%%%%%%%%%%
%%%%%%%%%%%%%%%%%%%%%%%%%%%%%%%%%%%%%%%%
%
%
We consider scattering processes for hadrons that are back-to-back with momenta.
%%%%%%%%%%%%%%%%%%%%%%%%%%%%%%%%%%%%%%%%
\begin{equation}
\pP^\mu = (\EP,0,0,\EP)
\quad,\quad
\pM^\mu = (\EM,0,0,-\EM)
\quad,\quad
\sTot = 2\lop{\pP}{\pM}=4\EP\EM
~.
\label{Eq:defPPbar}
\end{equation}
%%%%%%%%%%%%%%%%%%%%%%%%%%%%%%%%%%%%%%%%
%
The Sudakov decomposition of a general momentum $K^\mu$ in terms of $\pP^\mu,\pM^\mu$ is expressed with variables $\xP_K,\xM_K,\Kperp^\mu$ as
%
%%%%%%%%%%%%%%%%%%%%%%%%%%%%%%%%%%%%%%%%
\begin{equation}
K^\mu = \xP_K\pP^\mu + \xM_K\pM^\mu + \Kperp^\mu
\;\;,\;\;
\xP_K = \frac{\lop{K}{\pM}}{\lop{\pP}{\pM}}
\;\;,\;\;
\xM_K = \frac{\lop{K}{\pP}}{\lop{\pP}{\pM}}
\;\;,\;\;
\Kperp^\mu = K^\mu - \xP_K\pP^\mu - \xM_K\pM^\mu
~.
\end{equation}
%%%%%%%%%%%%%%%%%%%%%%%%%%%%%%%%%%%%%%%%
%
We will use the same symbol $\Kperp$ for both the two-dimensional object and its embedding in Minkowski space.
Wether a square is positive will always be made explicit with absolute value symbols.
The rapidity of a momentum can be expressed directly in terms of the Sudakov variables via
%
%%%%%%%%%%%%%%%%%%%%%%%%%%%%%%%%%%%%%%%%
\begin{equation}
%[\mathrm{rapidity}]_{K} = \yrap_K + \yLab
y^{\mathrm{lab}}_{K} = \yrap_K + \frac{1}{2}\ln\frac{\EP}{\EM}
%= \frac{1}{2}\ln\frac{K^+}{K^-}
\quad,\quad
y_K = \frac{1}{2}\ln\frac{\xP_K}{\xM_K}
~.
\end{equation}
%%%%%%%%%%%%%%%%%%%%%%%%%%%%%%%%%%%%%%%%
%
We will rather use the center-of-mass frame rapidity $y_K$ than the laboratory frame rapidity.
The most relevant expression for light-like momenta will be
%
%%%%%%%%%%%%%%%%%%%%%%%%%%%%%%%%%%%%%%%%
\begin{equation}
y_K = \ln\frac{\sqrtS\xP_K}{|\Kperp|}
\quad\textrm{for}\quad
K^2=0
~.
\end{equation}
%%%%%%%%%%%%%%%%%%%%%%%%%%%%%%%%%%%%%%%%
%
The partonic initial states are labelled $\inlbl,\inbar$, such that
the initial state momenta
%
%%%%%%%%%%%%%%%%%%%%%%%%%%%%%%%%%%%%%%%%
\begin{equation}
%\textrm{initial state momenta}\quad
\kin^\mu
\quad,\quad
\kinBar^\mu
\end{equation}
%%%%%%%%%%%%%%%%%%%%%%%%%%%%%%%%%%%%%%%%
%
have components
%
%%%%%%%%%%%%%%%%%%%%%%%%%%%%%%%%%%%%%%%%
\begin{equation}
\xP_{\inlbl}=\xP \;\;,\;\; \xM_{\inlbl}=0
\quad\textrm{and}\quad
\xP_{\inbar}=0 \;\;,\;\; \xM_{\inbar}=\xM
~,
\end{equation}
%%%%%%%%%%%%%%%%%%%%%%%%%%%%%%%%%%%%%%%%
%
and thus have rapidities
%
%%%%%%%%%%%%%%%%%%%%%%%%%%%%%%%%%%%%%%%%
\begin{equation}
\yrap_{\inlbl} \to \infty
\quad,\quad
\yrap_{\inbar} \to -\infty
~.
\end{equation}
%%%%%%%%%%%%%%%%%%%%%%%%%%%%%%%%%%%%%%%%
%
Final-state momenta are listed as
\begin{align}
\pSetN = \{p_1,p_2,\ldots,p_n\}
~.
\end{align}
%%%%%%%%%%%%%%%%%%%%%%%%%%%%%%%%%%%%%%%%
%
This list is used as the last argument for several functions, and is separated from previous arguments by a semicolon, as in  $\big(\ldots\,;\pSetN\big)$.
We can then take momenta out of the final-state list explicitly, by denoting
\begin{align}
\big(\ldots\,;q,\pSetN\big) &= \big(\ldots\,;\{p_1,p_2,\ldots,p_n,q\}\big)
~,
\\
\big(\ldots\,;r,q,\pSetN\big) &= \big(\ldots\,;\{p_1,p_2,\ldots,p_n,r,q\}\big)
~.
\end{align}
%%%%%%%%%%%%%%%%%%%%%%%%%%%%%%%%%%%%%%%%
%
Next before the final-state momenta, initial-state momenta are listed as arguments, for example in the final-state phase space:
%
%%%%%%%%%%%%%%%%%%%%%%%%%%%%%%%%%%%%%%%%
\begin{align}
d\Phi\big(Q\,;\pSetN\big)
&=
\bigg(\prod_{j=1}^n\frac{d^4p_j}{(2\pi)^3}\delta_+(p_j^2-m_j^2)\bigg)
(2\pi)^4\,\delta^4\bigg(Q-\sum_{j=1}^np_j\bigg)
~.
\end{align}
%%%%%%%%%%%%%%%%%%%%%%%%%%%%%%%%%%%%%%%%
%
The ``$d$'' in $d\Phi$ indicates that it is differential in the final-state variables.
We use this notation for all functions that depend on final-state variables.
For example, including matrix element, either on-shell or with a space-like gluon, and flux factor, we denote
%
%%%%%%%%%%%%%%%%%%%%%%%%%%%%%%%%%%%%%%%%
\begin{align}
d\Sigma_{\inlbl\inbar}\big(k_{\inlbl},\kinBar\,;\pSetN\big)
&=
d\Phi\big(k_{\inlbl}+\kinBar\,;\pSetN\big)
\,\frac{\Matel{\inlbl\inbar}\big(k_{\inlbl},\kinBar\,;\pSetN\big)}{2\xP\xM\sTot}
~.
\label{Eq:defSigma}
\end{align}
%%%%%%%%%%%%%%%%%%%%%%%%%%%%%%%%%%%%%%%%
%
The subscripts $\inlbl\inbar$ refer to the type of initial-state partons.
We will want to separate differential phase space volumes of single momenta from this formula, for which we introduce the following notation:
%
%%%%%%%%%%%%%%%%%%%%%%%%%%%%%%%%%%%%%%%%
\begin{align}
\frac{d\Sigma_{\inlbl\inbar}}{dq}\big(k_{\inlbl},\kinBar\,;q,\pSetN\big)
&=
d\Phi\big(k_{\inlbl}+\kinBar-q\,;\pSetN\big)
\,\frac{\Matel{\inlbl\inbar}\big(k_{\inlbl},\kinBar\,;q,\pSetN\big)}{2\xP\xM\sTot}
~,\\
\frac{d\Sigma_{\inlbl\inbar}}{dqdr}\big(k_{\inlbl},\kinBar\,;r,q,\pSetN\big)
&=
d\Phi\big(k_{\inlbl}+\kinBar-r-q\,;\pSetN\big)
\,\frac{\Matel{\inlbl\inbar}\big(k_{\inlbl},\kinBar\,;r,q,\pSetN\big)}{2\xP\xM\sTot}
~.
\end{align}
%%%%%%%%%%%%%%%%%%%%%%%%%%%%%%%%%%%%%%%%
%
With this notation, and light-like $q^\mu,r^\mu$ , we can write
%
%%%%%%%%%%%%%%%%%%%%%%%%%%%%%%%%%%%%%%%%
\begin{align}
d\Sigma_{\inlbl\inbar}\big(k_{\inlbl},\kinBar\,;q,\pSetN\big)
&=
\frac{d^4q}{(2\pi)^3}\delta_+(q^2)\frac{d\Sigma_{\inlbl\inbar}}{dq}\big(k_{\inlbl},\kinBar\,;q,\pSetN\big)
\label{Eq:dSigmadq}~,\\
d\Sigma_{\inlbl\inbar}\big(k_{\inlbl},\kinBar\,;r,q,\pSetN\big)
&=
\frac{d^4r}{(2\pi)^3}\delta_+(r^2)\frac{d^4q}{(2\pi)^3}\delta_+(q^2)\frac{d\Sigma_{\inlbl\inbar}}{dqdr}\big(k_{\inlbl},\kinBar\,;r,q,\pSetN\big)
~.
\end{align}
%%%%%%%%%%%%%%%%%%%%%%%%%%%%%%%%%%%%%%%%
%
Prepending an integral implies integration only over these momenta.
The next arguments before the initial-state variables are parameters like
%
%%%%%%%%%%%%%%%%%%%%%%%%%%%%%%%%%%%%%%%%
\begin{equation}
\vepv=\frac{4-\mathrm{dim}}{2}
\end{equation}
%%%%%%%%%%%%%%%%%%%%%%%%%%%%%%%%%%%%%%%%
%
in dimensional regularization.
For example for virtual contributions with light-like momentum $K_{\inlbl}$, we write
%
%%%%%%%%%%%%%%%%%%%%%%%%%%%%%%%%%%%%%%%%%
\begin{equation}
d\Sigma_{\inlbl\inbar}^{\LOOP}\big(\vepv\,;K_{\inlbl},\kinBar\,;\pSetN\big)
=
d\Phi\big(K_{\inlbl}+\kinBar\,;\pSetN\big)
\,\frac{2\mathrm{Re}\big\{\overline{M^{\dagger}M^{\LOOP}}\big\}\big(\vepv\,;K_{\inlbl},\kinBar\,;\pSetN\big)}{2\xP\xM\sTot}
~.
\end{equation}
%%%%%%%%%%%%%%%%%%%%%%%%%%%%%%%%%%%%%%%%%
%

The Born-level differential cross section in hadron scattering for a partonic final state with momenta $\pSetN$ within collinear factorization (CF) can be compactly written as
%
%%%%%%%%%%%%%%%%%%%%%%%%%%%%%%%%%%%%%%%%
\begin{equation}
d\sigma^{\mathrm{CF},\Born}\big(\pSetN\big)
=\intdLxx\,d\sigma^{\mathrm{CF},\Born}_{\inlbl\inbar}\big(\xP,\xM\,;\pSetN\big)
~,
\end{equation}
%%%%%%%%%%%%%%%%%%%%%%%%%%%%%%%%%%%%%%%%
%
with the abbreviation
%
%%%%%%%%%%%%%%%%%%%%%%%%%%%%%%%%%%%%%%%%
\begin{equation}
\intdLxx = \sum_{\inlbl,\inbar}\int_0^1d\xP\,f_{\inlbl}(\xP)\int_0^1d\xM\,f_{\inbar}(\xM)
~,
\end{equation}
%%%%%%%%%%%%%%%%%%%%%%%%%%%%%%%%%%%%%%%%
%
and
%
%%%%%%%%%%%%%%%%%%%%%%%%%%%%%%%%%%%%%%%%
\begin{equation}
d\sigma^{\mathrm{CF},\Born}_{\inlbl\inbar}\big(\xP,\xM\,;\pSetN\big)
= d\Sigma_{\inlbl\inbar}\big(\xP\pP,\xM\pM\,;\pSetN\big)\,\JetB\big(\pSetN\big)
~.
\end{equation}
%%%%%%%%%%%%%%%%%%%%%%%%%%%%%%%%%%%%%%%%
%
We omit possible scale arguments from the PDFs $f_{\inlbl},f_{\inbar}$, but we remind the reader that the cross section is differential in the final-state variables and we do not imply any restrictions on the scale choice.
The jet function $\JetB$ demands that the number of final-state jets is equal to the number of final-state partons.
We will use a similar compact notation for the integration over initial-state variables with a $k_T$-dependent PDF, as
%
%%%%%%%%%%%%%%%%%%%%%%%%%%%%%%%%%%%%%%%%
\begin{align}
\intdLxk &= \int_0^1d\xP\int\frac{d^2\kperp}{\pi}\,F(\xP,\kperp)
~,\label{Eq:intdLxk}\\
\intdLxkx &= \sum_{\inbar}\int_0^1d\xP\int\frac{d^2\kperp}{\pi}\,F(\xP,\kperp)\int_0^1d\xM\,f_{\inbar}(\xM)
\label{Eq:intdLxkx}
~.
\end{align}
%%%%%%%%%%%%%%%%%%%%%%%%%%%%%%%%%%%%%%%%
%

\section{\label{sec:ME-limits}Limits of auxiliary parton matrix elements}
In this section, we collect expressions for tree-level QCD matrix elements with an auxiliary high-energy parton and their relations with matrix elements with an off-shell parton. The LO case (Sec.~\ref{sec:aux-parton-MEs-LO}) was first considered in~\mycite{vanHameren:2012if}, where the auxiliary parton method was introduced. The limits of matrix elements with an additional real-emission in the final state, needed for NLO computations, are described in \Section{Sec:tripleLambda} -- \Section{sec:space-like-coll} and with the exception of the Multi-Regge limit (\Section{Sec:BFKLlimit}), these limits where already discussed in Ref.~\mycite{vanHameren:2022mtk} in more details, so we keep the discussion in this section relatively brief.

\subsection{Space-like tree-level matrix elements\label{sec:aux-parton-MEs-LO}}
As mentioned, the subscripts $\inlbl\inbar$ refer to the initial-state partons.
A space-like gluon is indicated with ``$\star$'' and has momentum
%
%%%%%%%%%%%%%%%%%%%%%%%%%%%%%%%%%%%%%%%%
\begin{equation}
\kstar^\mu = \xP\pP^\mu + \kperp^\mu
~.
\end{equation}
%%%%%%%%%%%%%%%%%%%%%%%%%%%%%%%%%%%%%%%%
%
A tree-level matrix element with a space-like initial-state gluon is understood to be defined with the help of auxiliary partons as~\mycite{vanHameren:2012if,vanHameren:2022mtk}
%%
%
%%%%%%%%%%%%%%%%%%%%%%%%%%%%%%%%%%%%%%%%
\begin{align}
&\Matel{\star\inbar}\big(\kstar,\kinBar\,;\pSetN\big)
=
\lim_{\Lambda\to\infty}
\frac{\xP^2|\kperp|^2}{\gQCD^2\ColorC{\inlbl}\Lambda^2}
\,\Matel{\inlbl\inbar}\big(k_\Lambda,\kinBar\,;q_\Lambda,\pSetN\big)
~,
\label{Eq:defAuxMethod}
\end{align}
%%%%%%%%%%%%%%%%%%%%%%%%%%%%%%%%%%%%%%%%
%
with
%
%%%%%%%%%%%%%%%%%%%%%%%%%%%%%%%%%%%%%%%%
\begin{equation}
k_\Lambda^2 = q_\Lambda^2 = 0
\quad,\quad
k_\Lambda^\mu-q_\Lambda^\mu \;\overset{\Lambda\to\infty}{=}\; \xP\pP^\mu + \kperp^\mu
~.
\end{equation}
%%%%%%%%%%%%%%%%%%%%%%%%%%%%%%%%%%%%%%%%
%
The following choice satisfies $k_\Lambda^\mu-q_\Lambda^\mu=\xP\pP^\mu + \kperp^\mu$ for any $\Lambda$ and does not require to deform the other momenta by $\Ord\big(\Lambda^{-1}\big)$ before the limit:
%
%%%%%%%%%%%%%%%%%%%%%%%%%%%%%%%%%%%%%%%%
\begin{align}
k_\Lambda^\mu &= \hspace{5.6ex}\Lambda\pP^\mu + \hspace{5.5ex}\alpha\kperp^\mu + \beta\pM^\mu~,
\\
q_\Lambda^\mu &= (\Lambda-\xP)\pP^\mu + (\alpha-1)\kperp^\mu + \beta\pM^\mu~,
\end{align}
%%%%%%%%%%%%%%%%%%%%%%%%%%%%%%%%%%%%%%%%
%
with
%
%%%%%%%%%%%%%%%%%%%%%%%%%%%%%%%%%%%%%%%%
\begin{equation}
\alpha = \frac{1}{1+\sqrt{1-\xP/\Lambda}}
\quad,\quad
\beta = \frac{\alpha^2|\kperp|^2}{\Lambda\sTot}
~.
\end{equation}
%%%%%%%%%%%%%%%%%%%%%%%%%%%%%%%%%%%%%%%%
%
The matrix element $\Matel{\star\inbar}$ is independent of the type of auxiliary parton $\inlbl$ used, partly thanks to the color correction factor $\ColorC{\inlbl}$, defined in \Appendix{Sec:constants}.
The factor $1/\gQCD^2$ corrects the power of the coupling constant.

\subsection{\label{Sec:tripleLambda}Triple-$\Lambda$ limit}
At NLO in $\alphaS$, an additional parton can be emitted and its momentum components can scale in various ways with $\Lambda$. In \mycite{vanHameren:2022mtk} the case when both the momentum of scattered auxiliary parton $q_\Lambda$ and momentum of an additional parton emitted at NLO $r_\Lambda$ scale in the same way had been studied. 
We introduce
%
%%%%%%%%%%%%%%%%%%%%%%%%%%%%%%%%%%%%%%%%
\begin{align}
k_\Lambda^\mu &= \hspace{12.2ex}\Lambda\pP^\mu~,
\\
r_\Lambda^\mu &= \hspace{5.5ex}z(\Lambda-\xP)\pP^\mu + \rperp^\mu \hspace{5.2ex}+  \xM_r\pM^\mu~,
\\
q_\Lambda^\mu &= (1-z)(\Lambda-\xP)\pP^\mu - \kperp^\mu -\rperp^\mu + \xM_q\pM^\mu~,
\end{align}
%%%%%%%%%%%%%%%%%%%%%%%%%%%%%%%%%%%%%%%%
%
where $\xM_q,\xM_r$ are such that $q^2=r^2=0$, and vanish as $1/\Lambda$.
These momenta satisfy.
%
%%%%%%%%%%%%%%%%%%%%%%%%%%%%%%%%%%%%%%%%
\begin{equation}
k_\Lambda^\mu - r_\Lambda^\mu - q_\Lambda^\mu
\;\overset{\Lambda\to\infty}{\longrightarrow}\;
\xP\pP^\mu + \kperp^\mu
~.
\end{equation}
%%%%%%%%%%%%%%%%%%%%%%%%%%%%%%%%%%%%%%%%
%
For the matrix element in this limit we have:
%
%%%%%%%%%%%%%%%%%%%%%%%%%%%%%%%%%%%%%%%%
\begin{align}
&
\frac{\xP^2|\kperp|^2}{\gQCD^4\ColorC{\inlbl}\Lambda^2}
\,\Matel{\inlbl\inbar}\big(k_\Lambda,\kinBar\,;r_\Lambda,q_\Lambda,\pSetN\big)
\;\overset{\Lambda\to\infty}{\longrightarrow}\;
2z(1-z)\,\myScript{Q}_{\inlbl}(z,\rperp)\,
\Matel{\star\inbar}\big(\kstar,\kinBar\,;\pSetN\big)
\end{align}
%%%%%%%%%%%%%%%%%%%%%%%%%%%%%%%%%%%%%%%%
%
where
%
%%%%%%%%%%%%%%%%%%%%%%%%%%%%%%%%%%%%%%%%
\begin{align}
\myScript{Q}_{\inlbl}(z,\rperp)
&=
\myScript{P}_{\inlbl}(z)
\bigg(\frac{c_q\,|\kperp|^2}{|\rperp|^2|\rperp+\kperp|^2}
      +\frac{c_q(1-z)^2\,|\kperp|^2}{|\rperp+\kperp|^2|\rperp+z\kperp|^2}
      +\frac{c_rz^2\,|\kperp|^2}{|\rperp|^2|\rperp+z\kperp|^2}\bigg)
~,
\end{align}
%%%%%%%%%%%%%%%%%%%%%%%%%%%%%%%%%%%%%%%%
%
with
%
%%%%%%%%%%%%%%%%%%%%%%%%%%%%%%%%%%%%%%%%
\begin{align}
\textrm{$k_\Lambda,q_\Lambda$ quarks, $r_\Lambda$ gluon:}&\quad
  \myScript{P}_{\inlbl}(z) = \frac{1}{z}+\frac{(1-\vepv)z-2}{2}
  \hspace{0.0ex}\quad,\quad c_q=\Nc\;,\; c_r=\frac{-1}{\Nc}
\\
\textrm{$k_\Lambda,q_\Lambda,r_\Lambda$ gluons:}&\quad
  \myScript{P}_{\inlbl}(z) = \frac{1}{z}+\frac{1}{1-z}-2+z(1-z)
  \quad,\quad c_q=c_r=\Nc
\\
\textrm{$k_\Lambda$ gluon, $q_\Lambda,r_\Lambda$ q-qbar pair:}&\quad
  \myScript{P}_{\inlbl}(z) = \frac{1}{2}-\frac{z(1-z)}{1-\vepv}
  \hspace{0.0ex}\quad,\quad c_q=\frac{1}{2}\;,\; c_r=\frac{-1}{2\Nc^2}
\end{align}
%%%%%%%%%%%%%%%%%%%%%%%%%%%%%%%%%%%%%%%%
%
The definitions of the constants and functions $\myScript{Q}_{\inlbl},\myScript{P}_{\inlbl}$ have been changed a bit compared to \mycite{vanHameren:2022mtk} to tidy up the formulas.

\subsection{\label{Sec:BFKLlimit}Multi-Regge limit}
Another important case, which has not been considered in \mycite{vanHameren:2022mtk}, is the situation when $q_\Lambda$ and $r_\Lambda$ both grow with $\Lambda$ but with different rates, such that $\bar{P}\cdot r_\Lambda \ll \bar{P}\cdot q_\Lambda $. This situation is called {\it Multi-Regge kinematics (MRK)} in BFKL physics and it is the region of real-emission phase-space from which large logarithmic corrections $\sim \ln\Lambda$ come. As an example of scaling of momenta, belonging to MRK let us consider:
%
%%%%%%%%%%%%%%%%%%%%%%%%%%%%%%%%%%%%%%%%
\begin{align}
k_\Lambda^\mu &= \hspace{12.5ex}\Lambda\pP^\mu~,
\\
r_\Lambda^\mu &= \hspace{10.0ex}\sqrt{\Lambda}\,\pP^\mu + \rperp^\mu \hspace{5.2ex}+  \xM_r\pM^\mu~,
\\
q_\Lambda^\mu &= \big(\Lambda-\sqrt{\Lambda}-\xP\big)\pP^\mu - \kperp^\mu -\rperp^\mu + \xM_q\pM^\mu~,
\end{align}
%%%%%%%%%%%%%%%%%%%%%%%%%%%%%%%%%%%%%%%%
%
where $\xM_q,\xM_r$ are again such that $q^2=r^2=0$ and vanish as $1/\Lambda,1/\sqrt{\Lambda}$ respectively.
The MRK limit does not depend on the exact behavior of $r^\mu$, and we can choose any $t(\Lambda)$ instead of $\sqrt{\Lambda}$, as long as $t(\Lambda)\to\infty$ while $t(\Lambda)/\Lambda\to0$.
For the matrix element in this limit we have:
%
%%%%%%%%%%%%%%%%%%%%%%%%%%%%%%%%%%%%%%%%
\begin{align}
&
\frac{\xP^2|\kperp|^2}{\gQCD^4\ColorC{\inlbl}\Lambda^2}
\,\Matel{\inlbl\inbar}\big(k_\Lambda,\kinBar\,;r_\Lambda,q_\Lambda,\pSetN\big)
\;\overset{\Lambda\to\infty}{\longrightarrow}\;
4\Nc\,\frac{|\kperp|^2}{|\rperp|^2|\rperp+\kperp|^2}\,
\Matel{\star\inbar}\big(\kstar,\kinBar\,;\pSetN\big)
~,
\label{Eq:BFKLME}
\end{align}
%%%%%%%%%%%%%%%%%%%%%%%%%%%%%%%%%%%%%%%%
%
for either case $k_\Lambda,q_\Lambda$ quarks, $r_\Lambda$ gluon, and $k_\Lambda,q_\Lambda,r_\Lambda$ gluons. The universal factor appearing in (\ref{Eq:BFKLME}) is just a square of Lipatov's vertex for the gluon emission in MRK. Unsurprisingly, the limit $z\to 0$ of expressions given in the previous subsection also gives the same result, since the previous limit overlaps with MRK if \eg\ $z\sim 1/\sqrt{\Lambda}$.
%
%This limit directly gives the $z=0$ case for the previous limit.
%
We give some more details to the derivation of \Equation{Eq:BFKLME} in \Appendix{Sec:BFKLdetails}.

\subsection{Space-like collinear limit\label{sec:space-like-coll}}
Finally, one should also consider the case when the real emission momentum $r$ does not scale with $\Lambda$. For this case in~\mycite{vanHameren:2022mtk,Giachino:2023loc}, it was shown that all single collinear and soft limits for matrix elements with a space-like gluon are very similar to those of completely on-shell matrix elements.
They have the same structure of singular factors times color-or spin-correlated matrix elements with a final-state parton fewer, where the only difference is that these matrix elements have the space-like initial state gluon again.
The singular factors are the same.
Only for the limit in which a final-state gluon becomes collinear to $\pP$, that is collinear to the longitudinal component of the space-like gluon, the singular factor is different, in fact simpler, than if the initial-state gluon were on-shell.
We have
%
%%%%%%%%%%%%%%%%%%%%%%%%%%%%%%%%%%%%%%%%
\begin{equation}
\Matel{\star\inbar}\big(\kstar,\kinBar\,;r,\pSetN\big)
\;\overset{\xM_r\to0}{\longrightarrow}\;
%\frac{1}{\lop{\xP\pP}{r}}\,\frac{2\Nc\,\xP^3}{\xP_r(\xP-\xP_r)^2}\,
%\Matel{\star\inbar}\big(\kstar-\xP_r\pP,\kinBar\,;\pSetN\big)
\gQCD^2\,\frac{4\Nc}{|\rperp|^2(1-\xP_r/\xP)^2}\,
\Matel{\star\inbar}\big(\kstar-\xP_r\pP-\rperp,\kinBar\,;\pSetN\big)
~,
\label{Eq:collim}
\end{equation}
%%%%%%%%%%%%%%%%%%%%%%%%%%%%%%%%%%%%%%%%
%
where of course $\xM_r\to0$ implies $|\rperp|\to0$.
Besides the limit $\xP_r\pP^\mu$ of $r^\mu$, we also subtracted the ``recoil'' $\rperp^\mu$ from the momentum of the space-like gluon.
This way, we make the deformation of $\Ord\big(|\rperp|\big)$ explicit, rather than {\em implying} it to be applied to the other momenta in order to ensure momentum conservation and on-shellness.
For completely on-shell matrix elements, the deformation cannot be as simple as a subtraction from a single momentum.
As a result, the implicit left-over deformation is now only of $\Ord\big(|\rperp|^2\big)$.
We could also have subtracted $\xM\pM^\mu$, as is done in the subtraction scheme presented in~\mycite{Giachino:2023loc}, making the deformation completely explicit, but for the following discussion it is better not to.

\section{\label{sec:CF-HEF}High-Energy Factorization from Collinear Factorization}
The hybrid high-energy factorization (\ie\ HEF with one off-shell parton) can be embedded into the following observable, which is infrared and collinear safe and therefore can be computed in collinear factorization (CF).
Let us consider the collision of two hadrons, with production of the final state of interest $\myScript{H}$:
\begin{equation}
    h(\lambda \pP) + h(\pM) \to \myScript{H} + \myScript{X}
~,
\end{equation}
where the final state of interest $\myScript{H}$ is defined by the jet-definition function $\JetB\big(\pSetN\big)$ at LO in CF, which is correspondingly generalized to NLO.
Eventually, we want to consider the high-energy limit regarding the first hadron, which we will achieve by making the parameter $\lambda$ large.
%

%\newpage
\subsection{\label{sec:x-and-kT}Definition of $\xP$ and $\kperp$}
We assume that there is a natural rapidity $Y_\mu$ associated with $\myScript{H}$, which separates the event into ``target'' and ``projectile'' parts (see Fig.~\ref{fig:CF-process}).
The projectile is moving with negative rapidity, and the target with positive rapidity.
%
%Then the following kinematic variables characterizing the event
The event can be characterized by the kinematic variables:
\noindent
\begin{minipage}[l]{0.45\linewidth}
\centering
\epsfig{figure=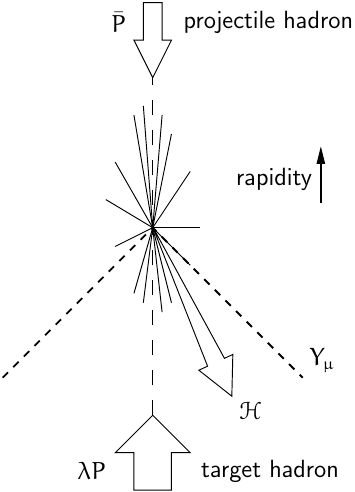,width=0.80\linewidth}
\captionof{figure}{High-energy hadronic collision event with production of the system of interest $\myScript{H}$.}
\label{fig:CF-process}
\end{minipage}
\hspace{\fill}
\begin{minipage}[r]{0.50\linewidth}
\begin{align}
    \xP &=\sum_{j}\theta\big(y_j<Y_\mu\big)\,\frac{\lop{p_j}{\pM}}{\lop{\pP}{\pM}} 
\label{Eq:x-def-exp}~, \\
    \kperp&=-\sum_{j}\theta\big(y_j<Y_\mu\big)\, p_{j\theperp}
\label{Eq:kperp-def-exp}~,
\end{align}
where the summation includes all momenta of particles $p_j$ with rapidity $y_j<Y_\mu$, including those belonging to $\myScript{H}$~\footnotemark.
The definitions of variables $\xP$ and $\kperp$ are infrared and collinear (IRC) safe because collinear particles with low $p_{\theperp}$ (which may not be captured by the detector) give small contribution to $\kperp$, and the contribution of particles with large negative rapidity (which may not be captured by the detector) to the variable $\xP$ is suppressed as $e^{-|y_j|}$, while contribution of large positive rapidities is removed by the cut $y_j<Y_\mu$, so the main contribution to $x$ comes from high-$\pperp{}$ particles with rapidities close to $Y_\mu$.
\end{minipage}
\footnotetext{%
The definition of $\xP$ is similar to the definition of variable $z=(p_{J/\psi} P)/(p_\gamma P)$, which was used by experiments on $J/\psi$-photoproduction at HERA~\cite{H1:2010udv}, where they had no access to $p_\gamma$ because the final-state electron was flying to the beam-pipe for photoproduction events with $Q^2<1$ GeV$^2$ and they had to reconstruct photon momentum from other particles in the event.
}%
\vspace{0.5ex}
Knowing $\xP$, one can associate rapidity scale $\muY$ (or Collins-Soper scale, often denoted by $\zeta=\muY^2$) to the rapidity $Y_\mu$ as
\begin{equation}
\muY = \sqrtS\xP\,e^{-Y_\mu}
\quad\Leftrightarrow\quad
Y_\mu = \ln\frac{\sqrtS\xP}{\muY}
~.\label{Eq:mu_Y-def}
\end{equation}
where variable $\sqrtS$ was defined in \Equation{Eq:defPPbar}.

Due to IRC-safety of variables $x$ and $\kperp$, the corresponding hadronic differential cross section:
\begin{equation}
    \frac{d\sigma^{\ColFac}_{\lambda}}{d\xP d^2\kperp}
~,
\end{equation}
which can also be differential further in variables associated with $\myScript{H}$, should be computable in CF, at least up to NLO, and in the limit:
\begin{equation}
    \lambda\to \infty
\quad,\quad
\xP, \kperp\textrm{-- fixed}~.
\label{Eq:lam-limit}
\end{equation}
The corresponding cross section is equal to:
\begin{equation}
\frac{d\sigma^{\ColFac}_{\lambda}}{d\xP d^2\kperp}\big(\xP,\kperp,\ldots\big)
 = \sum_{\inlbl,\inbar}\int_0^1dX\,f_{\inlbl}(X) \int_0^1d\xM\,f_{\inbar}(\xM)\,
   \frac{d\sigmahat^{\mathrm{CF}}_{\inlbl\inbar}}{d\xP d^2\kperp}\big(\lambda X,\xM\,;\xP,\kperp,\ldots\big)
~,
\end{equation}
where by ``$\ldots$'' we denote final-state kinematic variables defining $\myScript{H}$, which are independent from $\xP$ and $\kperp$.
For example, if $\myScript{H}=$ (1 jet), then it will be $p_{\theperp}^{\mathrm{jet}}$, $\xM^{\mathrm{jet}}$, and  $\xP^{\mathrm{jet}}\leq \xP$.
From now on, we will not explicate dependence on $\xP,\kperp$ via arguments anymore if is explicit via the derivative notation already.

In the auxiliary parton approach, the parton with momentum $\lambda X\pP^\mu$ is the initial-state auxiliary parton, and the momentum difference $(\lambda X-\xP)\pP^\mu$ is carried by final-state auxiliary partons.
Differentiating the cross section with respect to $\xP,\kperp$ means undoing part of the integration over the final-state auxiliary parton momenta.
For example the Born level formula for the cross section then can be more explicitly written as
%
%%%%%%%%%%%%%%%%%%%%%%%%%%%%%%%%%%%%%%%%
\begin{align}
\frac{d\sigma^{\ColFac,\Born}_{\lambda}\big(\pSetN\big)}{d\xP d^2\kperp}
&=\sum_{\inlbl,\inbar}\int_0^1dX\,f_{\inlbl}(X)\int_0^1d\xM\,f_{\inbar}(\xM)
  \,\frac{d\sigmahat^{\Born}_{\inlbl\inbar}\big(\lambda X,\xM\,;\pSetN\big)}{d\xP d^2\kperp}
~,
\end{align}
%%%%%%%%%%%%%%%%%%%%%%%%%%%%%%%%%%%%%%%%
%
with
%
%%%%%%%%%%%%%%%%%%%%%%%%%%%%%%%%%%%%%%%%
\begin{align}
\frac{d\sigmahat^{\Born}_{\inlbl\inbar}\big(\Lambda,\xM\,;\pSetN\big)}{d\xP d^2\kperp}
&=
 \int\frac{d^4q}{(2\pi)^3}\,\delta_+(q^2)
 \,2\xP\,\delta(\xP+\xP_q-\Lambda)\,\delta^2(\kperp+\qperp)
\label{Eq:defBorn}
\\&\hspace{20ex}\times
 \frac{d\Sigma_{\inlbl\inbar}}{dq}\big(\Lambda\pP,\xM\pM\,;q,\pSetN\big)
\,\JetB\big(\pSetN\big)
~,
\notag\end{align}
%%%%%%%%%%%%%%%%%%%%%%%%%%%%%%%%%%%%%%%%
%
where $q$ is the momentum of the final-state auxiliary parton.
We made use of the notation introduced in \Equation{Eq:dSigmadq}.
The arguments of $d\sigma^{\ColFac,\Born}_{\lambda}$ and $d\sigma^{\Born}_{\inlbl\inbar}$ are separated by a semicolon, where the ones before are (variables related to) initial-state momenta, and ones after the semicolon are final-state momenta.
The $\delta(\xP+\xP_q-\Lambda)$ forces the final-state auxiliary parton momentum $q$ to absorb most of the large momentum component
%%%%%%%%%%%%%%%%%%%%%%%%%%%%%%%%%%%%%%%%
\begin{equation}
 \Lambda=\lambda X
~.
\end{equation}
%%%%%%%%%%%%%%%%%%%%%%%%%%%%%%%%%%%%%%%%
%
This momentum $q$ does not enter the jet function, and its transverse momentum is exactly $-\kperp$.

\subsection{The $k_T$-factorizable contribution \label{sec:kT-fact-contr-def}}
We want to take the limit $\lambda\to\infty$ on the level of CF partonic cross section, but the integration over $X$ involves values for which $\lambda X$ is not large.
To address this issue, first of all we need to introduce the sequences of symbols
%
%%%%%%%%%%%%%%%%%%%%%%%%%%%%%%%%%%%%%%%%
\begin{equation}
\left.\begin{aligned}
%%
%&\lambda \succ \lambdaZRO \succ \lambdaONE \succ \cdots \succ \lambdaINF = \frac{\muY}{|\kperp|}
%\\
%&\DELTA \succ \DELTAZRO \succ \DELTAONE \succ \cdots \succ \DELTAINF = \frac{\xP}{\lambda}
%\end{aligned}\right\rbrace
%\quad\textrm{related via}\quad
%\DELTAI = \frac{\lambdaI\xP|\kperp|}{\lambda\muY}
%%%
%%&\lambda \succ \lambdaZRO \succ \lambdaONE \succ \cdots \succ \lambdaINF = 0
%%\\
%%&\DELTA \succ \DELTAZRO \succ \DELTAONE \succ \cdots \succ \DELTAINF = \frac{\xP}{\lambda}
%%\end{aligned}\right\rbrace
%%\quad\textrm{related via}\quad
%%\DELTAI = \frac{\lambdaI\xP|\kperp|}{\lambda\muY} + \frac{\xP}{\lambda}
%
&\lambda \succ \lambdaZRO \succ \lambdaONE \to \infty
\\
&\hspace{4.2ex}\DELTAZRO \succ \DELTAONE \succ \xP/\lambda
\end{aligned}\right\rbrace
\quad\textrm{related via}\quad
\DELTAI = \frac{\lambdaI\xP|\kperp|}{\lambda\muY} %+ \frac{\xP}{\lambda}
~.
\end{equation}
%%%%%%%%%%%%%%%%%%%%%%%%%%%%%%%%%%%%%%%%
%
The notation $b\succ a$ means $a/b\to0$ for both cases $a,b\to0$ and $a,b\to\infty$. 
The value
%%%%%%%%%%%%%%%%%%%%%%%%%%%%%%%%%%%%%%%%
\begin{equation}
\xP/\lambda\;\;\textrm{is the absolute minimum for}\;\;X
\nonumber
\end{equation}
%%%%%%%%%%%%%%%%%%%%%%%%%%%%%%%%%%%%%%%%
%
dictated by the demand that the $\xP$-fraction of the auxiliary partons is positive.
The parameters relate separators in rapidity to separators in $X$ following
%
%%%%%%%%%%%%%%%%%%%%%%%%%%%%%%%%%%%%%%%%
\begin{equation}
y_q = \ln\frac{\sqrtS(\lambda X-\xP)}{|\kperp|} \;>\; Y_\mu + \ln\lambdaI
\quad\Rightarrow\quad
X > \DELTAI
~.
\end{equation}
%%%%%%%%%%%%%%%%%%%%%%%%%%%%%%%%%%%%%%%%
%
For the converse to be true, the definition of $\DELTAI$ in terms of $\lambdaI$ requires the former to be multiplied by a factor $\big(1+\muY/(\lambdaI|\kperp|)\big)$, which we however neglect.

We write the cross section in the most rudimentary form still involving the relevant variables.
Sticking to leading order now, we write
%
%%%%%%%%%%%%%%%%%%%%%%%%%%%%%%%%%%%%%%%%
\begin{equation}
d{\sigma}^\LO=\int_0^1dX\,f(X)\,d\hat{\sigma}^\LO(\lambda X)
~.
\end{equation}
%%%%%%%%%%%%%%%%%%%%%%%%%%%%%%%%%%%%%%%%
%
The essential ingredient of our approach is that we can take $\lambda\to\infty$ inside the partonic cross section $d\hat{\sigma}(\lambda X)$ and neglect higher powers of $1/(\lambda X)$.
The integration over $X$, however, includes the region where $1/(\lambda X)$ is not small, and indeed neglecting higher powers in the integrand is incorrect if $f(X)$ is not integrable down to $X\to0$.
To cure this problem, we decompose
%%%%%%%%%%%%%%%%%%%%%%%%%%%%%%%%%%%%%%%%
\begin{equation}
f(X) = f^{>}(X) + f^{<}(X)
~,
\label{Eq:kTfactorizable1}
\end{equation}
%%%%%%%%%%%%%%%%%%%%%%%%%%%%%%%%%%%%%%%%
%
such that $f^{>}(X)$ is safe to use.
We could demand that $f^{>}(X)$ is integrable down to $X\to0$ for this purpose, but instead we put the {\em weaker} definition
%
%%%%%%%%%%%%%%%%%%%%%%%%%%%%%%%%%%%%%%%%
\begin{equation}
f^{>}(X) = f(X)\,\theta(X>\DELTAZRO)
\quad,\quad
f^{<}(X) = f(X)\,\theta(X<\DELTAZRO)
~,
\end{equation}
%%%%%%%%%%%%%%%%%%%%%%%%%%%%%%%%%%%%%%%%
%
with $\DELTAZRO\to0$, but such that $\DELTAZRO\succ\xP/\lambda$.
The \myem{$k_T$-factorizable} contribution is given by
%
%%%%%%%%%%%%%%%%%%%%%%%%%%%%%%%%%%%%%%%%
\begin{equation}
\int_0^1dX\,f^{>}(X)\,d\hat{\sigma}^\LO(\lambda X)
\;\overset{\lambda\to\infty}{\longrightarrow}\;
\bigg[\int_0^1dX\,f^{>}(X)\bigg]d\hat{\sigma}^\LO_{k_T\mathrm{-fact}}
~,
\end{equation}
%%%%%%%%%%%%%%%%%%%%%%%%%%%%%%%%%%%%%%%%
%
where the exact form of $d\hat{\sigma}^\LO_{k_T\mathrm{-fact}}$ will be derived in \Section{sec:Born}.
The contribution with $f^{<}(X)$ is {\it non-$k_T$-factorizable}.
By its kinematics, this contribution is similar to the diffractive contribution with the break-up of the hadron and a large rapidity gap.
But the gap in this case arises not due to the color-singlet exchange in $t$-channel but just as a (relatively rare) fluctuation in a normal inclusive event.

The issue of separation of the $\kperp$-factorizable contribution requires more care at NLO, due to collinear divergences. We know according to collinear factorization that all divergences cancel in
%
%%%%%%%%%%%%%%%%%%%%%%%%%%%%%%%%%%%%%%%%
\begin{equation}
d{\sigma}^\NLO
=
\int_0^1dX\,f(X)\,d\hat{\sigma}^\NLO(\lambda X)
+
\int_0^1dX\,\big[\myScript{Z}^{(1)}\otimes f\big](X)\,d\hat{\sigma}^\LO(\lambda X)
~,
\end{equation}
%%%%%%%%%%%%%%%%%%%%%%%%%%%%%%%%%%%%%%%%
%
where $d\hat{\sigma}^\NLO$ represents the NLO corrections to the partonic cross section, and the second term represents the collinear counter term for the ``target side'', as prescribed by the collinear factorization theorem~\mycite{Collins:2011zzd}, with $\myScript{Z}^{(1)}$ encapsulating the $1/\epsilon$-poles and LO DGLAP splitting functions.
The equivalent of this term regarding the ``projectile side'' is irrelevant for this discussion and assumed to be inside the first term.
This formula of course faces the same obstruction for directly taking $\lambda\to\infty$, and we decompose it again as
%
%%%%%%%%%%%%%%%%%%%%%%%%%%%%%%%%%%%%%%%%
\begin{align}
d{\sigma}^\NLO
=
&\int_0^1dX\,f^{>}(X)\,d\hat{\sigma}^\NLO(\lambda X)
+
\int_0^1dX\,\big[\myScript{Z}^{(1)}\otimes f^{>}\big](X)\,d\hat{\sigma}^\LO(\lambda X)
\\
+&\int_0^1dX\,f^{<}(X)\,d\hat{\sigma}^\NLO(\lambda X)
+
\int_0^1dX\,\big[\myScript{Z}^{(1)}\otimes f^{<}\big](X)\,d\hat{\sigma}^\LO(\lambda X)
~.
\end{align}
%%%%%%%%%%%%%%%%%%%%%%%%%%%%%%%%%%%%%%%%
%
One could think that the first line gives the $k_T$-factorizable NLO contribution, however, it turns out that not all poles in $\vepv$ cancel there.
We need to decompose further, and write
%
%%%%%%%%%%%%%%%%%%%%%%%%%%%%%%%%%%%%%%%%
\begin{align}
d{\sigma}^\NLO
&=
\int_0^1dX\,f^{>}(X)\,d\hat{\sigma}^\NLO(\lambda X)
+
\int_{\DELTAONE}^1\!\!dX\,\big[\myScript{Z}^{(1)}\otimes f^{>}\big](X)\,d\hat{\sigma}^\LO(\lambda X)
\notag\\&+
\int_0^1dX\,f^{<}(X)\,d\hat{\sigma}^\NLO(\lambda X)
+
\int_0^1dX\,\big[\myScript{Z}^{(1)}\otimes f^{<}\big](X)\,d\hat{\sigma}^\LO(\lambda X)
\label{Eq:kTfactorizable3}\\&+
\int_{0}^{\DELTAONE}\!\!dX\,\big[\myScript{Z}^{(1)}\otimes f^{>}\big](X)\,d\hat{\sigma}^\LO(\lambda X)
\notag
\end{align}
%%%%%%%%%%%%%%%%%%%%%%%%%%%%%%%%%%%%%%%%
%
with $\DELTAZRO\succ\DELTAONE$.
The first line of (\ref{Eq:kTfactorizable3}) proves to be the $k_T$-factorizable NLO contribution, and in particular we find
%
%%%%%%%%%%%%%%%%%%%%%%%%%%%%%%%%%%%%%%%%
\begin{equation}
\int_{\DELTAONE}^1\!\!dX\,\big[\myScript{Z}^{(1)}\otimes f^{>}\big](X)\,d\hat{\sigma}^\LO(\lambda X)
\;\overset{\lambda\to\infty}{\longrightarrow}\;
\bigg[\int_0^1dX\,f^{>}(X)\,p(\DELTAONE/X)\bigg]d\hat{\sigma}^\LO_{k_T\mathrm{-fact}}
~,
\end{equation}
%%%%%%%%%%%%%%%%%%%%%%%%%%%%%%%%%%%%%%%%
%
where $p$ is a function which can safely be expanded in $\DELTAONE$ underneath the integral, and is expressed in terms of DGLAP splitting functions inside $\myScript{Z}^{(1)}$.
It does produce a term proportional to $\ln\DELTAONE$, which is exactly necessary to cancel against a similar contribution inside the first term of (\ref{Eq:kTfactorizable3}).
The latter one appears because, as we will see, the proper definition of the first term of (\ref{Eq:kTfactorizable3}) requires the introduction of a rapidity separator $Y_\mu+\ln\lambdaONE$ for the radiation.
%
%{\color{red}%
%{\bf There are two remarks at order.}
%%
%Firstly, we want to be able to take $\lambdaONE\to1$, and including the factor $\big(1+\muY/(\lambdaONE|\kperp|)\big)$ in the definition of $\DELTAONE$ would then produce an extra term $\ln\big(1+\muY/|\kperp|\big)$ coming from $\ln\DELTAONE$.
%%
%In order to avoid this, $\DELTAONE$ must be, and can be, chosen without that factor.
%%
%}%\color{red}

\begin{figure}
    \centering
    \epsfig{figure=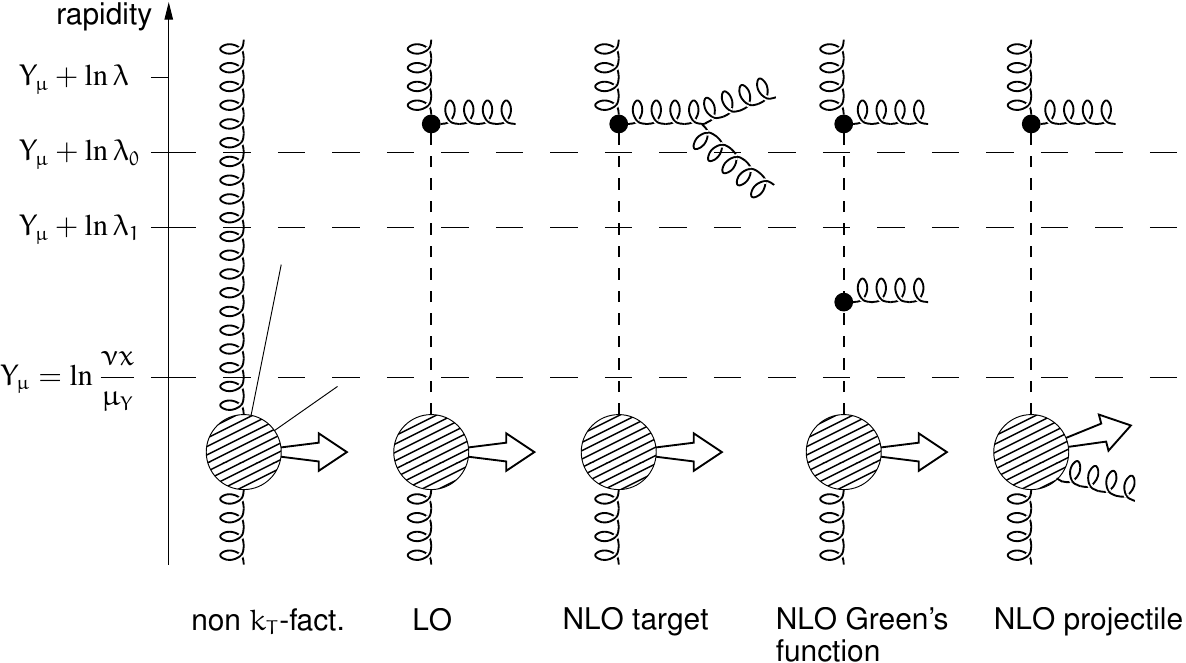,width=0.90\linewidth}
    \caption{Various contributions to the $k_T$-factorizable part of the cross section in the $\lambda\to\infty$ limit. The vertical dashed lines designate the gauge-invariant factorization of projectile, target and the contribution of an emission in MRK, which can be formulated in terms of the t-channel exchange of the off-shell ``Reggeized'' gluon.
}
    \label{fig:contrs-kT}
\end{figure}

In a bit more detail, 
we split the LO and NLO CF contributions to the {\bf $k_T$-factorizable part} of the cross section into the following contributions which have definite interpretation in terms of HEF, see Fig.~\ref{fig:contrs-kT}.
The contributions are the following:
\begin{enumerate}
    \item {\bf The LO contribution:} the rapidity of auxiliary parton is $y_q > Y_\mu + \ln\lambdaZRO$
    \item {\bf The NLO target impact-factor (IF) contribution:} corresponds to the {\it unfamiliar real} contribution from~\mycite{vanHameren:2022mtk}, but with the cuts on both partons $y_{q,r} > Y_\mu + \ln\lambdaONE$ .
    \item {\bf The Green's function contribution:} corresponds to $y_q > Y_\mu + \ln\lambdaONE$ while $Y_\mu> y_r > Y_\mu + \ln\lambdaONE$.
    \item {\bf The NLO projectile contribution:} has $y_r < Y_\mu$. This restriction is implemented by subtraction of the corresponding ``collinear'' asymptotics of the matrix element, with $y_r > Y_\mu$, from what in~\mycite{vanHameren:2022mtk} was called the {\it familiar real} contribution. 
\end{enumerate}
The parameter $\lambdaONE$ separates the target IF and Green's function contributions.
We will also use this parameter to remove energy logarithms from the IF contribution and resum them into the Green's function.

\section{\label{sec:Born}Born contribution}
Let us repeat the Born level formula for the cross section, now with $f^{>}_{\inlbl}(X)$ instead of $f_{\inlbl}(X)$ in order to ensure $k_T$-factorizability:
%
%%%%%%%%%%%%%%%%%%%%%%%%%%%%%%%%%%%%%%%%
\begin{align}
\frac{d\sigma^{\ColFac,\Born}_{\lambda}\big(\pSetN\big)}{d\xP d^2\kperp}
&=\sum_{\inlbl,\inbar}\int_0^1dX\,f^{>}_{\inlbl}(X)\int_0^1d\xM\,f_{\inbar}(\xM)
  \,\frac{d\sigmahat^{\Born}_{\inlbl\inbar}\big(\lambda X,\xM\,;\pSetN\big)}{d\xP d^2\kperp}~,
\end{align}
%%%%%%%%%%%%%%%%%%%%%%%%%%%%%%%%%%%%%%%%
%
with $d\sigmahat^{\Born}_{\inlbl\inbar}/d\xP/d^2\kperp$ defined in \Equation{Eq:defBorn}.
We also repeat that we omit possible scale arguments from the PDFs $f_{\inlbl},f_{\inbar}$, but that the cross section is differential in the final-state variables and we do not imply any restrictions on the scale choice.
Thanks to the lower limit on $X$ implied by $f^{>}_{\inlbl}$, taking $\lambda\to\infty$ is equivalent to $\Lambda\to\infty$, and following the same route as~\mycite{vanHameren:2022mtk}, we find that
%
%
%%%%%%%%%%%%%%%%%%%%%%%%%%%%%%%%%%%%%%%%
\begin{equation}
\frac{d\sigmahat^{\Born}_{\inlbl\inbar}\big(\Lambda,\xM\,;\pSetN\big)}{d\xP d^2\kperp}
\;\overset{\Lambda\to\infty}{\longrightarrow}\;
\frac{\alphaS\ColorC{\inlbl}}{2\pi^2|\kperp|^2}
\,d\Born_{\star\inbar}\big(\xP,\kperp,\xM\,;\pSetN\big)
\label{Eq:defBorn1}
\end{equation}
%%%%%%%%%%%%%%%%%%%%%%%%%%%%%%%%%%%%%%%%
%
with
%
%%%%%%%%%%%%%%%%%%%%%%%%%%%%%%%%%%%%%%%%
\begin{equation}
d\Born_{\star\inbar}\big(\xP,\kperp,\xM\,;\pSetN\big)
=
d\Phi\big(\xP\pP+\kperp+\xM\pM\,;\pSetN\big)
\,\frac{\Matel{\star \inbar}\big(\xP\pP+\kperp,\xM\pP\,;\pSetN\big)}{2\xP\xM\sTot}
\,\JetB\big(\pSetN\big)
~,
\end{equation}
%%%%%%%%%%%%%%%%%%%%%%%%%%%%%%%%%%%%%%%%
%
and where $\Matel{\star \inbar}$ is the tree-level matrix element with a space-like gluon.
We do not use the notation with differentiation with respect to $\xP,\kperp$ for $d\Born_{\star\inbar}$, because they became initial-state variables for this object.
The off-shell matrix element is independent of $\inlbl$, due to auxiliary-parton universality at the LO, thus we get the following result for the high-energy limit of the $k_T$-factorizable part of the cross section:
%
%%%%%%%%%%%%%%%%%%%%%%%%%%%%%%%%%%%%%%%%
\begin{equation}
\frac{d\sigma^{\ColFac,\Born}_{\lambda\to\infty}\big(\pSetN\big)}{d\xP d^2\kperp}
=
%\myScript{N}(\DELTAZRO)\times
F^{\LO}(\DELTAZRO,\kperp)
\sum_{\inbar}\int_0^1d\xM\,f_{\inbar}(\xM)
\,d\Born_{\star\inbar}\big(\xP,\kperp,\xM\,;\pSetN\big)
~,
\label{Eq:lamLimBorn}
\end{equation}
%%%%%%%%%%%%%%%%%%%%%%%%%%%%%%%%%%%%%%%%
%
with the LO in $\alphaS$ unintegrated PDF being defined as
%
%%%%%%%%%%%%%%%%%%%%%%%%%%%%%%%%%%%%%%%%
\begin{equation}
%\myScript{N}(\DELTAZRO)
F^{\LO}(\DELTAZRO,\kperp)
 = \sum_{\inlbl}\frac{\alphaS\ColorC{\inlbl}}{2\pi^2|\kperp|^2}\,\myScript{N}_{\inlbl}(\DELTAZRO)
\quad\textrm{and}\quad
\myScript{N}_{\inlbl}(\DELTAZRO) = 
\int_{0}^1dX\,f^{>}_{\inlbl}(X)
~,
\label{Eq:finlblcnst}
\end{equation}
%%%%%%%%%%%%%%%%%%%%%%%%%%%%%%%%%%%%%%%%
%
and where $\DELTAZRO$ is the lower limit on $X$ implied by $f^{>}_{\inlbl}(X)$.
The factor $\alphaS\ColorC{\inlbl}$ appear because we did not include the color and coupling corrections in \Equation{Eq:defBorn} compared to \Equation{Eq:defAuxMethod}.

At NLO, there are the real and virtual contributions, which have their own definition as the equivalent of \Equation{Eq:defBorn}.
However, these contributions individually will not all exhibit the simple factorized form as in \Equation{Eq:lamLimBorn}.
%, because their equivalent of \Equation{Eq:defBorn1} will still depend logarithmically on $\Lambda$.  
%
We will identify contributions
%
%%%%%%%%%%%%%%%%%%%%%%%%%%%%%%%%%%%%%%%%
\begin{align}
\frac{d\sigma^{\ColFac,\Xi,\xi}_{\lambda}\big(\vepv,\ldots;\pSetN\big)}{d\xP d^2\kperp}
=\sum_{\inlbl,\inbar}\int_0^1dX\,f^{>}_{\inlbl}(X)\int_0^1d\xM\,f_{\inbar}(\xM)
\,\frac{d\sigmahat_{\inlbl\inbar}^{\Xi,\xi}\big(\vepv,\ldots;\lambda X,\xM\,;\pSetN\big)}{d\xP d^2\kperp}
\end{align}
%%%%%%%%%%%%%%%%%%%%%%%%%%%%%%%%%%%%%%%%
%
that individually will have poles in $\vepv$ and depend on a number of extra scales and parameters indicated by the ellipsis.
The symbol $\Xi$ can be $\Virt$ for virtual, $\Real$ for real, and $\Coll$ for counter term.
The symbol $\xi$ stands for extra labels that will be attached to the various contributions.
The equivalent of $d\Born_{\star\inbar}$, obtained after the limit,
%
%%%%%%%%%%%%%%%%%%%%%%%%%%%%%%%%%%%%%%%%
\begin{align}
\frac{d\sigmahat_{\inlbl\inbar}^{\Xi,\xi}\big(\vepv,\ldots;\Lambda,\xM\,;\pSetN\big)}{d\xP d^2\kperp}
\;\overset{\Lambda\to\infty}{\longrightarrow}\;
\frac{\alphaS\ColorC{\inlbl}}{2\pi^2|\kperp|^2}
\,d\Xi_{\inlbl\inbar}^{\xi}\big(\vepv,\Lambda,\ldots;\xP,\kperp,\xM\,;\pSetN\big)
~,
\end{align}
%%%%%%%%%%%%%%%%%%%%%%%%%%%%%%%%%%%%%%%%
%
will for some of those contributions still exhibit a logarithmic dependence on $\Lambda=\lambda X$, and depend on the type of auxiliary parton.
Only after summing them, they will cancel, and the dependence on $\Lambda$ will be attributed to the projectile/Green's function.

\section{\label{Sec:virtual}Virtual NLO contribution}
Regarding phase space, the virtual contribution is equivalent to the Born contribution, and is obtained by replacing the tree-level matrix element in \Equation{Eq:defBorn} with the interference of tree-level with one-loop amplitudes:
%
%%%%%%%%%%%%%%%%%%%%%%%%%%%%%%%%%%%%%%%%
\begin{align}
\frac{d\sigmahat^{\Virt}_{\inlbl\inbar}\big(\vepv\,;\Lambda,\xM\,;\pSetN\big)}{d\xP d^2\kperp}
&=
 \int\frac{d^4q}{(2\pi)^3}\,\delta_+(q^2)
 \,2\xP\,\delta(\xP+\xP_q-\Lambda)\,\delta^2(\kperp+\qperp)
\label{Eq:virtual}\\&\hspace{16ex}\times
\frac{d\Sigma^{\LOOP}_{\inlbl\inbar}}{dq}\big(\vepv\,;\Lambda\pP,\xM\pM\,;q,\pSetN\big)
\,\JetB\big(\pSetN\big)
~.
\notag\end{align}
%%%%%%%%%%%%%%%%%%%%%%%%%%%%%%%%%%%%%%%%
%
One-loop amplitudes with auxiliary partons do not have a smooth limit for $\Lambda\to\infty$, and following~\mycite{vanHameren:2022mtk,Blanco:2022iai} we find
%
%%%%%%%%%%%%%%%%%%%%%%%%%%%%%%%%%%%%%%%%
\begin{equation}
\frac{d\sigmahat^{\Virt}_{\inlbl\inbar}\big(\vepv\,;\Lambda,\xM\,;\pSetN\big)}{d\xP d^2\kperp}
\;\overset{\Lambda\to\infty}{\longrightarrow}\;
\frac{\alphaS\ColorC{\inlbl}}{2\pi^2|\kperp|^2}
\Big[
d\Virt_{\star\inbar}^{\fam}(\vepv)
+d\Virt_{\inlbl\inbar}^{\unf}(\vepv,\Lambda)
\Big]
~.
\label{Eq:defVirt1}
\end{equation}
%%%%%%%%%%%%%%%%%%%%%%%%%%%%%%%%%%%%%%%%
%
The two contributions are called {\em familiar} and {\em unfamiliar}.
We already omitted the arguments $\big(\xP,\kperp,\xM\,;\pSetN\big)$ from them.

\subsection{Unfamiliar virtual contribution}
The \myem{unfamiliar} contribution is given by
%
%%%%%%%%%%%%%%%%%%%%%%%%%%%%%%%%%%%%%%%%
\begin{align}
d\Virt_{\inlbl\inbar}^{\unf}(\vepv,\Lambda)
=
d\Born_{\star\inbar}\times\aeps
\Nc\bigg(\frac{\mu^2}{|\kperp|^2}\bigg)^{\vepv}
\bigg[\frac{2}{\vepv}\ln\frac{\Lambda}{\xP}
      %-\imag\pi
      + \bar{\Vcoef}_{\inlbl}\bigg]
~,
%\VFF_{\inlbl}^{\unf}(\vepv,\Lambda\,;\xP,\kperp)
%\quad,\quad
%\VFF_{\inlbl}^{\unf}(\vepv,\Lambda\,;\xP,\kperp)
%=
%\Nc\bigg(\frac{\mu^2}{|\kperp|^2}\bigg)^{\vepv}
%\bigg[\frac{2}{\vepv}\ln\frac{\Lambda}{\xP}
%      %-\imag\pi
%      + \bar{\Vcoef}_{\inlbl}\bigg]
\label{Eq:unfVirt1}
\end{align}
%%%%%%%%%%%%%%%%%%%%%%%%%%%%%%%%%%%%%%%%
%
with
%
%%%%%%%%%%%%%%%%%%%%%%%%%%%%%%%%%%%%%%%%
\begin{align}
\bar{\Vcoef}_{q/\bar{q}} &=
   \frac{1}{\vepv}\,\frac{13}{6}
  + \frac{\pi^2}{3} +\frac{80}{18}
+\frac{1}{\Nc^2}\bigg[\frac{1}{\vepv^2}+\frac{3}{2}\frac{1}{\vepv}+4\bigg]
-\frac{n_f}{\Nc}\bigg[\frac{2}{3}\frac{1}{\vepv} + \frac{10}{9}\bigg]
~,
\\
\bar{\Vcoef}_{g} &=
-\frac{1}{\vepv^2} 
 + \frac{\pi^2}{3} 
~,
\end{align}
%%%%%%%%%%%%%%%%%%%%%%%%%%%%%%%%%%%%%%%%
%
and $\aeps$ is defined in \Appendix{Sec:constants}.
Besides depending on $\Lambda$, it also breaks the auxiliary parton universality, and also does not have a smooth limit for $|\kperp|\to0$.

\subsection{Familiar virtual contribution}
The \myem{familiar} contribution $d\Virt_{\star\inbar}^{\fam}(\vepv)$ is simply the rest of the virtual contribution.
It does exhibit auxiliary parton universality, does not depend on $\Lambda$, and has a smooth on-shell limit for $|\kperp|\to0$.
This holds for the finite part as well as the poles in $\vepv$.
In fact, the pole parts in $d\Virt_{\star\inbar}^{\fam}(\vepv)$ look exactly as if the space-like gluon were on-shell already.
More specifically, after UV subtraction in the $\overline{\mathrm{MS}}$ scheme on the familiar virtual contribution $d\Virt_{\star\inbar}^{\fam}$,
%%%%%%%%%%%%%%%%%%%%%%%%%%%%%%%%%%%%%%%%
\begin{equation}
d\Virt^{\fam,\UVsubt}_{\star\inbar}(\vepv)
=d\Virt_{\star\inbar}^{\fam}(\vepv)
-d\Born_{\star\inbar}\times\aeps\frac{\gamma_g}{\vepv}\times[\textrm{Born-level-power-of-}\alphaS]
~,
\end{equation}
%%%%%%%%%%%%%%%%%%%%%%%%%%%%%%%%%%%%%%%%
%
with $\gamma_g$ defined in \Appendix{Sec:constants},
the pole part of $d\Virt^{\fam,\UVsubt}_{\star\inbar}(\vepv)$ follows the well-known universal formula for one-loop amplitudes~\mycite{Kunszt:1994np,Catani:1998bh}.

\subsection{Definitions of virtual target IF, Green's function, and projectile contributions \label{sec:virt:targ-proj-GF}}
In this subsection we distribute the virtual NLO contribution between the projectile and the target. Compared to ~\mycite{vanHameren:2022mtk}, we realize that it is more convenient to take out the collinear divergence associated with the space-like gluon from the familiar(/projectile) contribution, and move it to the unfamiliar(/target IF) contribution, including the soft-collinear $1/\epsilon^2$ pole. 
The amount of non-collinear soft divergence we move is set by the scale $\muY$, which parametrizes the ambiguity of the projectile/target separation. Such separation of divergences is motivated by the divergence structure of one-loop corrections to impact-factors in the high-energy EFT approach~\mycite{Chachamis:2012cc,Nefedov:2019mrg,Nefedov:2024swu}, but differs from it in a sense that in high-energy EFT, instead of introducing an arbitrary scale $\muY$ one works in a specific scheme with $\muY=|\kperp|$.
We also identify the {\em BFKL Green's function} contribution with the help of the parameter $\lambdaONE$.
The reason for this will become clear when we define the real contributions.
Thus we define
%
%%%%%%%%%%%%%%%%%%%%%%%%%%%%%%%%%%%%%%%%
\begin{align}
&d\Virt^{\target}_{\inlbl\inbar}(\vepv,\Lambda,\lambdaONE,\muY)
\notag\\&\hspace{8ex}=
d\Virt^{\unf}_{\inlbl\inbar}(\vepv,\Lambda)
- d\Born_{\star\inbar}\times\aeps\bigg[\bigg(\frac{\mu^2}{\muY^2}\bigg)^{\!\vepv}
  \frac{\Nc}{\vepv^2}+\frac{\gamma_g}{\vepv}
   +\bigg(\frac{\mu^2}{|\kperp|^2}\bigg)^{\!\vepv}\frac{2\Nc\ln\lambdaONE}{\vepv}
  \bigg] \label{eq:dV-targ-def}
\\
&d\Virt^{\Green}_{\star\inbar}(\vepv,\lambdaONE)
\notag\\&\hspace{8ex}=\hspace{13.3ex} 
   d\Born_{\star\inbar}\times\aeps\bigg[\hspace{18ex}
    \bigg(\frac{\mu^2}{|\kperp|^2}\bigg)^{\!\vepv}\frac{2\Nc\ln\lambdaONE}{\vepv}\bigg]
\label{Eq:unfVirt2:Green}\\
&d\Virt_{\star\inbar}^{\project}(\vepv,\muY)
\notag\\&\hspace{8ex}=
d\Virt_{\star\inbar}^{\fam}(\vepv)
\hspace{2.4ex}
+ d\Born_{\star\inbar}\times\aeps\bigg[\bigg(\frac{\mu^2}{\muY^2}\bigg)^{\!\vepv}
  \frac{\Nc}{\vepv^2}+\frac{\gamma_g}{\vepv}\bigg]
~.
\label{Eq:unfVirt2}
\end{align}
%%%%%%%%%%%%%%%%%%%%%%%%%%%%%%%%%%%%%%%%
%
We want to stress that we do not add or subtract anything here, only reshuffle.
We will see in the following that for the real contribution the projectile/target separation can be achieved via rapidity restrictions depending on the parameters $\muY$ and $\lambdaONE$.

\section{\label{sec:NLO-proj}Projectile contribution at NLO}
\begin{figure}
\centering
\begin{subfigure}{\linewidth}
\epsfig{figure=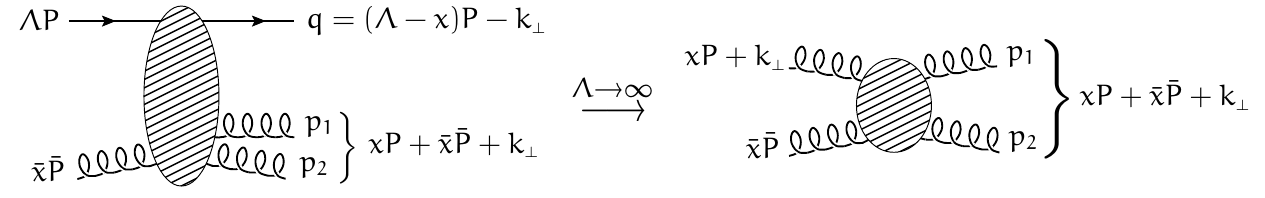,width=0.99\linewidth}
\caption{Born contribution.}
\label{Fig:Born}
\end{subfigure}

\vspace{4ex}

\begin{subfigure}{\linewidth}
\epsfig{figure=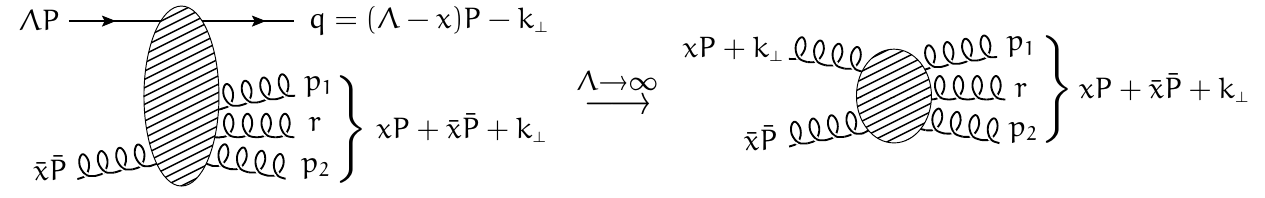,width=0.99\linewidth}
\caption{The familiar real contribution, requires a subtraction to become the real projectile contribution, see \Figure{Fig:realProj}.}
\label{Fig:realFam}
\end{subfigure}

\vspace{2ex}

\begin{subfigure}{\linewidth}
\epsfig{figure=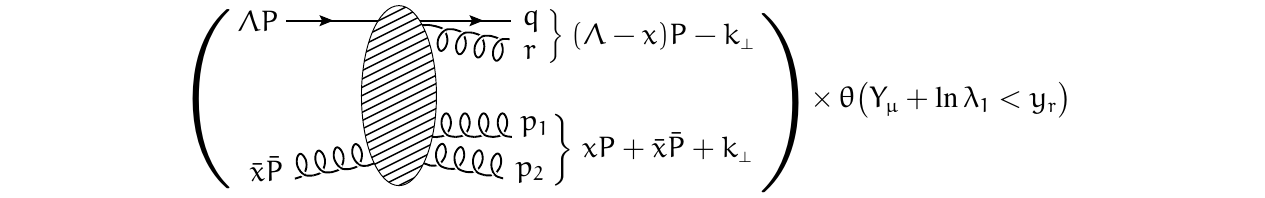,width=0.99\linewidth}
\caption{Real target contribution, with the triple-$\Lambda$ limit of \Section{Sec:tripleLambda}.}
\label{Fig:realTarget}
\end{subfigure}

\vspace{4ex}

\begin{subfigure}{\linewidth}
\epsfig{figure=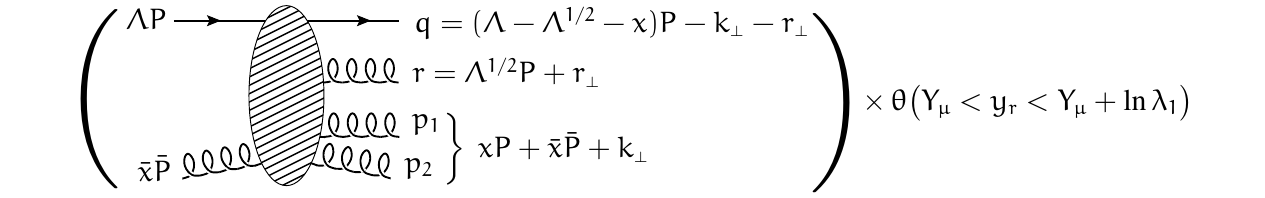,width=0.99\linewidth}
\caption{Real Green's function contribution, with the multi-Regge limit of \Section{Sec:BFKLlimit}.}
\label{Fig:realGreen}
\end{subfigure}

\vspace{4ex}

\begin{subfigure}{\linewidth}
\epsfig{figure=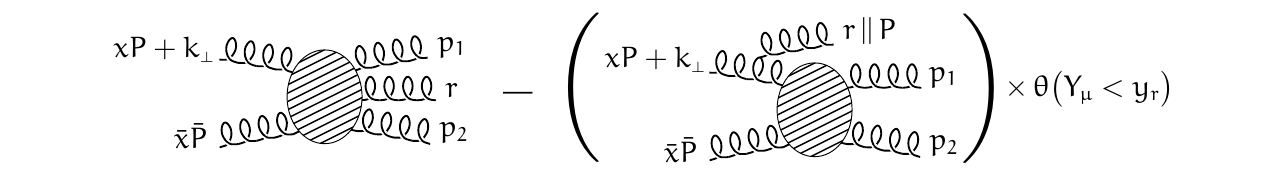,width=0.99\linewidth}
\caption{Real projectile contribution. Rather than restricting the familiar contribution of \Figure{Fig:realFam} to a maximum rapidity $\theta\big(y_r<Y_\mu\big)$ for the radiation, the collinear contribution above a minimum rapidity is subtracted.}
\label{Fig:realProj}
\end{subfigure}

\vspace{2ex}

\caption{Born and real contributions using auxiliary quarks, for the example of a pair of gluons contributing to the case where the system of interest $\myScript{H}$ refers to a dijet: $\myScript{H}=g(p_1)+g(p_2)$. Momentum components of $\Ord\big(\Lambda^{-1}\big)$ and $\Ord\big(\Lambda^{-1/2}\big)$ are omitted.}
\end{figure}

In the present section we begin the discussion of the real-emission NLO contribution to the $d\sigmahat_{\inlbl\inbar}$. The real contribution cannot be written with a single starting formula like \Equation{Eq:virtual}.
It consists of a number of terms, each with different limits on the auxiliary parton radiative matrix elements.
Each limit happens in a certain rapidity range for the radiation, defined via the scale $\muY$ and the parameter $\lambdaONE$.
They are the target, Green's function, and projectile contribution, and are depicted in \Figure{Fig:realTarget}, \Figure{Fig:realGreen}, and \Figure{Fig:realProj}.
%

%\newpage
%
%
%\clearpage
%

\subsection{\label{Sec:famreal}Real projectile contribution}
One contribution for which the simple factorized form of \Equation{Eq:lamLimBorn} still holds is the {\em familiar} real contribution, \Figure{Fig:realFam}.
It is the na{\"\i}ve extrapolation from the Born HEF contribution to a real contribution.
An expression is obtained from \Equation{Eq:defBorn} by replacing $\pSetN$ with $\pSetNpls$ and $\JetB\big(\pSetN\big)$ with appropriate IRC-safe jet definition with an additional parton -- $\JetR\big(\pSetNpls\big)$:
\begin{align}
\frac{d\sigmahat^{\Real,\fam}_{\inlbl\inbar}\big(\vepv\,;\Lambda,\xM\,;\pSetNpls\big)}{d\xP d^2\kperp}
%\notag\\&\hspace{16ex}=
&=
 \int\frac{d^4q}{(2\pi)^3}\,\delta_+(q^2)
 \,2\xP\,\delta(\xP+\xP_q-\Lambda)\,\delta^2(\kperp+\qperp)
% \DELTA(\Lambda\,;\xP,\kperp;\xP_q,\qperp)
\label{Eq:famReal}\\&\hspace{16ex}\times
\frac{d\Sigma_{\inlbl\inbar}}{dq}\big(\vepv\,;\Lambda\pP,\xM\pM\,;q,\pSetNpls\big)
\,\JetR\big(\pSetNpls\big)
~.
\notag\end{align}
%%%%%%%%%%%%%%%%%%%%%%%%%%%%%%%%%%%%%%%%
%
The jet function $\JetR$ allows for one jet fewer than the number of final-state partons.
The divergences resulting from this are regularized within dimensional regularization, so the phase space integral is now in $4-2\vepv$ dimensions, and the expression requires a factor $\mu^{2\vepv}$ where $\mu$ is the renormalization scale.
We have 
%
%%%%%%%%%%%%%%%%%%%%%%%%%%%%%%%%%%%%%%%%
\begin{equation}
\frac{d\sigmahat^{\Real,\fam}_{\inlbl\inbar}\big(\vepv\,;\Lambda,\xM\,;\pSetNpls\big)}{d\xP d^2\kperp}
\;\overset{\Lambda\to\infty}{\longrightarrow}\;
\frac{\alphaS\ColorC{\inlbl}}{2\pi^2|\kperp|^2}
\,d\Real_{\star\inbar}^{\fam}\big(\vepv\,;\xP,\kperp,\xM\,;\pSetNpls\big)
~,
\end{equation}
%%%%%%%%%%%%%%%%%%%%%%%%%%%%%%%%%%%%%%%%
%
with
%
%%%%%%%%%%%%%%%%%%%%%%%%%%%%%%%%%%%%%%%%
\begin{align}
&d\Real_{\star\inbar}^{\fam}\big(\vepv\,;\xP,\kperp,\xM\,;\pSetNpls\big)
\\&\hspace{4ex}= 
d\Phi\big(\vepv\,;\xP\pP+\kperp+\xM\pM\,;\pSetNpls\big)
\,\frac{\Matel{\star\inbar}\big(\xP\pP+\kperp,\xM\pP\,;\pSetNpls\big)}{2\xP\xM\sTot}
\,\JetR\big(\pSetNpls\big)
~.
\label{Eq:defReal1}
\end{align}
%%%%%%%%%%%%%%%%%%%%%%%%%%%%%%%%%%%%%%%%
%

\subsubsection{Resolved and divergent (unresolved) real familiar contribution}
In this section, we collect the divergences which arise in the familiar real contribution described above as a result of integration over the phase-space of an additional parton. In \mycite{Giachino:2023loc} an approach is presented how to calculate
%
%%%%%%%%%%%%%%%%%%%%%%%%%%%%%%%%%%%%%%%%
\begin{align}
d\sigma^{\mathrm{HF},\Real,\fam}\big(\vepv\,;\pSetNpls\big)
&=
\intdLxkx
\,d\Real_{\star\inbar}^{\fam}\big(\vepv\,;\xP,\kperp,\xM\,;\pSetNpls\big)
~,
\end{align}
%%%%%%%%%%%%%%%%%%%%%%%%%%%%%%%%%%%%%%%%
%
that is how to calculate the $1/\vepv^2,1/\vepv$ coefficients and the finite coefficient.
More specifically, it gives a constructive prescription to calculate terms in the decomposition
%
%%%%%%%%%%%%%%%%%%%%%%%%%%%%%%%%%%%%%%%%
\begin{equation}
d\sigma^{\mathrm{HF},\Real,\fam}\big(\vepv\,;\pSetNpls\big)
=
d\sigma^{\mathrm{HF},\Real,\fam,\unresolv}\big(\vepv\,;\pSetN\big)
+d\sigma^{\mathrm{HF},\resolved}\big(\pSetNpls\big)
+\Ord(\vepv)
~.
\end{equation}
%%%%%%%%%%%%%%%%%%%%%%%%%%%%%%%%%%%%%%%%
%
Rather than using the labels ``resolved vs.\ unresolved'' or ``finite vs.\ divergent'', we use a mixed labelling ``resolved vs.\ divergent'' to highlight that one term includes {\em all} resolved phase space (with the $n+1$ jets), and the other {\em all} divergences.
The separation is not unique and eventually depends on a choice of distribution of finite unresolved contributions.
The choice is, however, well-defined and is a pure technicality, since in the present paper we are essentially interested only in the structure of divergences.

Only the familiar real contribution has a resolved part.
It can directly be designated to the projectile, and we will not address it any further.
The divergent part $d\sigma^{\mathrm{HF},\Real,\fam,\unresolv}$ lives in Born phase space, and can be written as
%
%%%%%%%%%%%%%%%%%%%%%%%%%%%%%%%%%%%%%%%%
\begin{align}
d\sigma^{\mathrm{HF},\Real,\fam,\unresolv}\big(\vepv\,;\pSetN\big)
=
\intdLxkx
\,d\Real_{\star\inbar}^{\fam,\unresolv}\big(F,\{f_{\inbar}\}\,;\vepv\,;\xP,\kperp,\xM\,;\pSetN\big)
~.
\notag\end{align}
%%%%%%%%%%%%%%%%%%%%%%%%%%%%%%%%%%%%%%%%
%
By construction of the familiar real and virtual contributions, most of the divergences in $d\Real_{\star\inbar}^{\fam,\unresolv}$ cancel against the corresponding divergences in $d\Virt_{\star\inbar}^{\fam}$ by the same mechanisms as in the usual NLO CF computation. In the present paper we are mostly interested in those divergences which are associated with initial-state off-shell parton and do not cancel in a usual way in HEF. The non-cancelling divergences in the familiar-real contribution are encapsulated in the following term, depending on initial-state PDFs:
%
%%%%%%%%%%%%%%%%%%%%%%%%%%%%%%%%%%%%%%%%
\begin{align}
d\Real_{\star\inbar}^{\fam,\unresolv} \supset d\Born_{\star\inbar}\big(\xP,\kperp,\xM\,;\pSetN\big)\times\aeps\Bigg[
{-}\frac{1}{\vepv}\sum_{\inbar'}
   \frac{\big[\myScript{P}_{\inbar\inbar'}^{\Real}\otimes f_{\inbar'}\big](\xM)}{f_{\inbar}(\xM)}
  -\frac{1}{\vepv}
   \frac{\big[\myScript{P}_{\star}^{\Real}\otimes \Fstar\big](\xP,\kperp)}{\Fstar(\xP,\kperp)}
\Bigg]
\label{Eq:colldiv}
~.
\end{align}
%%%%%%%%%%%%%%%%%%%%%%%%%%%%%%%%%%%%%%%%
%
The splitting functions $\myScript{P}_{\star}^{\Real}$ and  $\myScript{P}_{\inbar\inbar'}^{\Real}$ are given in \Appendix{Sec:splitting}.
The convolutions indicated with ``$\otimes$'' are addressed in \Appendix{Sec:convolutions}.
We stress here that this is only the real contribution, and the splitting functions  do not contain virtual terms proportional to $\delta(1-z)$.

The first term is the usual collinear divergence which is removed by the renormalization of the collinear PDF of the projectile.
This left-over is removed in the $\overline{\mathrm{MS}}$ scheme by subtracting 
%
%%%%%%%%%%%%%%%%%%%%%%%%%%%%%%%%%%%%%%%%
\begin{equation}
d\Coll^{\project}_{\star\inbar}\big(\{f\}\,;\vepv,\muFbar\,;\xP,\kperp,\xM\,;\pSetN\big)
=
d\Born_{\star\inbar}\big(\xP,\kperp,\xM\,;\pSetN\big)\times
\aeps\,\CFF^{\project}_{\inbar}(\{f\}\,;\vepv,\muFbar\,;\xM)
\end{equation}
%%%%%%%%%%%%%%%%%%%%%%%%%%%%%%%%%%%%%%%%
%
with
%
%%%%%%%%%%%%%%%%%%%%%%%%%%%%%%%%%%%%%%%%
\begin{equation}
\CFF^{\project}_{\inbar}(\{f\}\,;\vepv,\muFbar\,;\xM)
=
{-}
\frac{1}{\vepv}
\bigg(\frac{\mu^2}{\muFbar^2}\bigg)^{\!\vepv}
\sum_{\inbar'}\frac{\big[\myScript{P}_{\inbar\inbar'}\otimes f_{\inbar'}\big](\xM)}{f_{\inbar}(\xM)}
\label{Eq:colinbar}
\end{equation}
%%%%%%%%%%%%%%%%%%%%%%%%%%%%%%%%%%%%%%%%
for some factorization scale $\muFbar$, and now the splitting function $\myScript{P}_{\inbar\inbar'}$ includes the virtual contribution. 

\subsubsection{Divergent real projectile contribution}
The second term in \Equation{Eq:colldiv} comes from the phase space region where the radiation becomes collinear to the momentum $\pP$, or in other words, when it is moving in the direction in rapidity of the target.
Like in the virtual case, discussed in Sec. \ref{sec:virt:targ-proj-GF}, we would like to remove the contribution of this momentum-region from the projectile, because it overlaps with the contribution of the target, and as mentioned, we can achieve this with a rapidity restriction.
However, our goal is to remove only the logarithmic contribution, which belongs to the
target, from the projectile contribution, so rather than just restricting the phase space on \Equation{Eq:famReal}, we subtract a contribution with the matrix element at the collinear limit of \Equation{Eq:collim}.
We define
\begin{align}
&d\Real^{\fam,\coll}_{\star\inbar}\big(\vepv,\muY\,;\xP,\kperp,\xM\,;\pSetN\big)
\notag\\&\hspace{8ex}=
 \int\frac{\mu^{2\vepv}d^{4-2\vepv}r}{(2\pi)^{3-2\vepv}}\,\delta_+(r^2)
  \,\theta(\xP_r<\xP)
  \,\theta\bigg(
    \yrap_r > \ln\frac{\sqrtS(\xP-\xP_r)}{\muY}
  \bigg)
\label{Eq:Rfamcoll}
\\&\hspace{16ex}\times
  \gQCD^2\,\frac{4\Nc}{|\rperp|^2(1-\xP_r/\xP)}
  \,d\Sigma_{\star\inbar}\big(\kstar-\xP_r\pP-\rperp,\xM\pM\,;\pSetN\big)
  \,\JetB\big(\pSetN\big)
~.
\notag\end{align}
There is a factor $(1-\xP_r/\xP)$ instead of its square in the denominator, because $d\Sigma_{\star\inbar}$ includes the flux factor, for which we want to include the radiation.
The rapidity threshold on the other hand is defined in terms of $(\xP-\xP_r)$, that is with the radiation subtracted.
Notice that the restriction is defined with the same scale $\muY$ as in \Section{sec:virt:targ-proj-GF}, but for the real-emission contribution its relation with rapidity separator $Y_\mu$ is explicit.
After some manipulations, we can write
%
%%%%%%%%%%%%%%%%%%%%%%%%%%%%%%%%%%%%%%%%
%
\begin{align}
&
%\int_0^1d\xP\int d^2\kperp\,F(\xP,\kperp)
\intdLxk
\,d\Real^{\fam,\coll}_{\star\inbar}\big(\vepv,\muY\,;\xP,\kperp,\xM\,;\pSetN\big)
%\notag\\&\hspace{16ex}=
%\int_0^1d\xP\int d^2\kperp\,F(\xP,\kperp)
=\intdLxk
\,d\Coll^{\star}_{\star\inbar}\big(F\,;\vepv,\muY\,;\xP,\kperp,\xM\,;\pSetN\big)
\end{align}
%%%%%%%%%%%%%%%%%%%%%%%%%%%%%%%%%%%%%%%%
%
where
%
%%%%%%%%%%%%%%%%%%%%%%%%%%%%%%%%%%%%%%%%
\begin{equation}
d\Coll^{\star}_{\star\inbar}\big(F\,;\vepv,\muY\,;\xP,\kperp,\xM\,;\pSetN\big)
=
d\Born_{\star\inbar}\big(\xP,\kperp,\xM\,;\pSetN\big)\times
\aeps\,\CFF_{\star}(F\,;\vepv,\muY\,;\xP,\kperp)
~,
\end{equation}
%%%%%%%%%%%%%%%%%%%%%%%%%%%%%%%%%%%%%%%%
%
with, omitting the arguments $(\xP,\kperp)$,% from now on,
%
%%%%%%%%%%%%%%%%%%%%%%%%%%%%%%%%%%%%%%%%
\begin{align}
%&\CFF_{\star}\big(\Fstar\,;\vepv,\muY\,;\xP,\kperp,\xM\big)
\CFF_{\star}(\Fstar\,;\vepv,\muY)
=
  \frac{2\Nc\mu^{2\vepv}}{\piep}
  \int_{\xP}^1\frac{dz}{z(1-z)}
  \int\frac{d^{2-2\vepv}\rperp}{|\rperp|^2}
  \,\frac{\Fstar\big(\xP/z,\kperp+\rperp\big)}{z\,\Fstar(\xP,\kperp)}
  \,\theta\bigg(|\rperp|<\muY\,\frac{1-z}{z}\bigg)
~.
\label{Eq:Ftilde}
\end{align}
%%%%%%%%%%%%%%%%%%%%%%%%%%%%%%%%%%%%%%%%
%
Now, we define the real divergent projectile contribution as, omitting arguments $\big(\xP,\kperp,\xM\,;\pSetN\big)$, 
%
%%%%%%%%%%%%%%%%%%%%%%%%%%%%%%%%%%%%%%%%
\begin{align}
&d\Real_{\star\inbar}^{\project,\unresolv}\big(F,\{f_{\inbar}\}\,;\vepv,\muY\big)
=
d\Real_{\star\inbar}^{\fam,\unresolv}\big(F,\{f_{\inbar}\}\,;\vepv\big)
-d\Coll^{\star}_{\star\inbar}\big(F\,;\vepv,\muY\big)
~. \label{eq:fam-div-subtr}
\end{align}
%%%%%%%%%%%%%%%%%%%%%%%%%%%%%%%%%%%%%%%%
%
See \Figure{Fig:realProj} for a pictorial representation.
To see that this indeed removes the collinear divergence, we expand
%
%%%%%%%%%%%%%%%%%%%%%%%%%%%%%%%%%%%%%%%%
\begin{equation}
\CFF_{\star}(\Fstar\,;\vepv,\muY)
=
\CFF_{\star}^{\DIV}(\Fstar\,;\vepv,\muY)
+\CFF_{\star}^{\FIN}(\Fstar\,;\muY)+\Ord(\vepv)
~,
\end{equation}
%%%%%%%%%%%%%%%%%%%%%%%%%%%%%%%%%%%%%%%%
%
where we find
%
%%%%%%%%%%%%%%%%%%%%%%%%%%%%%%%%%%%%%%%%
\begin{equation}
\CFF_{\star}^{\DIV}(\Fstar\,;\vepv,\muY)=
 \bigg(\frac{\mu^2}{\muY^2}\bigg)^{\!\vepv}
\bigg[
   \frac{\Nc}{\vepv^2}
   -\frac{1}{\vepv}\frac{\big[\myScript{P}_{\star}^{\Real}\otimes \Fstar\big](\xP,\kperp)}
                        {\Fstar(\xP,\kperp)}
%  -\frac{2\Nc}{\vepv}
%  \int_{0}^1dz\,\frac{1}{[1-z]_+}
%  \,\frac{F\big(\xP/z,\kperp\big)}{z^2F(\xP,\kperp)}
%  \,\theta(z>\xP)
\bigg]
~,
\label{Eq:FtildeDiv}
\end{equation}
%%%%%%%%%%%%%%%%%%%%%%%%%%%%%%%%%%%%%%%%
%
and
%
%%%%%%%%%%%%%%%%%%%%%%%%%%%%%%%%%%%%%%%%
\begin{align}
\CFF_{\star}^{\FIN}(\Fstar\,;\muY)&= 
  \frac{2\Nc}{\pi}
  \int_{\xP}^1\frac{dz}{z(1-z)}
  \int\frac{d^{2}\rperp}{|\rperp|^2}
  \bigg[\frac{\Fstar\big(\xP/z,\kperp+\rperp\big)}{z\,\Fstar(\xP,\kperp)}
        -\frac{\Fstar\big(\xP/z,\kperp\big)}{z\Fstar(\xP,\kperp)}\bigg]
  \,\theta\bigg(|\rperp|<\muY\frac{1-z}{z}\bigg)
\notag\\&\hspace{0ex}
  +4\Nc
  \int_{0}^1dz
  \Bigg\{\bigg[\frac{\ln(1-z)}{1-z}\bigg]_+
        -\frac{\ln(z)}{1-z}\Bigg\}
  \,\frac{\Fstar\big(\xP/z,\kperp\big)}{z^2\Fstar(\xP,\kperp)}
  \,\theta(z>\xP)
~.
\label{Eq:FtildeFin}
\end{align}
%%%%%%%%%%%%%%%%%%%%%%%%%%%%%%%%%%%%%%%%
%
So indeed, the second term of \Equation{Eq:colldiv} is cancelled, while the term with $1/\epsilon^2$ is cancelled by the corresponding term in $d\Virt_{\star\inbar}^{\project}$, Eq. (\ref{Eq:unfVirt2}).

\subsection{Complete (unresolved) finite projectile contribution at NLO}
In conclusion, we have the finite unresolved projectile contribution
%
%%%%%%%%%%%%%%%%%%%%%%%%%%%%%%%%%%%%%%%%
\begin{align}
&d\Real_{\star\inbar}^{\project,\unresolv}\big(F,\{f_{\inbar}\}\,;\vepv,\muY\big)
+d\Virt_{\star\inbar}^{\project,\UVsubt}(\vepv,\muY)
-d\Coll^{\project}_{\star\inbar}\big(\{f\}\,;\vepv,\muFbar\big)
\notag\\&\hspace{2.0ex}
=d\Born_{\star\inbar}\times\aeps\bigg[
\ln\frac{\mu^2}{\muFbar^2}
\sum_{\inbar'}\frac{\big[\myScript{P}_{\inbar\inbar'}\otimes f_{\inbar'}\big](\xM)}{f_{\inbar}(\xM)}
+\ln\frac{\mu^2}{\muY^2}
  \frac{\big[\myScript{P}_{\star}\otimes \Fstar\big](\xP,\kperp)}{\Fstar(\xP,\kperp)}
 -\CFF_{\star}^{\FIN}(\Fstar\,;\muY)
\bigg]
\label{Eq:famCancelled}\\&\hspace{2.0ex}
+\big[\textrm{independent of $\muFbar,\muY$}\big]
~,
\notag\end{align}
%%%%%%%%%%%%%%%%%%%%%%%%%%%%%%%%%%%%%%%%
%
which does not depend on the structure of the UPDF $F(x,\kperp)$. This formula is one of the main results of the present paper, since it provides a practical prescription for the NLO computations for the arbitrary process with one off-shell initial-state parton in HEF. What is left to do now is to study the consistency of the NLO target contribution with this prescription, which is done in the following sections.

\section{\label{sec:UPDF-evol}UPDF evolution from the projectile point of view}
\subsection{The $\muY$-evolution equation for UPDF from the projectile point of view}
The inclusive cross section of the production of the system of interest $\myScript{H}$ in HEF should not depend on the arbitrary scale $\muY$. Physically it is easy to understand, since the only thing which depends on $\muY$ are the definitions of variables $x$ and $\kperp$ (Sec.~\ref{sec:x-and-kT}), which depend on the value of the rapidity cut $Y_\mu$. In the inclusive cross section one just integrates over all possible values of $x$ and $\kperp$ and therefore all dependence from their definition and from the scale $\muY$ should disappear.  On the level of the cross section formula, the dependence on $\muY$ cancels between the off-shell coefficient function (projectile impact-factor) and the UPDF, order-by-order in $\alphaS$.

We obtain the $\muY$-evolution equation for the $k_T$-dependent PDF $\Fstar(\xP,\kperp;\muY)$ by demanding that the differential cross section is independent of $\muY$ order by order in $\alphaS$.
The PDF is expanded as
%
%%%%%%%%%%%%%%%%%%%%%%%%%%%%%%%%%%%%%%%%
\begin{equation}
F(\xP,\kperp;\muY) = \Fstar^{(0)}(\xP,\kperp) + \aeps\,\Fstar^{(1)}(\xP,\kperp;\muY)
~.
\end{equation}
%%%%%%%%%%%%%%%%%%%%%%%%%%%%%%%%%%%%%%%%
%
From the Born contribution and from the NLO contribution given by \Equation{Eq:famCancelled} we get
%
%%%%%%%%%%%%%%%%%%%%%%%%%%%%%%%%%%%%%%%%
\begin{equation}
0 = \frac{d}{d\ln\muY^2}\bigg[
\Fstar^{(1)}(\xP,\kperp;\muY) + \ln\frac{\mu^2}{\muY^2}\big[\myScript{P}_{\star}\otimes \Fstar^{(0)}\big](\xP,\kperp) - \CFF_{\star}^{\FIN}\big(\Fstar^{(0)}\,;\muY\,;\xP,\kperp\big)\Fstar^{(0)}(\xP,\kperp)
\bigg]
~.
\end{equation}
%%%%%%%%%%%%%%%%%%%%%%%%%%%%%%%%%%%%%%%%
%
Instead of the PDF, we prefer to consider
%
%%%%%%%%%%%%%%%%%%%%%%%%%%%%%%%%%%%%%%%%
\begin{equation}
\hat{F}(\xP,\kperp;\muY) = \xP\,\Fstar(\xP,\kperp;\muY)
~.
\end{equation}
%%%%%%%%%%%%%%%%%%%%%%%%%%%%%%%%%%%%%%%%
%
Using that
%
%%%%%%%%%%%%%%%%%%%%%%%%%%%%%%%%%%%%%%%%
\begin{equation}
\frac{d}{d\ln\muY^2}\,\theta\bigg(|\rperp|<\muY\,\frac{1-z}{z}\bigg)\,\theta(z>x)
=
\frac{z(1-z)}{2}\,\delta\bigg(z-\frac{\muY}{\muY+|\rperp|}\bigg)
\,\theta\bigg(|\rperp|<\muY\,\frac{1-x}{x}\bigg)
\end{equation}
%%%%%%%%%%%%%%%%%%%%%%%%%%%%%%%%%%%%%%%%
%
we get
%
%%%%%%%%%%%%%%%%%%%%%%%%%%%%%%%%%%%%%%%%
\begin{align}
&
\frac{d\hat{F}^{(1)}(\xP,\kperp;\muY)}{d\ln\muY^2}
=
2\Nc\int_0^1dz\bigg[\frac{1}{(1-z)_+}+\frac{1}{z}\bigg]\hat{F}^{(0)}(\xP/z,\kperp)\,\theta(z>\xP)
\\&\hspace{0ex}+
\frac{\Nc}{\pi}\int\frac{d^2\rperp}{|\rperp|^2}
\Bigg\{\hat{F}^{(0)}\bigg(\xP\bigg[1+\frac{|\rperp|}{\muY}\bigg],\kperp+\rperp\bigg) 
       - \hat{F}^{(0)}\bigg(\xP\bigg[1+\frac{|\rperp|}{\muY}\bigg],\kperp\bigg)\Bigg\}
\theta\bigg(|\rperp|<\muY\,\frac{1-x}{x}\bigg)
\notag
~.
\end{align}
After some manipulations we find
%
%%%%%%%%%%%%%%%%%%%%%%%%%%%%%%%%%%%%%%%%
\begin{align}
\frac{d\hat{F}^{(1)}(\xP,\kperp;\muY)}{d\ln\muY^2}
\label{eq:evol-eq-mu0}&=
\frac{\Nc}{\pi}\int\frac{d^2\rperp}{|\rperp|^2}
\Bigg\{\hat{F}^{(0)}\bigg(\xP\bigg[1+\frac{|\rperp|}{\muY}\bigg],\kperp+\rperp\bigg) 
\theta\bigg(|\rperp|<\muY\,\frac{1-x}{x}\bigg)
\\&\hspace{20ex}
       - \frac{\mu_0^\beta}{\mu_0^\beta+|\rperp|^\beta}\,\hat{F}^{(0)}(\xP,\kperp)\Bigg\}
+\Nc\,\hat{F}^{(0)}(\xP,\kperp)\,\ln\frac{\mu_0^2}{\muY^2} 
~,\notag
\end{align}
%%%%%%%%%%%%%%%%%%%%%%%%%%%%%%%%%%%%%%%%
%
which is independent of the arbitrary scale $\mu_0$ and parameter $\beta>0$.
Taking $\beta\to\infty$ corresponds to $\mu_0^\beta/\big(\mu_0^\beta+|\rperp|^\beta\big)\to\theta\big(\mu_0-|\rperp|\big)$.
We promote the obtained evolution equation to all orders in $\alphaS$ as:
%
%%%%%%%%%%%%%%%%%%%%%%%%%%%%%%%%%%%%%%%%
\begin{align}
\frac{d\hat{F}(\xP,\kperp;\muY)}{d\ln\muY^2}
&=
\frac{\alphaS\Nc}{2\pi} \int\frac{d^2\rperp}{\pi |\rperp|^2}
\label{eq:evol-eq-muY-fin}
\Bigg\{\hat{F}\bigg(\xP\bigg[1+\frac{|\rperp|}{\muY}\bigg],\kperp+\rperp\,;\muY \bigg) 
\,\theta\bigg(|\rperp|<\muY\,\frac{1-x}{x}\bigg)
\\&\hspace{20ex}
       - \theta\big(\mu_0-|\rperp|\big)\,\hat{F}(\xP,\kperp;\muY)\Bigg\}
+\frac{\alphaS\Nc}{2\pi}\,\hat{F}(\xP,\kperp;\muY)\ln\frac{\mu_0^2}{\muY^2} 
~.\notag
\end{align}
%%%%%%%%%%%%%%%%%%%%%%%%%%%%%%%%%%%%%%%%
%
Due to its connection to the Collins-Soper-Sterman equation, discussed below, we propose to call the \Equation{eq:evol-eq-muY-fin}) and its equivalents -- {\it the generalized CSS equation}.
However unlike the CSS equation, this equation mixes $x$ and $\kperp$-dependence, which as it will be shown below, is necessary to reproduce DGLAP.

We can also derive a $2-2\epsilon$-dimensional version of the equation which is equivalent to \Equation{eq:evol-eq-muY-fin} but not manifestly IR finite, from \Equation{Eq:unfVirt2} and \Equation{Eq:Ftilde}:
%
%%%%%%%%%%%%%%%%%%%%%%%%%%%%%%%%%%%%%%%%
\begin{equation}
0 = \frac{d}{d\ln\muY^2}\bigg[
\Fstar^{(1)}(\xP,\kperp;\muY)
+ \bigg(\frac{\mu^2}{\muY^2}\bigg)\frac{\Nc}{\vepv^2}\,\Fstar^{(0)}(\xP,\kperp)
- \CFF_{\star}\big(\Fstar^{(0)}\,;\vepv,\muY\,;\xP,\kperp\big)\Fstar^{(0)}(\xP,\kperp)
\bigg]
~.
\end{equation}
%%%%%%%%%%%%%%%%%%%%%%%%%%%%%%%%%%%%%%%%
%
It leads to
%
%
%%%%%%%%%%%%%%%%%%%%%%%%%%%%%%%%%%%%%%%%
\begin{align}
\frac{d\hat{F}(\xP,\kperp;\muY)}{d\ln\muY^2}
&=
\aeps \frac{\Nc\mu^{2\vepv}}{\piep}\int\frac{d^{2-2\vepv}\rperp}{|\rperp|^2}
\,\hat{F}\bigg(\xP\bigg[1+\frac{|\rperp|}{\muY}\bigg],\kperp+\rperp\,;\muY\bigg)
\theta\bigg(|\rperp|<\muY\,\frac{1-x}{x}\bigg)
\notag\\&\hspace{0ex}
+ \aeps \Nc\,\hat{F}(\xP,\kperp;\muY)\bigg[\frac{1}{\vepv} + \ln\frac{\mu^2}{\muY^2} 
    \bigg]
~,
\label{eq:evol-eqn-eps}
\end{align}
%%%%%%%%%%%%%%%%%%%%%%%%%%%%%%%%%%%%%%%%
%
which is more convenient for the analysis below.

\subsubsection{$\muY$-evolution in impact-parameter space\label{sec:xT-space-evol}}
The form of the $\muY$-evolution \Equation{eq:evol-eqn-eps} is convenient to perform the diagonalization of transverse-momentum convolutions. To this end we move to the $\xperp$-space Fourier-conjugate to $\kperp$:
\begin{equation}
    \hat{F}(x,\kperp;\muY)=\int d^{2-2\vepv}\xperp\, e^{\imag\kperp\xperp} \tilde{F}(x,\xperp;\muY)
~. \label{eq:F-xT-def}
\end{equation}
In \Appendix{App:impact-parameter}, we show that \Equation{eq:evol-eqn-eps} in the limit $\epsilon\to 0$ leads to the following equation for $\tilde{F}(x,\xperp;\muY)$:
\begin{align}
\frac{d\tilde{F}(\xP,\xperp ; \muY)}{d\ln\muY^2}
&=   \frac{\alphaS\Nc}{\pi} \int_x^1 \frac{dz}{z(1-z)_+} J_{0} \left( \muY |\xperp| \frac{1-z}{z} \right) \tilde{F}\left( \frac{\xP}{z},\xperp; \muY \right)  
~,
\label{eq:evol-eqn-xT-final}
\end{align}
where $J_{0}$ is the Bessel function.
The limit $|\xperp|\to 0$ of this equation corresponds to the evolution of the $\kperp$-integrated distribution, and we find that it evolves according to the following DGLAP-like equation:
\begin{align}
\frac{d\tilde{F}(\xP,0 ; \muY)}{d\ln\muY^2}
&=   \frac{\alphaS\Nc}{\pi} \int_x^1 \frac{dz}{z(1-z)_+} \tilde{F}\left( \frac{\xP}{z},0\,; \muY \right)  
~,
\label{eq:evol-eqn-xT0-DGLAP}
\end{align}
which in particular preserves the momentum sum rule $\int_0^1dx\,\tilde{F}(\xP,0 ; \muY)=1$. 

\subsubsection{Relation with the Collins-Soper-Sterman equation}
If we assume that $|\kperp|,|\rperp|\ll\mu_Y$ then the $x$-argument of $\hat{F}(x,\kperp;\muY)$ is not affected by the evolution. 
Substituting the $e^{\imag\kperp\xperp}$ into the real-emission part of the kernel of \Equation{eq:evol-eqn-eps} we get:
\begin{eqnarray}
    \mu^{2\vepv} \int \frac{d^{2-2\vepv}\rperp}{\piep\rperp^2}\,e^{\imag\xperp(\kperp+\rperp)}
%=-e^{\imag\xperp\kperp}  \frac{\Gamma^2(1-\vepv)}{\vepv} \left( \frac{\mu^2 \xperp^2}{4} \right)^\vepv
= -\frac{e^{\imag\xperp\kperp}}{\vepv} \left( \mu^2 \overline{x}_{\theperp}^2 \right)^\vepv + \Ord(\vepv)~,
\end{eqnarray}
where $\overline{x}_{\theperp}=\xperp/(2e^{-\gamma_E})$. 
Therefore, in $\xperp$-space \Equation{eq:evol-eqn-eps} becomes:
\begin{equation}
    \frac{d}{d\ln\mu_Y^2} \tilde{F}(x,\xperp;\mu_Y)=\frac{\alphaS}{2\pi} \Big[ {-}\Nc \ln\big(\mu_Y^2 \overline{x}_{\theperp}^2\big) \Big] \tilde{F}(x,\xperp;\mu_Y)
~. 
\end{equation}
The last equation is the LO Collins-Soper-Sterman (CSS) equation~\mycite{Collins:2011zzd} for TMD evolution with respect to rapidity scale $\mu_Y^2$ often denoted by $\zeta$, see \eg~\mycite{Scimemi:2018xaf}, and the expression in the square brackets is the rapidity anomalous dimension at LO in $\alphaS$.
The choice of $\mu_0=\mu_Y$ in \Equation{eq:evol-eq-muY-fin} in the TMD limit gives the $\kperp$-space form of the CSS equation:
\begin{align}
&
\frac{d\hat{F}(\xP,\kperp; \muY)}{d\ln\muY^2}=
\frac{\alphaS\Nc}{2\pi} \int\frac{d^2\rperp}{\pi |\rperp|^2}
\Bigg\{\hat{F}\big(\xP,\kperp+\rperp ; \muY\big) 
       - \theta\big(|\rperp|^2<\muY^2\big)\,\hat{F}\big(\xP,\kperp; \muY\big)\Bigg\}
~,
\label{eq:CSS-kT-space-theta}
\end{align}
compare with the Eqns.~(5.29), (5.56) and definitions in Appendix~F of~\mycite{Chiu:2012ir} with the identification of the scale $\nu^2=\mu_Y^2$.

\section{\label{sec:target-NLO}Target contribution at NLO}

\subsection{\label{Sec:unfreal}Real target IF contribution}
The real target IF contribution is closely related to what in \mycite{vanHameren:2022mtk} was called the {\em unfamiliar real} contribution, for which the radiation momentum $r$ is also allowed to grow with $\Lambda$ just like $q$.
We define
%
%%%%%%%%%%%%%%%%%%%%%%%%%%%%%%%%%%%%%%%%
\begin{align}
&\frac{d\sigmahat^{\Real,\target}_{\inlbl\inbar}\big(\vepv,\lambdaONE,\muY\,;\Lambda,\xM\,;\pSetN\big)}{d\xP d^2\kperp}
\label{Eq:defunfReal}
\\&\hspace{2ex}=
 \int\frac{d^4q}{(2\pi)^3}\,\delta_+(q^2)
 \int\frac{\mu^{2\vepv}d^{4-2\vepv}r}{(2\pi)^{3-2\vepv}}\,\delta_+(r^2)
 %\,\theta\bigg(y_q>\ln\frac{\sqrtS\xP}{\muY}+\ln\lambdaONE\bigg)
 %\,\theta\bigg(y_r>\ln\frac{\sqrtS\xP}{\muY}+\ln\lambdaONE\bigg)
 \,\theta\big(y_q>Y_\mu+\ln\lambdaONE\big)
 \,\theta\big(y_r>Y_\mu+\ln\lambdaONE\big)
\notag\\&\hspace{6ex}\times
 2\xP\,\delta(\xP+\xP_q+\xP_r-\Lambda)\delta^{2-2\vepv}(\kperp+\qperp+\rperp)
 \,\frac{d\Sigma_{\inlbl\inbar}}{dqdr}\big(\Lambda\pP,\xM\pM\,;q,r,\pSetN\big)
\,\JetB\big(\pSetN\big)
~.
\notag\end{align}
%%%%%%%%%%%%%%%%%%%%%%%%%%%%%%%%%%%%%%%%
%
See \Figure{Fig:realTarget} for a pictorial representation.
While the minimum value of $X$ implied by $f^{>}(X)$ now sets the minimum rapidity of the sum of the momenta $q$ and $r$ to
%
%%%%%%%%%%%%%%%%%%%%%%%%%%%%%%%%%%%%%%%%
\begin{equation}
y_{q+r} > Y_\mu + \ln\lambdaZRO
~,
\end{equation}
%%%%%%%%%%%%%%%%%%%%%%%%%%%%%%%%%%%%%%%%
%
the rapidities of the individual momenta are restricted following $\lambdaONE\prec\lambdaZRO$.
In reference to~\mycite{vanHameren:2022mtk}, we note that the restriction
%
%%%%%%%%%%%%%%%%%%%%%%%%%%%%%%%%%%%%%%%%
%
\begin{equation}
%\theta\bigg(y_r>\ln\frac{\sqrtS\xP}{\muY}+\ln\lambdaONE\bigg)
\theta\big(y_r>Y_\mu+\ln\lambdaONE\big)
=\theta\bigg(|\rperp|<\muY\frac{\Lambda z_r}{\lambdaONE\xP}\bigg)
\quad\textrm{replaces}\quad
\theta\bigg(\frac{|\rperp|}{\nu\sqrt{\Lambda}}< \zP_r < \frac{|\rperp|}{|\rperp+\kperp|}\bigg)
%\;\;\textrm{in~\mycite{vanHameren:2022mtk}}
\label{Eq:newrestriction}
\end{equation}
%%%%%%%%%%%%%%%%%%%%%%%%%%%%%%%%%%%%%%%%
%
in that paper, where $\xP_r=\Lambda\zP_r$.
This determines the difference between the real target IF contribution and what was called the unfamiliar real contribution in that paper.
The second condition, the upper limit on $z_r$, was included there in order to remove a collinear region that is clearly double counted when keeping both the whole familiar and unfamiliar real contribution.
In the present paper, we just follow the separation between the projectile and target regions in rapidity, which was defined in Sec.~\ref{sec:kT-fact-contr-def} in terms of scale $\muY$ and already implemented for the projectile contribution by the subtraction (\ref{eq:fam-div-subtr}).
For large $\lambda$ we get
%
%%%%%%%%%%%%%%%%%%%%%%%%%%%%%%%%%%%%%%%%
\begin{equation}
\frac{d\sigmahat^{\Real,\target}_{\inlbl\inbar}\big(\vepv,\lambdaONE,\muY\,;\Lambda,\xM\,;\pSetN\big)}{d\xP d^2\kperp}
\;\overset{\Lambda\to\infty}{\longrightarrow}\;
\frac{\alphaS\ColorC{\inlbl}}{2\pi^2|\kperp|^2}
\,d\Real_{\inlbl\inbar}^{\target}\big(\vepv,\Lambda,\lambdaONE,\muY\,;\xP,\kperp,\xM\,;\pSetN\big)
+
\Ord(\vepv)
~,
\end{equation}
%%%%%%%%%%%%%%%%%%%%%%%%%%%%%%%%%%%%%%%%
%
where, omitting the arguments $\big(\xP,\kperp,\xM\,;\pSetN\big)$,
% 
%%%%%%%%%%%%%%%%%%%%%%%%%%%%%%%%%%%%%%%%
\begin{align}
d\Real_{\inlbl\inbar}^{\target}(\vepv,\Lambda,\lambdaONE,\muY)
&=
d\Born_{\star\inbar}\times\aeps\Nc
\bigg(\frac{\mu^2}{|\kperp|^2}\bigg)^{\vepv}
\bigg[\frac{1}{\vepv^2} - \frac{4}{\vepv}\ln\frac{\Lambda\muY}{\lambdaONE\xP|\kperp|}
      %-\imag\pi
      + \bar{\myScript{R}}_{\inlbl}
\label{Eq:unfReal}
%\\&\hspace{36ex}
%   - \mathrm{Li}_2\bigg(\Big(\frac{\muY|\kperp|}{\lambdaONE\sTot\xP\xM}\Big)^2\bigg)\notag
\bigg]
~,
\end{align}
%%%%%%%%%%%%%%%%%%%%%%%%%%%%%%%%%%%%%%%%
%
with
%
%%%%%%%%%%%%%%%%%%%%%%%%%%%%%%%%%%%%%%%%
\begin{align}
\bar{\myScript{R}}_{q/\bar{q}} &=
\frac{3}{\vepv} -\frac{2\pi^2}{3} + \frac{7}{2}
-\frac{1}{\Nc^2}\,\bigg[\frac{1}{\vepv^2} + \frac{3}{2\vepv} + 4\bigg]
~,
\label{Eq:009}
\\
\bar{\myScript{R}}_{g} &=
\frac{1}{\vepv^2} 
+ \frac{1}{\vepv}\frac{11}{2}
%\notag\\&\hspace{24ex}
 - \frac{2\pi^2}{3}+ \frac{67}{9}
-\frac{n_f}{\Nc}\bigg[
   \frac{2}{3\vepv}
  +\frac{10}{9}
  -\frac{1}{\Nc^2}\bigg(\frac{1}{3\vepv}-\frac{1}{6}\bigg)
\bigg]
~.
\end{align}
%%%%%%%%%%%%%%%%%%%%%%%%%%%%%%%%%%%%%%%%
%
We see that the different phase space condition explicit in \Equation{Eq:newrestriction} only changes what in \mycite{vanHameren:2022mtk} was referred to as the ``universal pole contribution''.
The soft-collinear $1/\vepv^2$ appears now because the collinear region was not avoided, but it cancels against the $1/\epsilon^2$ pole in the $d\Virt^{\target}_{\inlbl\inbar}$, \Equation{eq:dV-targ-def}.

\subsubsection{Combined real and virtual target contribution before collinear subtraction}
Combining the virtual target IF contribution of \Equation{eq:dV-targ-def} and the real target IF contribution of \Equation{Eq:unfReal}, we get
%
%%%%%%%%%%%%%%%%%%%%%%%%%%%%%%%%%%%%%%%%
\begin{align}
d\Virt_{\inlbl\inbar}^{\target}\big(\vepv,\Lambda,\lambdaONE,\muY\big)+
d\Real_{\inlbl\inbar}^{\target}\big(\vepv,\Lambda,\lambdaONE,\muY\big)
=
d\Born_{\star\inbar}\times\aeps\,\Divunf\big(\vepv,\Lambda,\lambdaONE,\muY\big)
\end{align}
%%%%%%%%%%%%%%%%%%%%%%%%%%%%%%%%%%%%%%%%
%
with
%
%%%%%%%%%%%%%%%%%%%%%%%%%%%%%%%%%%%%%%%%
\begin{align}
\Divunf\big(\vepv,\Lambda,\lambdaONE,\muY\big)
&=
 \frac{1}{\vepv}\bigg(\frac{\mu^2}{|\kperp|^2}\bigg)^{\!\vepv}
\bigg[\myScript{J}_{\inlbl}
-2\Nc\ln\frac{\Lambda\muY}{\lambdaONE\xP|\kperp|}
\bigg] + \frac{\gamma_g}{\vepv}
+2\gamma_g\ln\frac{\mu^2}{|\kperp|^2}+2\Kcusp
\notag\\&\hspace{0ex}
-\frac{\Nc}{2}\ln^2\frac{\muY^2}{|\kperp|^2}
%-\Nc\mathrm{Li}_2\bigg(\Big(\frac{\muY|\kperp|}{\lambdaONE\sTot\xP\xM}\Big)^2\bigg)\notag
\label{Eq:divUnf}
\end{align}
%%%%%%%%%%%%%%%%%%%%%%%%%%%%%%%%%%%%%%%%
%
and
%
%%%%%%%%%%%%%%%%%%%%%%%%%%%%%%%%%%%%%%%%
\begin{equation}
\myScript{J}_{q/\bar{q}} = \frac{3\Nc}{2}+\frac{\Nc}{2}\vepv
\quad,\quad
\myScript{J}_{g} = \frac{11\Nc}{6}+\frac{n_f}{3\Nc^2}-\frac{n_f}{6\Nc^2}\vepv 
~.
\label{Eq:Jaux}
\end{equation}
%%%%%%%%%%%%%%%%%%%%%%%%%%%%%%%%%%%%%%%%
%
We changed the definition of $\myScript{J}_{\inlbl}$ by a factor $\Nc$ compared to~\mycite{vanHameren:2022mtk}, and wrote what is there denoted $\myScript{J}_{\univ}$ directly in terms of $\gamma_g$ and $\Kcusp$.
Also, due to the different restriction on the momentum $r$ in the unfamiliar real contribution, the $\Lambda$-dependence does not cancel anymore, which is to be expected for the target contribution. 
Due to the change expressed by \Equation{Eq:unfVirt2}, the $1/\vepv^2$ from the unfamiliar real contribution is cancelled again like in~\mycite{vanHameren:2022mtk}, but there is also a term $\gamma_g/\vepv$ fewer.

\subsubsection{\label{Sec:corrfinlbl}Target collinear counter term}
For a complete NLO calculation, we also need to take into account the correction on the collinear PDFs $f^{>}_{\inlbl}(X)$ on the target side, that is the second term on the first line of \Equation{Eq:kTfactorizable3}. 
This is achieved by replacing $\myScript{N}_{\inlbl}(\DELTAZRO)$ in \Equation{Eq:finlblcnst} with
%
%%%%%%%%%%%%%%%%%%%%%%%%%%%%%%%%%%%%%%%%
\begin{equation}
\myScript{N}_{\inlbl}(\DELTAZRO) \to \myScript{N}_{\inlbl}(\DELTAZRO)
      + \int_{\DELTAONE}^1\!\!dX\,\frac{\aeps}{\vepv}\bigg(\frac{\mu^2}{\muF^2}\bigg)^{\!\vepv}
        \sum_{\inlbl'}\big[\myScript{P}_{\inlbl\inlbl'}\otimes f^{>}_{\inlbl'}\big](X)
~,
\end{equation}
%%%%%%%%%%%%%%%%%%%%%%%%%%%%%%%%%%%%%%%%
%
where $\myScript{P}_{\inlbl\inlbl'}$ represent the appropriate collinear splitting functions, and we repeat that
%
%%%%%%%%%%%%%%%%%%%%%%%%%%%%%%%%%%%%%%%%
\begin{equation}
\DELTAONE\prec\DELTAZRO
~,
\end{equation}
%%%%%%%%%%%%%%%%%%%%%%%%%%%%%%%%%%%%%%%%
%
with $\DELTAZRO$ being the lower limit on $X$ implied by $f^{>}_{\inlbl}(X)$.
We want to include the extra contribution as a sum over terms with fixed $\inlbl$, so we define
%
%%%%%%%%%%%%%%%%%%%%%%%%%%%%%%%%%%%%%%%%
\begin{equation}
\DeltaN_{\inlbl}(\DELTAZRO,\DELTAONE)
=
\int_{\DELTAONE}^1\!\!dX\sum_{\inlbl'}\frac{\ColorC{\inlbl'}}{\ColorC{\inlbl}}\big[\myScript{P}_{\inlbl'\inlbl}\otimes f^{>}_{\inlbl}\big](X)
~,
\end{equation}
%%%%%%%%%%%%%%%%%%%%%%%%%%%%%%%%%%%%%%%%
%
and we can instead write
%
%%%%%%%%%%%%%%%%%%%%%%%%%%%%%%%%%%%%%%%%
\begin{align}
\myScript{N}_{\inlbl}(\DELTAZRO)
\to
\myScript{N}_{\inlbl}(\DELTAZRO)
 +\frac{\aeps}{\vepv}\bigg(\frac{\mu^2}{\muF^2}\bigg)^{\!\vepv}
\DeltaN_{\inlbl}(\DELTAZRO,\DELTAONE)
\end{align}
%%%%%%%%%%%%%%%%%%%%%%%%%%%%%%%%%%%%%%%%
%
while obtaining the same result when also summing over $\inlbl$, that is for $F^{\LO}(\DELTAZRO,\kperp)$ in \Equation{Eq:finlblcnst}.
For general splitting functions, we need to calculate
%
%%%%%%%%%%%%%%%%%%%%%%%%%%%%%%%%%%%%%%%%
\begin{align}
\DeltaN_{\inlbl}(\DELTAZRO,\DELTAONE)
&=
\int_{\DELTAONE}^1\!\!dX
\Bigg\{\bigg[\bigg(
\frac{\ColorC{\inlbl}}{[1-Z]_+}+\frac{c_{\inlbl}^{(-1)}}{Z} + c_{\inlbl}^{(0)} + c_{\inlbl}^{(1)}Z + c_{\inlbl}^{(2)}Z^2\bigg)
\otimes f^{>}_{\inlbl}
\bigg](X)
 + \gamma_{\inlbl} f^{>}_{\inlbl}(X)\Bigg\}
~.
\end{align}
%%%%%%%%%%%%%%%%%%%%%%%%%%%%%%%%%%%%%%%%
%
The coefficients $c_{\inlbl}^{(k)}$ contain ratios $\ColorC{\inlbl'}/\ColorC{\inlbl}$.
We find
%
%%%%%%%%%%%%%%%%%%%%%%%%%%%%%%%%%%%%%%%%
\begin{align}
\DeltaN_{\inlbl}(\DELTAZRO,\DELTAONE)
\;\overset{\lambda\to\infty}{\longrightarrow}\;
\int_{0}^1\!dX
  \,f^{>}_{\inlbl}(X)\bigg[
c_{\inlbl}^{(-1)}\ln\frac{X}{\DELTAONE}
+ c_{\inlbl}^{(0)} + \frac{1}{2}\,c_{\inlbl}^{(1)} + \frac{1}{3}\,c_{\inlbl}^{(2)}
+ \gamma_{\inlbl}
\bigg]
~.
\end{align}
%%%%%%%%%%%%%%%%%%%%%%%%%%%%%%%%%%%%%%%%
%
The term with $\ColorC{\inlbl}/[1-Z]_+$ turns out to vanish for $\DELTAONE/X\to0$, and can be put to $0$ in the integrand thanks to $\DELTAONE/\DELTAZRO\to0$, as do other terms with powers of $\DELTAONE/X$.
The term with $\ln(X/\DELTAONE)$ remains, and comparing the expression to \Equation{Eq:divUnf}, we see that we must choose
%
%%%%%%%%%%%%%%%%%%%%%%%%%%%%%%%%%%%%%%%%
\begin{equation}
\DELTAONE = \frac{\lambdaONE\xP|\kperp|}{\lambda\muY}
\end{equation}
%%%%%%%%%%%%%%%%%%%%%%%%%%%%%%%%%%%%%%%%
%
in order for all poles in $\vepv$ to cancel for the target contribution.
The relation $\DELTAONE\prec\DELTAZRO$ is then indeed consistent with $\lambdaONE\prec\lambdaZRO$.

Let us concentrate on $\inlbl=g$.
Then, we have $c_{\inlbl}^{(-1)}=\ColorC{\inlbl}=\ColorC{g}=2\Nc$, so indeed the cancellation for $\ln(X/\DELTAONE)$ happens.
Furthermore,
%
%%%%%%%%%%%%%%%%%%%%%%%%%%%%%%%%%%%%%%%%%
\begin{equation}
c_{\inlbl}^{(0)} = -4\Nc + \frac{\ColorC{q}}{\ColorC{g}}\,n_f
\;\;,\quad
c_{\inlbl}^{(1)} = 2\Nc - \frac{\ColorC{q}}{\ColorC{g}}\,2n_f
\;\;,\quad
c_{\inlbl}^{(2)} = -2\Nc + \frac{\ColorC{q}}{\ColorC{g}}\,2n_f
~,
\end{equation}
%%%%%%%%%%%%%%%%%%%%%%%%%%%%%%%%%%%%%%%%
%
so
%
%%%%%%%%%%%%%%%%%%%%%%%%%%%%%%%%%%%%%%%%
\begin{equation}
\gamma_{\inlbl} + c_{\inlbl}^{(0)} + \frac{1}{2}\,c_{\inlbl}^{(1)} + \frac{1}{3}\,c_{\inlbl}^{(2)}
%=\gamma_g - \frac{11\Nc}{3} + \frac{\Nc^2-1}{2\Nc^2}\frac{2n_f}{3}
=-\frac{11\Nc}{6} - \frac{n_f}{3\Nc^2}
\end{equation}
%%%%%%%%%%%%%%%%%%%%%%%%%%%%%%%%%%%%%%%%
%
which cancels against (the $1/\vepv$ part of) $\myScript{J}_{g}$.
For $\inlbl=q$, we have
%
%%%%%%%%%%%%%%%%%%%%%%%%%%%%%%%%%%%%%%%%
\begin{equation}
c_{\inlbl}^{(-1)} = \frac{\ColorC{g}}{\ColorC{q}}\,2\CF = 2\Nc
\;\;,\quad
c_{\inlbl}^{(0)} = -\CF - \frac{\ColorC{g}}{\ColorC{q}}\,2\CF
\;\;,\quad
c_{\inlbl}^{(1)} = -\CF + \frac{\ColorC{g}}{\ColorC{q}}\,\CF
~,
\end{equation}
%%%%%%%%%%%%%%%%%%%%%%%%%%%%%%%%%%%%%%%%
%
and $c_{\inlbl}^{(2)} = 0$.
So again, we observe that the $1/\vepv$ part of the term with the explicit logarithm cancels, and
%
%%%%%%%%%%%%%%%%%%%%%%%%%%%%%%%%%%%%%%%%
\begin{equation}
\gamma_{\inlbl} + c_{\inlbl}^{(0)} + \frac{1}{2}\,c_{\inlbl}^{(1)} + \frac{1}{3}\,c_{\inlbl}^{(2)}
%= \frac{3}{2}\CF-\CF-2\Nc - \frac{1}{2}\CF + \frac{1}{2}\Nc
= -\frac{3}{2}\Nc
\end{equation}
%%%%%%%%%%%%%%%%%%%%%%%%%%%%%%%%%%%%%%%%
%
cancels against (the $1/\vepv$ part of) $\myScript{J}_{q}$.

\subsection{Complete finite target IF contribution at NLO}
Summarizing, we can write the corrections coming form renormalization of the collinear PDF of the target hadron in the $k_T$-factorizable part of the cross section as follows:
%
%%%%%%%%%%%%%%%%%%%%%%%%%%%%%%%%%%%%%%%%
\begin{equation}
d\Coll^{\target}_{\inlbl\inbar}\big(\vepv,\Lambda,\lambdaONE,\muY,\muF\big)
=
d\Born_{\star\inbar}
\times\aeps\,\CFF^{\target}_{\inlbl}\big(\vepv,\Lambda,\lambdaONE,\muY,\muF\big)
\end{equation}
%%%%%%%%%%%%%%%%%%%%%%%%%%%%%%%%%%%%%%%%
%
where we omitted the arguments $\big(\xP,\kperp,\xM\,;\pSetN\big)$ and with
%
%%%%%%%%%%%%%%%%%%%%%%%%%%%%%%%%%%%%%%%%
\begin{equation}
\CFF^{\target}_{\inlbl}\big(\vepv,\Lambda,\lambdaONE,\muY,\muF\big)
=
\frac{1}{\vepv}\bigg(\frac{\mu^2}{\muF^2}\bigg)^{\!\vepv}
  \bigg[
\myScript{J}_{\inlbl}^{(0)}
-2\Nc\ln\frac{\Lambda\muY}{\lambdaONE\xP|\kperp|}
\bigg]
~,\label{Eq:coll-sub-result}
\end{equation}
%%%%%%%%%%%%%%%%%%%%%%%%%%%%%%%%%%%%%%%%
%
where we write $\myScript{J}_{\inlbl}=\myScript{J}_{\inlbl}^{(0)}+\vepv\myScript{J}_{\inlbl}^{(1)}$, with $\myScript{J}_{\inlbl}$ is defined in \Equation{Eq:Jaux}.
Combining with the target contribution (\Equation{Eq:divUnf}), we get, now with explicit arguments $(\xP,\kperp)$,
%
%%%%%%%%%%%%%%%%%%%%%%%%%%%%%%%%%%%%%%%%
\begin{align}
&\VplusR^{\target}_{\inlbl}\big(\vepv,\Lambda,\lambdaONE,\muY\,;\xP,\kperp\big)
- \CFF^{\target}_{\inlbl}\big(\vepv,\Lambda,\lambdaONE,\muY,\muF\,;\xP,\kperp\big)
\notag\\&\hspace{4ex}=
 \frac{\gamma_g}{\vepv}
+
\bigg[ \myScript{J}_{\inlbl}^{(0)} -2\Nc\ln\frac{\Lambda\muY}{\lambdaONE\xP|\kperp|} \bigg] 
\ln\frac{\muF^2}{|\kperp|^2}
+2\gamma_g\ln\frac{\mu^2}{|\kperp|^2}+2\Kcusp+\myScript{J}_{\inlbl}^{(1)}
-\frac{\Nc}{2}\ln^2\frac{\muY^2}{|\kperp|^2}
\notag\\&\hspace{4ex}\equiv
 \frac{\gamma_g}{\vepv}
+\VplusRminusC^{\target}_{\inlbl}\big(\Lambda,\lambdaONE,\muY,\muF\,;\xP,\kperp\big)
%-\Nc\mathrm{Li}_2\bigg(\Big(\frac{\muY|\kperp|}{\lambdaONE\sTot\xP\xM}\Big)^2\bigg)
~.\label{Eq:target}
\end{align}
%%%%%%%%%%%%%%%%%%%%%%%%%%%%%%%%%%%%%%%% 
%
The remaining divergence $\gamma_g/\vepv$ indicates that we counted a power of $\alphaS$ too few in the UV subtraction on the projectile side, namely, we should also renormalize the $\alphaS$ in \Equation{Eq:finlblcnst}. The rest of the expression is finite.

\subsection{Green's function contribution}
In this section we discuss the contribution of the real-emission in MRK. The corresponding limit of the matrix element is given by \Equation{Eq:BFKLME}. The contribution to the cross section is:
%
%%%%%%%%%%%%%%%%%%%%%%%%%%%%%%%%%%%%%%%%
\begin{align}
&\frac{d\sigmahat_{\inlbl\inbar}^{\Real,\Green}\big(\vepv,\lambdaONE,\muY\,;\Lambda,\xM\,;\pSetN\big)}{d\xP d^2\kperp}
\notag\\&\hspace{4ex}=
 \int\frac{d^4q}{(2\pi)^3}\,\delta_+(q^2)
 \int\frac{\mu^{2\vepv}d^{4-2\vepv}r}{(2\pi)^{3-2\vepv}}\,\delta_+(r^2)
 \,\theta\big(Y_\mu<y_r<Y_\mu+\ln\lambdaONE\big)
\notag\\&\hspace{8ex}\times
\int_0^1dz_q\,\delta\big(z_q-\xP_q/\Lambda\big)
\int_0^1dz_r\,\delta\big(z_r-\xP_r/\sqrt{\Lambda}\big)
\\&\hspace{8ex}\times
 2\xP|\kperp|^2\,\delta(\xP+\xP_q+\xP_r-\Lambda)\,\delta^2(\kperp+\qperp+\rperp)
 \,\frac{d\Sigma_{\inlbl\inbar}}{dqdr}\big(\Lambda\pP,\xM\pM\,;q,r,\pSetN\big)
\,\JetB\big(\pSetN\big)
\notag~.
\end{align}
%%%%%%%%%%%%%%%%%%%%%%%%%%%%%%%%%%%%%%%%
%
Note that due to definition of MRK, given in Sec.~\ref{Sec:BFKLlimit}, the $x_r$ is set to scale with $\sqrt{\Lambda}$ instead of $\Lambda$.
For our purpose, it can be any behavior $t(\Lambda)\to\infty$ with $t(\Lambda)/\Lambda\to0$, but we will stick to the square root for illustration.
Using \Equation{Eq:BFKLME}, we find
%
%%%%%%%%%%%%%%%%%%%%%%%%%%%%%%%%%%%%%%%%
\begin{align}
\frac{d\sigmahat_{\inlbl\inbar}^{\Real,\Green}\big(\vepv,\lambdaONE,\muY\,;\Lambda,\xM\,;\pSetN\big)}{d\xP d^2\kperp}
\overset{\lambda\to\infty}{\longrightarrow}
d\Born_{\star\inbar}\big(\xP,\kperp,\xM\,;\pSetN\big)
\times\aeps\,\myScript{R}^{\Green}(\vepv,\lambdaONE\,;\kperp)
\end{align}
with
\begin{align}
\myScript{R}^{\Green}(\vepv,\lambdaONE\,;\kperp)
&=
\frac{\mu^{2\vepv}}{2\piep}
\int_0^1\frac{dz_r}{z_r}\int d^{2-2\vepv}\rperp
\theta\bigg(\frac{\sqrt{\Lambda}\,z_r}{\lambdaONE\xP}<\frac{|\rperp|}{\muY}<\frac{\sqrt{\Lambda}\,z_r}{\xP}\bigg)
\,4\Nc\,\frac{|\kperp|^2}{|\rperp|^2|\rperp+\kperp|^2}
\notag\\&\hspace{0ex}=
4\Nc\,\frac{\mu^{2\vepv}}{2\piep}
\int_1^{\lambdaONE}\frac{dz_r}{z_r}\int d^{2-2\vepv}\rperp
\,\frac{|\kperp|^2}{|\rperp|^2|\rperp+\kperp|^2}
\label{Eq:GF-contr-0}\\&\hspace{0ex}=
\frac{1}{\vepv}\bigg(\frac{\mu^2}{|\kperp|^2}\bigg)^{\!\vepv}
\big[{-}4\Nc\ln\lambdaONE\big]
\label{Eq:BFKL01}
~.
\end{align}
%%%%%%%%%%%%%%%%%%%%%%%%%%%%%%%%%%%%%%%%
%
Remember that we should imagine that $\vepv$ is negative to regularize IR divergences, and the result is positive.

We see that adding this contribution to \Equation{Eq:defunfReal} simply removes $\lambdaONE$ from \Equation{Eq:unfReal}.
So if one adds the real Green's function contribution (\ref{Eq:BFKL01}) to the real-emission target contribution \Equation{Eq:unfReal}, the separator $\lambdaONE$ disappears from the computation and one should put $\lambdaONE=1$ in the collinear subtraction term (\ref{Eq:coll-sub-result}).
This way, the cancellation of divergences is not spoiled.
This result means that if one wants to keep the Green's function contribution separate, then the following piece of collinear subtraction term actually belongs to it:
\begin{align}
\CFF^{\Green}\big(\vepv,\Lambda,\lambdaONE,\muY,\muF\big)
&=
\CFF^{\target}_{\inlbl}\big(\vepv,\Lambda,\lambdaONE=1,\muY,\muF\,;\xP,\kperp\big)-\CFF^{\target}_{\inlbl}\big(\vepv,\Lambda,\lambdaONE,\muY,\muF\,;\xP,\kperp\big)
\nonumber\\&\hspace{0ex}=
\frac{1}{\vepv}\bigg(\frac{\mu^2}{\muF^2}\bigg)^{\!\vepv}\big[{-}2N_c \ln\lambdaONE\big] 
\nonumber\\&\hspace{0ex}=
4N_c \frac{\mu^{2\vepv}}{2\piep}
\int_{1}^{\lambdaONE}\frac{dz_r}{z_r} \int \frac{d^{2-2\vepv}\rperp}{|\kperp+\rperp|^2}\,\theta\big(\mu_F^2-|\kperp+\rperp|^2\big)
~.
%d\Born_{\star\inbar}\big(\xP,\kperp,\xM\,;\pSetN\big).
\end{align}
Subtracting this contribution form the Green's function contribution (\ref{Eq:GF-contr-0}) and adding the virtual part $d\Virt^{\Green}_{\star\inbar}$, \Equation{Eq:unfVirt2:Green}, we obtain the Green's function contribution which is IR-finite:
\begin{align}
&\VplusRminusC^{\Green}(\lambdaONE,\muF\,;\kperp)
\label{Eq:Green}\\&\hspace{4ex}=
4N_c\frac{\mu^{2\vepv}}{2\piep}
\int_1^{\lambdaONE}\frac{dz_r}{z_r} \int\frac{d^{2-2\vepv}\rperp}{|\kperp+\rperp|^2} 
\left[ \frac{|\kperp|^2}{|\rperp|^2} - \theta\big(\mu_F^2-|\kperp+\rperp|^2\big) + \delta^{(2-2\epsilon)}(\rperp)\frac{\pi_\epsilon}{\epsilon}\big|\kperp^2\big|^{1-\epsilon} \right]
\notag
~.
\end{align}

\section{\label{sec:UPDF}Structure of the unintegrated PDF at NLO and beyond}
In this section we will use the results for finite target IF and Green's function contributions to the cross section, obtained in the previous section, to derive the expression for UPDF at NLO in $\alphaS$ and to understand the structure of the resummation of high-energy logarithms. This will lead us to the (BFKL-Collins-Ellis) evolution equation for the Green's function (\ref{eq:G-evol-0}) and to the matching ansatz (\ref{eq:UPDF-PDF-match}), expressing the initial condition of UPDF evolution at the scale $\muY=|\kperp|$ in terms of collinear PDFs.

\subsection{NLO Matching ansatz for UPDF at $\muY=|\kperp|$}

Reminding the reader of the LO structure from \Equation{Eq:lamLimBorn} and \Equation{Eq:finlblcnst}, we can add-up the LO and NLO contributions from the target side to the NLO CF cross section in the $\lambda\to\infty$ limit, and  write it up to and including $\Ord\big(\alphaS\big)$ as:
\begin{equation}
    \frac{d\sigma^{\ColFac,\Born+\NLO}_{\lambda\to\infty,\target}\big(\pSetN\big)}{d\xP d^2\kperp}=   F^{\LO+\NLO}(\DELTAZRO,\kperp; \muY) \sum_{\inbar}\int_0^1d\xM\,f_{\inbar}(\xM)
\,d\Born_{\star\inbar}\big(\xP,\kperp,\xM\,;\pSetN\big)
~,
\end{equation}
with
%
%%%%%%%%%%%%%%%%%%%%%%%%%%%%%%%%%%%%%%%%
\begin{align}
&  F^{\LO+\NLO}(\DELTAZRO,\kperp; \muY)
\nonumber\\&\hspace{2ex}
= 
\sum_{\inlbl}\frac{\alphaS\ColorC{\inlbl}}{2\pi^2|\kperp|^2}\,\int_{\DELTAZRO}^1\!\!dX\,f_{\inlbl}(X,\muF) \bigg[ 1+\aeps\VplusRminusC^{\target}_{\inlbl}\big(\lambda X,\lambdaONE,\muY,\muF\,;\xP,\kperp\big)
\\& \hspace{33ex} 
+\aeps\VplusRminusC^{\Green}\big(\lambdaONE,\muF\,;\kperp\big) \bigg]
~.\notag
%
% \notag\\&\hspace{2ex}=
% %
%  \sum_{\inlbl}\frac{\alphaS\ColorC{\inlbl}}{2\pi^2}
% \Bigg\{\int_{\DELTAZRO}^1\!\!dX\,f_{\inlbl}(X,\muF) \bigg[1+\aeps\,\DeltaI_{\inlbl}\bigg(\frac{\lambdaONE\xP |\kperp|}{\lambda X \muY},\kperp,\muF\bigg)+\aeps\,\DeltaU(\kperp,\muY)\bigg]
% \\&\hspace{16ex}\times
% \bigg[
% \frac{1}{|\kperp|^2}
% +\aeps
% \int_1^{\lambdaONE}\frac{dz_r}{z_r}
% \int{d^{2-2\vepv}\rperp}
% \,\frac{1}{|\kperp + \rperp|^2}
% \,\kernel(\kperp+\rperp,\kperp,\muF)
% \bigg]+\Ord\big(\alphaS^2\big)
% \Bigg\}
% % %
\end{align}
The quantities $\VplusRminusC^{\target}_{\inlbl}$ and $\VplusRminusC^{\Green}$ were defined in \Equation{Eq:target} and \Equation{Eq:Green} respectively, leading to
\begin{align}
&F^{\LO+\NLO}(\DELTAZRO,\kperp; \muY)
\notag\\&\hspace{2ex}
=
\sum_{\inlbl}\frac{\alphaS\ColorC{\inlbl}}{2\pi^2}\,\int_{\DELTAZRO}^1\!\!dX\,f_{\inlbl}(X,\muF) \int{d^{2-2\vepv}\kperp'}\Bigg\{
\frac{1}{|\kperp'|^2}\bigg[1+\aeps\,\DeltaI_{\inlbl}\bigg(\frac{\lambdaONE\xP |\kperp'|}{\lambda X \muY},\kperp',\muF\bigg)\bigg]
\label{eq:UPDF-NLO-00}\\&\hspace{32ex}\times
\GreenF\big(\lambdaONE^{-1},\muF,\kperp',\kperp \big)
\bigg[ 1+\aeps\,\DeltaU(\kperp,\muY)\bigg]+\Ord\big(\alphaS^2\big) \Bigg\}
\nonumber~,
\end{align}
%%%%%%%%%%%%%%%%%%%%%%%%%%%%%%%%%%%%%%%%
%
where we abbreviate
%
%%%%%%%%%%%%%%%%%%%%%%%%%%%%%%%%%%%%%%%%
\begin{align}
\DeltaI_{\inlbl}(z,\kperp,\muF) &= \bigg[ \myScript{J}_{\inlbl}^{(0)} -2\Nc\ln\frac{1}{z} \bigg] 
\ln\frac{\muF^2}{|\kperp|^2}
+2\gamma_g\ln\frac{\mu^2}{|\kperp|^2}+2\Kcusp+\myScript{J}_{\inlbl}^{(1)}
~, \\
 \DeltaU(\kperp,\muY) &= -\frac{\Nc}{2}\ln^2\frac{\muY^2}{|\kperp|^2}
~,
\end{align}
%%%%%%%%%%%%%%%%%%%%%%%%%%%%%%%%%%%%%%%%
%
and where the Green's function $\GreenF$ at this order takes the form:
\begin{equation}
\GreenF\big(y,\muF,\kperp',\kperp\big) = \delta^{(2-2\epsilon)}(\kperp'-\kperp)
+\aeps \int_y^1 \frac{dz}{z}\, \kernel(\kperp',\kperp,\muF)  + \Ord\big(\alphaS^2\big)
~, \label{eq:G-NLO}
\end{equation}
with
%
%%%%%%%%%%%%%%%%%%%%%%%%%%%%%%%%%%%%%%%%
\begin{align}
\kernel(\kperp',\kperp,\muF)
=
4N_c\frac{\mu^{2\vepv}}{2\piep}
\bigg[
\frac{1}{|\kperp'-\kperp|^2}
      - \frac{\theta\big(\mu_F^2-|\kperp'|^2\big)}{|\kperp|^2}
      + \delta^{(2-2\epsilon)}\big(\kperp'-\kperp\big)\frac{\pi_\epsilon}{\epsilon}\big|\kperp^2\big|^{-\epsilon}
\bigg]
~.\label{eq:K-prelim}
\end{align}
%%%%%%%%%%%%%%%%%%%%%%%%%%%%%%%%%%%%%%%%
%
At this point one can observe that for the obtained expression for the UPDF (\ref{eq:UPDF-NLO-00}) the limit $\lambda\to\infty$ for $x$-fixed, is equivalent to the limit $x\to 0$ for $\lambda=1$.
Moreover, one can completely remove the large $\ln [(\xP|\kperp|)/(X\muY)]$ from the IF contribution $\DeltaI_\inlbl$ and move them to the Green's function $\GreenF$ by choosing $\lambdaONE=\big(X\muY\big)/\big(\xP|\kperp|\big)$.
Then, generalizing thus-obtained result to all orders in $\alphaS$ and setting $\muY=|\kperp|$, we conjecture the following  matching formula between the collinear PDF and unintegrated PDF at the scale $\muY=|\kperp|$:
\begin{align}
F\big(x,\kperp,\muY=|\kperp|\big)
&=
\sum_{\inlbl}\int_x^1\!dX  \, f_{\inlbl}(X,\mu_F) \int{d^{2}\kperp'}\Impact_{\inlbl}\big(\kperp',\muF\big) 
\,\GreenF\bigg(\kperp',\kperp,\frac{X}{x},\muF\bigg)
~,
\label{eq:UPDF-PDF-match}
\end{align}
where all we know about the partonic target impact-factor $\Impact_\inlbl$ so far is it's first two perturbative orders:
\begin{equation}
    \Impact_{\inlbl}(\kperp,\muF) = \frac{\alphaS C_{\inlbl}}{2\pi^2|\kperp|^2} \Big[ 1 + \aeps \DeltaI_\inlbl (1,\kperp,\muF) + \Ord\big(\alphaS^2\big) \Big]
\label{eq:IF-NLO}
~.
\end{equation}
\Equation{eq:UPDF-PDF-match}, together with the solution of the evolution equation for the Green's function $\GreenF$ derived in the next subsection, provides one with the initial condition for the evolution of UPDF at $\muY=|\kperp|$ in terms of the usual collinear PDFs.
With the help of evolution equation (\ref{eq:evol-eq-muY-fin}) the UPDF can be evolved up (or down) to the scale $\muY$ of the process under consideration.

One may notice, that in Eq.~(\ref{eq:UPDF-PDF-match}) we do not put any restrictions on the $X$-integration from below, which played such an important role in all arguments above.
Instead, we integrate all the way down to $x$, which is equivalent to the integration in $X$ down to $1/\lambda$ in the previous discussion.
In this way, we have come back to the point of view of traditional $k_T$-factorization, which ignores the existence of the non-$k_T$-factorizable contribution to the cross section.
A systematic approach to avoid this problem would be to perform matching between HEF and CF in the $X$-variable, in the spirit of Refs.~\mycite{Lansberg:2021vie} and \mycite{Lansberg:2023kzf}.
We plan to come back to this discussion in future works.

\subsection{The $x$-evolution of the Green's function: BFKL-Collins-Ellis equation}

The Green's function (\ref{eq:G-NLO}) satisfies the evolution equation, which can be written in the integral form as:
\begin{align}
\GreenF\big(\kperp',\kperp,x,\muF\big)
&= \delta^{(2-2\epsilon)}(\kperp'-\kperp) \nonumber \\
& + \aeps \int_x^1 \frac{dz}{z} \int d^{2-2\vepv}\qperp\,\kernel\big(\kperp',\qperp,\muF\big)
\,\GreenF\bigg(\qperp,\kperp,\frac{x}{z},\muF\bigg) %+ \Ord\big(\alphaS^2\big)
~. \label{eq:G-evol-0}
\end{align}
Let us note that the kernel (\ref{eq:K-prelim}) can be extended with the term $-\theta(\muF^2-|\kperp'|^2)$ $\delta^{(2-2\epsilon)}(\kperp)$ $\frac{\piep}{\epsilon}|\kperp|^{-2\epsilon}$, which is simply equal to zero for $\epsilon<0$.
After that, the evolution equation for the Green's function can be rewritten in the following way:
\begin{align}
&\GreenF\big(\kperp',\kperp,x,\muF\big)
= \delta^{(2-2\epsilon)}(\kperp'-\kperp)
\label{eq:BFKL-CE-0}\\&\hspace{3ex}+ 
\aeps\int_x^1 \frac{dz}{z} \int d^{2-2\vepv}\qperp\Big[\kernelBFKL\big(\kperp',\qperp\big)
-\theta\big(\muF^2-|\kperp'|^2\big)\,\kernelBFKL\big(0,\qperp\big) \Big]
\GreenF\bigg(\qperp,\kperp,\frac{x}{z},\muF\bigg)
~,
\notag
\end{align}
in terms of the ordinary LO BFKL kernel in dimensional regularization:
\begin{align}
\kernelBFKL\big(\kperp',\kperp\big)
=
4N_c\frac{\mu^{2\vepv}}{2\piep}
\bigg[
\frac{1}{|\kperp'-\kperp|^2}
      + \delta^{(2-2\epsilon)}\big(\kperp'-\kperp\big)\frac{\pi_\epsilon}{\epsilon}\big|\kperp^2\big|^{-\epsilon}
\bigg]
~,\label{eq:K-BFKL}
\end{align}
which can be equivalently written in a form where the IR-divergence is cancelled explicitly when the kernel is convoluted with a smooth function of transverse momentum as:
\begin{equation}
\kernelBFKLfin\big(\kperp,\rperp,\kperp'\big) =  \frac{2\Nc}{\pi|\rperp |^2} \Big[\delta^{(2)}\big(\kperp'-\kperp-\rperp\big) - \theta\big(|\rperp| < |\kperp|\big)\,\delta^{(2)}\big(\kperp'-\kperp\big) \Big]
~. \label{eq:K-BFKL-fin}
\end{equation}
The equation (\ref{eq:BFKL-CE-0}) is closely related to evolution equation (3.4) in Ref.~\mycite{Collins:1991ty} for the quantity $\tilde{X}$, defined in Eqn.~(3.2) of that paper.
In our notation this quantity in momentum-fraction space%
\footnote{In Ref.~\mycite{Collins:1991ty} the Mellin variable $j$, conjugate to the momentum fraction $x$ is used, see Eq. (3.1) there.} %
 can be expressed up to an overall numerical factor as:
\begin{align}
\tilde{X}(x,\kperp')
&\propto
\int d^2 \kperp \int_x^1 \frac{dz}{z} \myScript{G}\big(\kperp',\kperp,z/x,\muF \big)
\notag\\&\hspace{4ex}\times
\Big[ d\Born_{\star\inbar}\big(z,\kperp,\xM\,;\pSetN\big) - d\Born_{\star\inbar}\big(z,0,\xM\,;\pSetN\big) \theta\big(\muF^2- |\kperp|^2\big) \Big]
~,
\end{align}
where $\myScript{G}\big(\kperp',\kperp,x,\muF \big)=\aeps\int d^{2-2\epsilon}\qperp\,\kernelBFKL \big(\kperp', \qperp\big)\,\GreenF\big(\qperp,\kperp,x,\muF \big) $.
Indeed, the equation (\ref{eq:BFKL-CE-0}) can be rewritten%
\footnote{One needs to assume the existence of the inverse kernel $K_{\mathrm{BFKL}}^{-1}(\kperp,\qperp)$ on the intermediate step of this derivation, which is justified, because BFKL kernel has no zero eigenvalues.} %
 in terms of the Green's function $\myScript{G}$ as:
\begin{align}
\myScript{G}\big(\kperp',\kperp,\lambdaONE,\muF\big)
&= \kernelBFKL\big(\kperp',\kperp\big) +
\aeps \int d^{2-2\epsilon}\qperp \kernelBFKL\big(\kperp',\qperp\big)
\\&\hspace{8ex}\times
\Big[ \myScript{G}\big(\qperp,\kperp,\lambdaONE,\muF \big) - \theta\big(\muF^2-|\qperp|^2\big)\myScript{G}\big(0,\kperp,\lambdaONE,\muF \big) \Big]
~,
\notag
\end{align}
which directly leads to the Eq.~(3.4) for the quantity $\tilde{X}$ in Ref.~\mycite{Collins:1991ty}. This equation performs the resummation of higher-order corrections to the CF coefficient function, enhanced by logarithms of partonic energy, which is a main goal of HEF in the LLA. Strictly speaking, the resummed coefficient function, computed with the help of BFKL-Collins-Ellis equation is defined in a scheme which deviates form the standard $\overline{MS}$-scheme at N$^3$LO and beyond. To stay strictly within the $\overline{MS}$-scheme to all orders, the formalism of Refs.~\mycite{Catani:1990xk,Catani:1990eg,Catani:1994sq} should be used instead.

In the Ref.~\mycite{Collins:1991ty} it was shown, that together with $\ln (1/x)$, this evolution resums a large class of $\muF$-dependent corrections, in particular, all double logs of the form $\left(\alphaS\ln (1/x) \ln (\muF^2/\kperp^2)\right)^n$, so the $\mu_F$-dependence will cancel within Eq.~(\ref{eq:UPDF-PDF-match}) between PDF and the Green's function to some logarithmic accuracy. In this sense, the UPDF is $\mu_F$-independent, but in practice the cancellation will not be perfect, due to limited accuracy of various factors in this formula. In fact, if the usual NLO or NNLO PDFs are used for the phenomenological computation, then one has to truncate the resummation in $G$ down to the double-logarithmic accuracy, resumming only terms $\propto\left(\alphaS\ln (1/x) \ln (\muF^2/\kperp^2)\right)^n$, to avoid the significant mismatch in the $\muF$-dependence between the PDF and $G$, as it was done \eg\ in Refs.~\mycite{Lansberg:2021vie} and \mycite{Lansberg:2023kzf}.

\subsection{Cross-check of the $\muY$-evolution at NLO}
In this subsection we are checking if the UPDF (\ref{eq:UPDF-NLO-00}), which we have derived from the target side  up to NLO in $\alphaS$, with $\lambda=1$ and $\lambdaONE=\big(X\muY\big)/\big(\xP|\kperp|\big)$:
\begin{align}
&F^{\LO+\NLO}(x,\kperp; \muY)
\notag\\&\hspace{2ex}=
\sum_{\inlbl}\frac{\alphaS\ColorC{\inlbl}}{2\pi^2}\,\int_{x}^1 dX\,f_{\inlbl}(X,\muF)
%\\&\hspace{20ex}\times
\Bigg\{ \frac{1}{|\kperp|^2} \Big[1+\aeps\,\Delta I_{\inlbl}\big(1,\kperp,\muF\big)+\aeps\,\Delta U\big(\kperp,\muY\big)\Big]
\\&\hspace{28ex}+
\aeps
\ln \bigg(\frac{X\mu_Y}{x|\kperp|}\bigg)
\int{d^{2-2\vepv}\rperp}
\,\frac{1}{|\kperp + \rperp|^2} 
\,\kernel(\kperp+\rperp,\kperp,\muF) \Bigg\} ,  \notag   
\end{align}
satisfies the $\muY$-evolution equation (\ref{eq:evol-eq-muY-fin}), as it necessarily should for the consistency of our matching ansatz (\ref{eq:UPDF-PDF-match}).
The derivative of this result w.r.t.\ $\ln \muY^2$ is
\begin{align}
&\frac{d}{d \ln \muY^2}\,F^{\LO+\NLO}(x,\kperp; \muY)
\\&\hspace{2ex}=
%=
\sum_{\inlbl}\frac{\alphaS\ColorC{\inlbl}}{2\pi^2}\,\int_{x}^1 dX\,f_{\inlbl}(X,\muF)
\bigg\{ \frac{\aeps}{|\kperp|^2} \bigg[{-}\Nc \ln \frac{\muY^2}{|\kperp|^2} \bigg]
%\notag\\&\hspace{28ex}+
+
\frac{\aeps}{2}
\int
\frac{d^{2-2\vepv}\rperp}{|\kperp + \rperp|^2} 
\,\kernel(\kperp+\rperp,\kperp,\muF) \bigg\}
\notag\\&\hspace{2ex}=
- \aeps \Nc \ln \frac{\muY^2}{|\kperp|^2}\,F^{\LO}(x,\kperp; \muY) + \frac{\aeps}{2} \int{d^{2}\rperp} {d^{2}\kperp'}
\, F^{\LO}(x,\kperp'; \muY)\,\kernelBFKLfin(\kperp,\rperp,\kperp')
\notag\\&\hspace{4ex}-
\frac{\aeps}{2} \int{d^{2}\rperp} {d^{2}\kperp'}
\, F^{\LO}(x,\kperp'; \muY)\,\theta(\muF-|\kperp'|) \kernelBFKLfin(\kperp,\rperp,0)
~, \label{eq:cross-check2}
\end{align}
with $F^{\LO}$ defined in Eq.~(\ref{Eq:finlblcnst}).
Using the definition of the finite kernel (\ref{eq:K-BFKL-fin}) one can see, that the first two terms of Eq.~(\ref{eq:cross-check2}) formally coincide with the right-hand side of the Eq.~(\ref{eq:evol-eq-muY-fin}) for $\mu_0=|\kperp|$ in the asymptotic limit $x\ll 1$ and $\mu_Y\gg |\kperp|$, where one can neglect the $\theta$-function in Eq.~(\ref{eq:evol-eq-muY-fin}).
The reason why the projectile contribution satisfies the evolution equation (\ref{eq:evol-eq-muY-fin}) only in the low-$x$ limit is that this limit corresponds to the limit $\lambda\to\infty$ which was strictly taken in the derivation of the projectile contribution.
The last term in Eq. (\ref{eq:cross-check2}) is responsible for the subtraction of the collinear singularity, which arises in the next-to-last term, and can be related to the renormalization of the LO PDF.

This last term can be nullified if we go back to $(2-2\epsilon)$-dimensional transverse space and introduce the following $O(\alphaS^0)$ term to the UPDF\footnote{This term lives at $\kperp=0$ and therefore was invisible to the analysis above, since we had been working at finite $\kperp$ all the way up to this point.}:
\begin{align}
&F^{\LO+\NLO}(x,\kperp,\muY)
\notag\\&\hspace{8ex}\to
%\to
\delta^{(2-2\epsilon)}(\kperp)\,x\sum\limits_{\inlbl}\int_x^1 \frac{dz}{z}\,Z_{g\inlbl}(z)\,f_{\inlbl}\bigg(\frac{x}{z},\muF\bigg)\;+\; F^{\LO+\NLO}(x,\kperp,\muY)
~, \label{eq:UPDF+as0}
\end{align}
where one should take into account that the collinear renormalization factors $Z_{ij}(z)=\delta_{ij}\delta(1-z)+O(\alphaS)$. Then the term in the last line of \Equation{eq:cross-check2} up to $O\big(\alphaS^2\big)$ becomes:
\begin{align}
&  -\frac{\aeps \Nc}{2 \pi |\kperp|^2} \int {d^{2-2\epsilon}\kperp'}
\, F^{\Ord(\alphaS^0)+\LO}(x,\kperp'; \muY)\,\theta(\muF-|\kperp'|)
\notag\\&\hspace{4ex}=
-\frac{\aeps \Nc}{2 \pi |\kperp|^2}\,x\,f_g(x,\muF) - \frac{\aeps\Nc}{2\pi |\kperp|^2} \Bigg\{ \frac{\alphaS}{2\pi} \frac{1}{\epsilon} \left(\frac{\mu^2}{\muF^2} \right)^\epsilon x \sum\limits_{\inlbl} \int_x^1 \frac{dz}{z}\,P_{g\inlbl}\bigg(\frac{x}{z}\bigg)\,f_{\inlbl}(z)
\\&\hspace{28ex}
+ \sum\limits_{\inlbl} \frac{\alphaS\ColorC{\inlbl}}{2\pi} \int_x^1 dX\, f_{\inlbl}(X,\muF) \int \frac{\mu^{2\epsilon} d^{2-2\epsilon} \kperp'}{\piep (\kperp')^2}\,\theta(\muF-|\kperp'|)  \Bigg\}
~.\notag
\end{align}
The first term in curly brackets has come form the $\Ord(\alphaS)$ term in $Z_{ij}(z)$ while the second term came from the $F^{\LO}$ of\Equation{Eq:finlblcnst}.
These two terms cancel each-other up to $\Ord(x)$-corrections for $x\ll 1$.
The remaining first term will be cancelled against the second term in the next-to-last line of \Equation{eq:cross-check2} if we replace $F^{\LO}\to F^{\Ord(\alphaS^0)+\LO}$ there.

The new $O(\alphaS^0)$ term in \Equation{eq:UPDF+as0} suggests the (approximate) equality between the $\kperp$-integral of the UPDF and the momentum-density PDF.
This connection is further supported by the evolution equation for the integral of the UPDF (\ref{eq:evol-eqn-xT0-DGLAP}) derived above, which closely resembles the DGLAP equation for the momentum-density gluon PDF $xf_g(x,\muY^2)$. 

One can argue, that by including this $O(\alphaS^0)$ term into the UPDF we are effectively including the $\kperp$-non-factorizable contribution into the HEF computation. This is certainly true at LO, since the on-shell limit $|\kperp|\to 0$ of the off-shell matrix element of HEF coincides with the ordinary LO matrix element with the on-shell initial-state parton. However at NLO this question is more subtle and requires further analysis.

\section{\label{sec:conclusions}Conclusions and outlook}

In the present paper we have derived a scheme for NLO computations in HEF for arbitrary processes. It turned out, that the ambiguity of the separation between projectile and target contributions necessitates the introduction of a rapidity scale $\muY$, the evolution of UPDF with respect to which is similar to the CSS evolution. This result brings the notion of UPDF of HEF formalism closer to the notion of the TMD PDF in the standard TMD formalism. We have derived the matching formula between UPDF and collinear PDF at the NLO in $\alphaS$ and generalized it to all orders at the scale $\muY=|\kperp|$, thus providing a first-principle initial condition for the UPDF evolution. The BFKL-Collins-Ellis evolution of the Green's function in this initial condition is resumming the logarithms of partonic center of mass energy $\ln (1/x) \sim \ln (\hat{s}/\mu^2)$, which corresponds to the original formulation of HEF within CF~\mycite{Collins:1991ty,Catani:1990eg,Catani:1992rn,Catani:1994sq}.

There are several possible directions of future development. First of all we emphasize, that in the present paper we where mostly concerned about cancellation of IR/collinear divergences and by the removal of high-energy logarithms from the projectile part of the calculation. The finite part of the projectile contribution may still contain dangerous large logarithms, in particular $\ln^2 |\kperp|$ and $\ln |\kperp|$, taking care of which might require an ``optimal'' choice of $\muY$ or even the modification of the evolution equation for UPDF. We where also not able to determine the scale of $\alphaS$ in the kernels of evolution equations (\ref{eq:evol-eq-muY-fin}) and (\ref{eq:G-evol-0}). Doing this rigorously will require the continuation of our program to NNLO, but various heuristic arguments can be put forward, which we will discuss in future publications. Finally, staying at NLO in $\alphaS$ one can also continue our program beyond leading power of ``energy'' $\lambda$, \ie\ to go beyond the eikonal accuracy, which also aligns with the recent trends in low-$x$ physics~\mycite{Altinoluk:2024tyx,Altinoluk:2024dba,Borden:2024bxa,Chirilli:2020gol}.  The next-to-eikonal HEF will in particular include contributions with off-shell ``Reggeized'' quarks in the $t$-channel, which where studied in the past in context of LO calculations in the auxiliary-parton method~\mycite{vanHameren:2013csa} and Parton Reggeization Approach~\mycite{Nefedov:2012cq,Nefedov:2013ywa,Kniehl:2014qva, Nefedov:2015ara}. However, besides these quark-induced contributions, also the next-to-eikonal gluon-induced contributions in the HEF will appear, which where not taken into account in those studies even at LO in $\alphaS$.

\appendix

\section{\label{Sec:constants}Constants}
%%%%%%%%%%%%%%%%%%%%%%%%%%%%%%%%%%%%%%%%
\begin{equation}
\aeps = \frac{\alphaS}{2\pi}\,\oeps
\quad,\quad
\oeps = \frac{(4\pi)^\vepv}{\Gamma(1-\vepv)}
\quad,\quad
\piep = \frac{\pi^{1-\vepv}}{\Gamma(1-\vepv)}
\end{equation}
\begin{equation}
\ColorC{g} = 2\Nc
\quad,\quad
\ColorC{q} = 2\CF=\frac{\Nc^2-1}{\Nc}
\label{Eq:defCinlbl}
~.
\end{equation}
\begin{equation}
\gamma_g 
= \frac{\beta_0}{2}
= \frac{11\Nc}{6} - \frac{2T_Rn_f}{3}
\quad,\quad
\gamma_q = \frac{3\CF}{2}
\quad,\quad
\Kcusp = \Nc\bigg(\frac{67}{18}-\frac{\pi^2}{6}\bigg) -\frac{5n_f}{9}
~.
\end{equation}
%%%%%%%%%%%%%%%%%%%%%%%%%%%%%%%%%%%%%%%%
%%%%%%%%%%%%%%%%%%%%%%%%%%%%%%%%%%%%%%%%

\section{\label{Sec:splitting}Splitting functions}
%%%%%%%%%%%%%%%%%%%%%%%%%%%%%%%%%%%%%%%%
\begin{align}
\myScript{P}_{gg}^{\Real}(z) &= 2\Nc\bigg[\frac{1}{(1-z)_+}+\frac{1}{z} - 2 + z(1-z)\bigg]
%+\gamma_g\delta(1-z)
\hspace{1ex},\;\;\myScript{P}_{gg}(z) = \myScript{P}_{gg}^{\Real}(z) + \gamma_g\delta(1-z)
\\
%\myScript{P}_{qg}(z) &= 2T_R\bigg[\frac{1}{2}-\frac{z(1-z)}{1-\vepv}\bigg] \quad,\quad T_R=\frac{1}{2}
\myScript{P}_{qg}^{\Real}(z) &= 2T_R\bigg[\frac{1}{2}-z(1-z)\bigg] \quad,\quad T_R=\frac{1}{2}
\hspace{4.2ex},\;\;\myScript{P}_{qg}(z) = \myScript{P}_{qg}^{\Real}(z)
\\
%\myScript{P}_{qq}(z) &= 2\CF\bigg[\frac{1}{(1-z)_+} - \frac{1+\vepv+(1-\vepv)z}{2} \bigg]
\myScript{P}_{qq}^{\Real}(z) &= 2\CF\bigg[\frac{1}{(1-z)_+} - \frac{1+z}{2} \bigg]
%+\gamma_q\delta(1-z)
\hspace{11.7ex},\;\;\myScript{P}_{qq}(z) = \myScript{P}_{qq}^{\Real}(z) + \gamma_q\delta(1-z)
\\
%\myScript{P}_{gq}(z) &= 2\CF\bigg[\frac{1}{z} - \frac{2-(1-\vepv)z}{2} \bigg]
\myScript{P}_{gq}^{\Real}(z) &= 2\CF\bigg[\frac{1}{z} - \frac{2-z}{2} \bigg]
\hspace{18.5ex},\;\;\myScript{P}_{gq}(z) = \myScript{P}_{gq}^{\Real}(z)
\\
\myScript{P}_{\star}^{\Real}(z) &= 2\Nc\bigg[\frac{1}{(1-z)_+}+\frac{1}{z}\bigg]
\hspace{15.5ex},\;\;\myScript{P}_{\star}(z) = \myScript{P}_{\star}^{\Real}(z)
\end{align}
%%%%%%%%%%%%%%%%%%%%%%%%%%%%%%%%%%%%%%%%
%

\section{\label{Sec:convolutions}Convolutions}
For the plus-distribution:
%
%%%%%%%%%%%%%%%%%%%%%%%%%%%%%%%%%%%%%%%%
\begin{align}
\bigg[\bigg(\frac{g(Z)}{1-Z}\bigg)_{\!\!+}\otimes f\bigg](x) 
&\equiv\int_0^1dz\;\bigg(\frac{g(z)}{1-z}\bigg)_{\!\!+}\;\frac{1}{z}f\bigg(\frac{x}{z}\bigg)\,\theta(z>x)
\\
&\equiv\int_0^1dz\,\frac{g(z)}{1-z}\bigg[\frac{1}{z}f\bigg(\frac{x}{z}\bigg)\theta(z>x)-f(x)\bigg]
%\\
%&=\int_x^1\frac{dz}{1-z}\bigg[\frac{1}{z}f\bigg(\frac{x}{z}\bigg)-f(x)\bigg] 
%  + f(x)\ln(1-x)
%\\
%&=\int_x^1\frac{dz}{z}\,\frac{1}{1-z}\bigg[f\bigg(\frac{x}{z}\bigg)-f(x)\bigg] 
%  + f(x)\ln\frac{1-x}{x}
\\
&=\int_x^1dz\,g\bigg(\frac{x}{z}\bigg)\frac{f(z)-f(x)}{z-x}
  + f(x)\bigg[\int_0^xdz\,\frac{g(z)}{1-z}+\int_x^1dz\,\frac{g(z)}{z}\bigg]
~.
\end{align}
%%%%%%%%%%%%%%%%%%%%%%%%%%%%%%%%%%%%%%%%
For regular functions:
%%%%%%%%%%%%%%%%%%%%%%%%%%%%%%%%%%%%%%%%
\begin{align}
\big[g\otimes f\big](x) 
&\equiv\int_x^1\frac{dz}{z}\,g(z)\,f\bigg(\frac{x}{z}\bigg)
=\int_x^1\frac{dz}{z}\,g\bigg(\frac{x}{z}\bigg)\,f(z)
~.
\end{align}
%%%%%%%%%%%%%%%%%%%%%%%%%%%%%%%%%%%%%%%%%
%Rule for the change of integration order:
%%%%%%%%%%%%%%%%%%%%%%%%%%%%%%%%%%%%%%%%%
%\begin{align}
%\int_a^1dy\int_y^1\frac{dz}{z}\,F(y,z)
%%&=\int_0^1dy\int_y^1\frac{dz}{z}\,F(y,z)\,\theta(y>a)
%%=\int_0^1\frac{dz}{z}\int_0^zdy\,F(y,z)\,\theta(y>a)
%%\\
%%&=\int_0^1dz\int_0^1dy\,F(zy,z)\,\theta(zy>a)
%=\int_a^1dy\int_{a/y}^1dz\,F(zy,y)
%\end{align}
%%%%%%%%%%%%%%%%%%%%%%%%%%%%%%%%%%%%%%%%%
Integrated convolutions:
%%%%%%%%%%%%%%%%%%%%%%%%%%%%%%%%%%%%%%%%
\begin{align}
\int_{a}^1dy\big[g\otimes f\big](y)
%&=\int_{a}^1dy\int_y^1\frac{dz}{z}\,g\bigg(\frac{y}{z}\bigg)f(z)
=\int_{a}^1dy\,f(y)\int_{a/y}^1dz\,g(z)
~,
\end{align}
%%%%%%%%%%%%%%%%%%%%%%%%%%%%%%%%%%%%%%%%
%
and
%
%%%%%%%%%%%%%%%%%%%%%%%%%%%%%%%%%%%%%%%%
\begin{align}
\int_{a}^1dy\bigg[\frac{1}{(1-Z)_+}\otimes f\bigg](y)
&=\int_{a}^1dy
\int_0^1\frac{dz}{1-z}\bigg[\frac{1}{z}f\bigg(\frac{y}{z}\bigg)\theta(z>y)-f(y)\bigg]
\\&=
\int_0^1\frac{dz}{1-z}\bigg[\int_{a}^zdy\,\frac{1}{z}f\bigg(\frac{y}{z}\bigg)-\int_{a}^1dy\,f(y)\bigg]
\\&=
\int_0^1\frac{dz}{1-z}\bigg[\int_{a/z}^1dy\,f(y)-\int_{a}^1dy\,f(y)\bigg]
\\&=
-\int_{a}^1dy\,f(y)\int_0^1\frac{dz}{1-z}\,\theta(y<a/z)
\\
&=
\int_{a}^1dy\,f(y)\ln\bigg(1-\frac{a}{y}\bigg)
~.
\end{align}
%%%%%%%%%%%%%%%%%%%%%%%%%%%%%%%%%%%%%%%%
%
Useful relation: partial fractioning still works as usual for
%
%%%%%%%%%%%%%%%%%%%%%%%%%%%%%%%%%%%%%%%%
\begin{equation}
\int_0^1dz\bigg[\frac{1}{(1-z)_+} + \frac{1}{z}\bigg]\,f\bigg(\frac{x}{z}\bigg)\,\theta(z>x)
=
\int_0^1dz\,\frac{1}{(1-z)_+}\,\frac{1}{z}\,f\bigg(\frac{x}{z}\bigg)\,\theta(z>x)
~.
\end{equation}
%%%%%%%%%%%%%%%%%%%%%%%%%%%%%%%%%%%%%%%%

\section{\label{Sec:BFKLdetails}Multi-Regge limit details}
Starting from Appendix~D of \mycite{vanHameren:2022mtk}, the spinors of equations (D.12-14) change in such a way that $z_q\to1$ and $z_r\Lambda\to\Lambda^{\beta}$ for some positive $\beta<1$.
The spinor products of (D.15) then become
%
%%%%%%%%%%%%%%%%%%%%%%%%%%%%%%%%%%%%%%%%
\begin{align}
\langle q\bar{q}\rangle&=-\kappa^*_q & [\bar{q}q]&=\kappa_q
\notag\\
\langle r\bar{q}\rangle&= -\Lambda^{(1-\beta)/2}\kappa^*_r & [\bar{q}r]&=\Lambda^{(1-\beta)/2}\kappa_r
\label{Eq:BFKLdetail1}\\
\langle qr\rangle&=-\Lambda^{(1-\beta)/2}\kappa^*_r & [qr]&=-\Lambda^{(1-\beta)/2}\kappa_r
\notag
\end{align}
%%%%%%%%%%%%%%%%%%%%%%%%%%%%%%%%%%%%%%%%
%
Amplitudes surviving the large $\Lambda$ limit are those with the maximum possible power of $\Lambda$ in the numerator, and minimum power in the denominator, such that the overall power is $\Lambda^1$.
With some short considerations one can convince oneself that those are the same amplitudes, wether one applies the limits of (D.15) in \mycite{vanHameren:2022mtk} or the limits above.
For example for MHV amplitudes with an auxiliary quark pair, in both cases the radiative gluon momentum must not be in the numerator, since it would reduce the maximum possible power of $(\Lambda^{1/2})^4=\Lambda^2$.
Any amplitude has at least a $\Lambda^{1/2}$ in the denominator coming from the quarks.
Furthermore, any amplitude has at least 2 $r$-spinors in the denominator as $\langle ar\rangle\langle rb\rangle$ or with square brackets. 
The minimum power is $1\times\Lambda^{1/2}$ from \eg\ $\langle qr\rangle\langle rb\rangle$ in the limits of \mycite{vanHameren:2022mtk}, and $\Lambda^{(1-\beta)/2}\Lambda^{\beta/2}=\Lambda^{1/2}$ from the same combination here.

Continuing with the auxiliary quarks plus a radiative gluon, the formulas in Appendix~D.1 of \mycite{vanHameren:2022mtk} then all still hold with the substitutions $z_q,z_r\to1$ and
%
%%%%%%%%%%%%%%%%%%%%%%%%%%%%%%%%%%%%%%%%
\begin{align}
\langle r\bar{q}\rangle&\to -\kappa^*_r & [\bar{q}r]&\to\kappa_r
\\
\langle qr\rangle&\to-\kappa^*_r & [qr]&\to-\kappa_r
\notag
\end{align}
%%%%%%%%%%%%%%%%%%%%%%%%%%%%%%%%%%%%%%%%
%
We note that strictly speaking there is a mistake in equation (D.27), where $c_rz_r$ should be $c_r(1-z_q)$.
This can easily be inferred from the fact that the formulas (D.19-22) do not have a $z_r$ in the numerator.
The formula as presented is only correct keeping in mind that eventually $z_q=1-z_r$.
With this correction, the substitutions above lead directly to \Equation{Eq:BFKLME} in this write-up.

For the case of auxiliary gluons and a radiative gluon in Appendix D.2 of \mycite{vanHameren:2022mtk} the substitutions cannot be that straightforward, since the amplitude-level formulas have $z_r$ in the numerator.
One must carefully re-evaluate the color sum of the squared matrix element to arrive at the same conclusion.
Alternatively, one keeps \Equation{Eq:BFKLdetail1}, and substitutes $z_q\to1$ and $z_r\to\Lambda^{\beta-1}$, effectively taking $z_r\to0$ and the residue of the formulas in \mycite{vanHameren:2022mtk}.

\section{\label{App:impact-parameter}Derivation of $\muY$-evolution in impact-parameter space}
It is convenient to make a step back and restore the longitudinal integration in the real-emission part of \Equation{eq:evol-eqn-eps} by re-introducing the $\delta$-function:
\[
\aeps \frac{\Nc\mu^{2\vepv}}{\piep} \int\limits_{\xP}^1 dz \int\frac{d^{2-2\vepv}\rperp}{|\rperp|^2}
\,\hat{F}\bigg(\frac{\xP}{z},\kperp+\rperp; \muY\bigg) \delta \bigg( z- \frac{\muY}{\muY+|\rperp|} \bigg).
\]
Substituting the Fourier-transform (\ref{eq:F-xT-def}) in place of $\hat{F}$ one obtains:
\begin{eqnarray}
    \aeps \frac{\Nc\mu^{2\vepv}}{\piep} \int\limits_{\xP}^1 dz \int d^{2-2\vepv}\xperp 
\,\tilde{F}\bigg(\frac{\xP}{z},\xperp; \muY\bigg) \int \frac{d^{2-2\epsilon} \rperp}{|\rperp|^2}\,\delta \bigg( z- \frac{\muY}{\muY+|\rperp|} \bigg) e^{\imag \xperp (\kperp+\rperp)}
~. \label{eq:ev-eq-real-xT}
\end{eqnarray}
Let us consider the integral:
\begin{align}
  &  \frac{\mu^{2\vepv}}{\piep}  \int \frac{d^{2-2\epsilon} \rperp}{|\rperp|^2}\,\delta \bigg( z- \frac{\muY}{\muY+|\rperp|} \bigg) e^{\imag \xperp \rperp}
\nonumber\\& = 
\int\limits_0^\infty \frac{d|\rperp|}{\piep\rperp^2} \left( \frac{\mu^2}{\rperp^2} \right)^{\epsilon} \frac{\muY+|\rperp|}{z} \,\delta \bigg( |\rperp|-\muY \frac{1-z}{z} \bigg)
% \nonumber \\ & \times
\frac{2\pi^{1/2-\epsilon}}{\Gamma(1/2-\epsilon)} \int\limits_0^\pi d\theta\, \sin^{-2\epsilon} \theta \, e^{\imag|\xperp| |\rperp| \cos\theta}
\nonumber \\& = 
\frac{(2\pi)^{1-\epsilon}}{\piep} \int\limits_0^\infty \frac{d|\rperp|}{\rperp^2} \left( \frac{\mu^2}{\rperp^2} \right)^{\epsilon} \frac{\muY+|\rperp|}{z} \,\delta \bigg( |\rperp|-\muY \frac{1-z}{z} \bigg) \big(|\xperp| |\rperp|\big)^\epsilon \,J_{-\epsilon}\big(|\xperp| |\rperp|\big)
\nonumber \\ & = 
\frac{(2\pi)^{1-\epsilon}}{\piep} \left( \frac{\mu^2}{\muY^2} \right)^\epsilon \frac{1}{z^{1-2\epsilon} (1-z)^{1+2\epsilon}} \left[ \left( \muY |\xperp| \frac{1-z}{z}  \right)^{\epsilon} J_{-\epsilon} \left( \muY |\xperp| \frac{1-z}{z} \right) \right].
\end{align}
To expand this expression in $\epsilon$ we need to use the following well-known expansion of $(1-z)^{-1-2\epsilon}$ in terms of distributions in $z$:
\[
(1-z)^{-1-2\epsilon} = -\frac{1}{2\epsilon} \delta(1-z) + \frac{1}{(1-z)_+}+\Ord(\epsilon)
~,
\]
together with the expansion for the Bessel-function factor:
\[
x^\epsilon J_{-\epsilon}(x)=J_0(x) + \epsilon \left( J_0(x) \ln x  - \frac{\pi}{2} Y_0(x) \right)+\Ord(\epsilon^2)
~,
\]
where in fact, the only thing we need to know is the limit $\lim\limits_{x\to 0}  \left(  J_0(x) \ln x - \frac{\pi}{2} Y_0(x) \right) = \ln 2 -\gamma_E $, since the $\Ord(\epsilon)$-term gets multiplied by $\delta(1-z)/\epsilon$.
The final result for the expansion of the integral in question is:
\[
\frac{(2\pi)^{1-\epsilon}}{\piep} \left( \frac{\mu^2}{\muY^2} \right)^\epsilon \left[ -\left( \frac{1}{\epsilon} + \ln2-\gamma_E \right) \frac{\delta(1-z)}{2} + \frac{1}{z(1-z)_+} \,J_{0} \left( \muY |\xperp| \frac{1-z}{z} \right) + O(\epsilon)  \right]~.
\]
Substituting this result back to (\ref{eq:ev-eq-real-xT}) we find that the pole in $\epsilon$ cancels against the virtual part and the evolution equation in $\xperp$-space takes the form of \Equation{eq:evol-eqn-xT-final}.

\section{UPDF evolution in Mellin-$\xperp$ space}
In this appendix we study the UPDF $\muY$-evolution equation in $(N,\xperp)$-space instead of $(x,\kperp)$-space, which may be useful for the solution of the evolution equation, since in this space it reduces to the ordinary differential equation.
The relation between distributions in $(N,\xperp)$ and $(x,\kperp)$ space is:
\begin{equation}
    \hat{F}(x,\kperp,\mu_Y) = \int\frac{d^{2-2\vepv} \xperp}{(2\pi)^{1-\vepv}} \int\frac{dN}{2\pi\imag}\,x^{-N} e^{\imag\xperp \kperp} \tilde{F}(N,\xperp,\mu_Y). 
\end{equation}
Then, substituting $x^{-N}e^{\imag\xperp \kperp}$ to the real-emission part of the kernel of the Eq.~(\ref{eq:evol-eqn-eps}) we get:
\begin{align}
\mathrm{RHS}&\equiv
    \int \frac{d^{2-2\vepv} \rperp}{\piep\rperp^2} \left[x \left(1+\frac{|\rperp|}{\mu_Y} \right) \right]^{-N} e^{\imag\xperp (\kperp+\rperp)} 
\\&\hspace{0ex}= x^{-N} e^{\imag\xperp \kperp} \int \frac{d^{2-2\vepv} \rperp}{\piep\rperp^2} \left( 1+ \frac{|\rperp|}{\mu_Y}\right)^{-N} e^{\imag\xperp\rperp}
~.
\notag
\end{align}
To compute the integral over $\rperp$ in a convenient form, we use the ``Mellin-Barnes'' identity:
\[
\left( 1+ \frac{|\rperp|}{\mu_Y}\right)^{-N} = \int \frac{dz}{2\pi\imag} \frac{\Gamma(-z)\Gamma(N+z)}{\Gamma(N)} \left(\frac{|\rperp|}{\mu_Y} \right)^{z},
\]
where the contour in $z$-plane goes in between poles of $\Gamma(-z)$ and $\Gamma(z+N)$. After that, we just get the power $(\rperp^2)^{-1+z/2}$ under $\rperp$-integral and the Fourier-transform can be computed with the help of standard Fourier-transform of a power:
\[
\int \frac{d^{2-2\vepv} \rperp}{(\rperp^2)^{\alpha}}\,e^{\imag\xperp \rperp} = \frac{\pi^{1-\vepv} \Gamma(1-\alpha-\vepv)}{\Gamma(\alpha)} \left( \frac{\xperp^2}{4} \right)^{\alpha+\vepv-1},
\]
so we obtain:
\begin{align}
\mathrm{RHS}
&=
   x^{-N} e^{\imag\xperp \kperp} \int \frac{dz}{2\pi\imag}\,\frac{e^{-2\gamma_E \vepv} \Gamma(1-\vepv) \Gamma(-z) \Gamma(N+z) \Gamma(z/2-\vepv)}{\Gamma(N)\Gamma(1-z/2)}
\\&\hspace{32ex}\times
\left( \frac{\mu_Y^2 \xperp^2}{4} \right)^{-z/2} \left(\frac{\mu^2 \xperp^2}{4e^{-2\gamma_E \vepv}} \right)^{\vepv}
~.
\notag
\end{align}
It turns out that only poles with ${\rm Re} z<0$ lead to the convergent series, so we may think of $z$ as having negative real part. If we move the contour to the left of the pole at $z=2\vepv$, then we can safely take the limit $\vepv\to 0$ in the rest of the expression. It is also convenient to move the contour to the left of the pole at $z=-N$, to facilitate the study of $N\to 0$ behavior, so we rewrite the previous result as:
\begin{align}
\mathrm{RHS}
&=
x^{-N} e^{\imag\xperp \kperp} \bigg[ -\frac{1}{\vepv} + \ln \frac{\mu_Y^2}{\mu^2} -2\gamma_E -2\psi(N) - \frac{2}{N} \frac{\Gamma(1-N/2)}{\Gamma(1+N/2)} \bigg(\frac{\xperp^2 \mu_Y^2}{4} \bigg)^{N/2} 
\\&\hspace{46ex}
+ f\bigg(N,\frac{\xperp^2 \mu_Y^2}{4}\bigg) + O(\vepv) \bigg]
~,
\notag
\end{align}
where
\begin{equation}
    f(N,X)= \int\limits_{{\rm Re} z< -N} \frac{dz}{2\pi\imag}\,\frac{\Gamma(-z) \Gamma(N+z) \Gamma(z/2)}{\Gamma(N)\Gamma(1-z/2)}\,X^{-z/2}
~, 
\end{equation}
with the contour being located to the left of the point $z=-N$. We note, that both for $\xperp^2\to 0$ and $N\to 0$ the function $f(N,\xperp^2\mu_Y^2/4) \to 0$. Substituting this results back to Eq.~(\ref{eq:evol-eqn-eps}) we see that $1/\vepv$ cancels and we obtain:
\begin{align}
\frac{d}{d\ln\mu_Y^2} \tilde{F}(N,\xperp,\mu_Y) 
&=
\frac{\alphaS}{2\pi}\bigg[{-}2\Nc \left( \gamma_E + \psi(N)  \right) -\frac{2\Nc}{N} \frac{\Gamma(1-N/2)}{\Gamma(1+N/2)} \left(\frac{\xperp^2 \mu_Y^2}{4} \right)^{N/2} \label{eq:evol-N-xT}
\\&\hspace{36ex}
 + \Nc f\bigg(N,\frac{\xperp^2 \mu_Y^2}{4}\bigg) \bigg]\tilde{F}(N,\xperp,\mu_Y)
\notag~.
\end{align}
From this equation we can pull-out two limits:
\begin{itemize}
    \item $\xperp^2 \to 0$, corresponds to the evolution of integrated momentum-density PDF, since $\tilde{F}(x,\xperp=0)=xf_g(x)$. In this limit $f(N,\xperp^2\mu_Y^2/4) \to 0$ and $(\xperp^2 \mu_Y^2)^{N/2}\to 0$ if ${\rm Re} N>0$. Then we obtain:
    \begin{equation}
        \frac{d}{d\ln\mu_Y^2} \tilde{F}(N,0,\mu_Y) = \frac{\alphaS}{2\pi}\Big[{-}2\Nc \left( \gamma_E + \psi(N)  \right) \Big]\tilde{F}(N,0,\mu_Y)~.
    \end{equation}
\end{itemize}
The combination $-\gamma_E - \psi(N)$ is a Mellin transform of a plus-distribution:
\begin{equation}
    \int\limits_0^1 \frac{dz\, z^{N-1}}{(1-z)_+} = \int\limits_0^1 dz\,\frac{z^{N-1}-1}{1-z} = -\gamma_E-\psi(N)~,
\end{equation}
so we conclude that our {integrated} PDF evolves with the scale according to the splitting function:
\[
2\Nc \left[\frac{1}{z} + \frac{1}{(1-z)_+}\right],
\]
which is the same result as \Equation{eq:evol-eqn-xT0-DGLAP}.
\begin{itemize}
    \item $N\to 0$, in this limit poles $1/N^k$ correspond to $\ln^{k-1}(1/z)$ in momentum-fraction space, so it is related with high-energy resummation. For $N\to 0$ in \Equation{eq:evol-N-xT}  we have  $f(N,\xperp^2\mu_Y^2/4)$ $\to$ $0$, and the pole at $N\to 0$ cancels so one obtains:
    \begin{equation}
       \frac{d}{d\ln\mu_Y^2} \tilde{F}(N\to 0,\xperp,\mu_Y) = \frac{\alphaS}{2\pi}\left[{-}\Nc\ln \left(\frac{\mu_Y^2 \xperp^2}{4e^{-2\gamma_E}}\right) \right]\tilde{F}(N\to 0,\xperp,\mu_Y)~, 
    \end{equation}
    \ie\ we get the same CSS equation without any $1/N$-poles in the kernel.
\end{itemize}
The last result means that there is no double-counting of high-energy logarithms between resummations performed by the $\muY$-evolution (\ref{eq:evol-eq-muY-fin}) and the BFKL-Collins-Ellis evolution (\ref{eq:G-evol-0}).

\subsection*{Acknowledgments}
This work was supported by grant no.\ 2019/35/B/ST2/03531 of the Polish National Science Centre.
The work of MN had been supported by European Union's Horizon 2020
research and innovation programme under grant agreement No.~101065263
for the Marie Sk{\l}odowska-Curie action ``RadCor4HEF'', the  Binational Science Foundation grants \#2012124 and 
\#2021789,  by the ISF grant \#910/23, and  by  MSCA RISE 823947 “Heavy ion collisions: collectivity and precision in saturation physics” (HIEIC).
The authors acknowledge the stimulating environment from the collaboration of IFJPAN and IJClab.

%\bibliography{refs}{}\bibliographystyle{JHEP}
\providecommand{\href}[2]{#2}\begingroup\raggedright\endgroup

\begin{appendix}
\addtocontents{toc}{\protect\setcounter{tocdepth}{1}}
\end{appendix}

\end{document}